\documentclass[twocolumn,aps,prd,superscriptaddress,preprintnumbers,nofootinbib,10pt,floatfix]{revtex4-1}

\usepackage{amssymb}
\usepackage{amsmath}
\usepackage{amsfonts}
\usepackage{dsfont}
\usepackage{slashed}
\usepackage[colorlinks]{hyperref}
\usepackage{tablefootnote}
\usepackage{verbatim}
\usepackage{graphicx}
\usepackage{tikz,tikz-qtree}
\usepackage[compat=1.1.0]{tikz-feynman}
\usepackage{slashed}
\usepackage{ulem}

\begin{document}

\title{No planar degeneracy for the Landau gauge quark-gluon vertex}

\author{Georg Wieland}
\email{georg.wieland@uni-graz.at}
\affiliation{Institute of Physics, University of Graz, NAWI Graz, Universit\"atsplatz 5, 8010 Graz, Austria}
\author{Reinhard Alkofer}
\email{reinhard.alkofer@uni-graz.at}
\affiliation{Institute of Physics, University of Graz, NAWI Graz, Universit\"atsplatz 5, 8010 Graz, Austria}

\begin{abstract}
Based on a suitable basis system for the quark-gluon vertex' transverse tensor structures and on
carefully chosen kinematical variables, the transverse part of the quark-gluon vertex in quenched
QCD in the Landau gauge is obtained from a system of Dyson-Schwinger equations. 
We demonstrate by analysing this solution that the angular dependence of these transverse quark-gluon
vertex form factors is seemingly weak. We nevertheless argue that this does not imply a planar
degeneracy for this vertex because even this mild dependence cannot be neglected when aiming for
reasonably precise results for derived quantities.  Last but not least, 
for a self-consistently coupled systems of 3PI Dyson-Schwinger equations for the quark propagator and 
the quark-gluon vertex we confirm that  the core ingredient to dynamical 
chiral symmetry breaking is the dynamically generated tensor coupling of glue to quarks 
which itself is only possible because of chiral symmetry breaking. Furthermore, we find
(i) a relation in between the calculated chirality violating  vertex form factors;
(ii) that the quark propagator is identical within numerical errors when obtained either from a 
decoupling solution or the scaling solution for the Yang-Mills propagators and vertex functions; and 
(iii) that the resulting quark propagator is consistent with possessing poles only on the real time-like 
half-axis. 
Furthermore, we provide high-precision fits for
the form factors based on sometimes astonishingly simple model functions.

\end{abstract}

\maketitle

\section{Introduction}
\label{sec:intro}
The study of quantum chromodynamics (QCD) has become a very mature field, see, e.g., the review \cite{Gross:2022hyw}. Nevertheless, there are many aspects of QCD for which there is not yet a 
satisfactory understanding.  Certainly, dynamical chiral symmetry breaking (D$\chi$SB) belongs to this 
class of QCD's features. Although some of its characteristics have been unveiled already in the seventies 
\cite{Pagels:1977xv,Marciano:1977su,Pagels:1978ba,Gusynin:1978tr} its intrinsic non-perturbative nature 
prevented and still prevents a detailed insight into the underlying mechanisms.

With non-perturbative techniques required a well-suited set of approaches for the study of D$\chi$SB
is given by functional methods, for recent synopses of the application of these methods to QCD see 
\cite{Eichmann:2025wgs,Huber:2025cbd}.
The respective central elements are the QCD correlation functions, in particular, the primitively divergent 
ones. As the Landau gauge offers, besides being Poincar\'e-invariant, the technical advantage that only
the tensor structures being transverse to gluon momenta need to be considered\footnote{ 
See, e.g., \cite{Braun:2025gvq,Eichmann:2026ttr} and references therein.}
most respective studies have been performed in this gauge. 

Amongst QCD's primitively divergent  correlation function it is the quark-gluon vertex (QGV) which provides
the link in between the Yang-Mills sector and the matter sector of QCD. Having this special role 
it comes with no surprise that it has been the focus of many studies, see, e.g., refs.\
\cite{Skullerud:1995uk,Skullerud:1997wc,Skullerud:2001bu,Skullerud:2002sk, Skullerud:2004pt,Lin:2005zd,Kizilersu:2006et,Furui:2008zm,Oliveira:2016muq,Sternbeck:2017ntv,Kizilersu:2021jen,Skullerud:2021pel,Marques:2023cmi,Davydychev:2000rt,Davydychev:2000yy,Gracey:2011vw,Huang:2020fiu,vonSmekal:1990kgh,Bender:2002as,Watson:2004kd,Bhagwat:2004kj,Holl:2004qn,Fischer:2004ym,Llanes-Estrada:2004hnb,Alkofer:2006gz,Matevosyan:2006bk,Matevosyan:2007cx,Alkofer:2008et,Alkofer:2008tt,He:2009sj,Fischer:2009jm,Windisch:2012de,Hopfer:2012cnq,Aguilar:2012pfj,Ayala:2012pb,Rojas:2013tza,Alkofer:2013qoc,Williams:2014iea,Aguilar:2014lha,Rojas:2014tya,Ayala:2014uua,Mitter:2014wpa,Chen:2015mda,Pelaez:2015tba,Binosi:2016wcx,Williams:2016zpc,Blum:2015lsa,Blum:2016fib,Gomez-Rocha:2014vsa,Gomez-Rocha:2015qga,Gomez-Rocha:2016cji,Bermudez:2017bpx,Cyrol:2017ewj,Contant:2018zpi,Oliveira:2018fkj,Oliveira:2018ukh,Sultan:2018tet,Vujinovic:2018nko,Oliveira:2020yac,Gao:2021wun,El-Bennich:2022obe,Aguilar:2023mam,Aguilar:2024ciu,Alkofer:2023lrl,Alkofer:2023syz,Guzman:2025qbq,Fu:2025hcm,Ferreira:2025wpu,Ferreira:2026gbe}.

Besides the QGV's special role in D$\chi$SB, see, e.g., the discussion in \cite{Alkofer:2023lrl}, there are indications that the chiral-symmetry-breaking tensor structures of the QGV change the 
analytic properties of the solution to the respective quark propagator's Dyson-Schwinger equation (DSE), see, e.g., 
\cite{Alkofer:2003jj,Pawlowski:2024kxc}. And, as a matter of fact, we will not only present here respective results based 
on highly precise solutions for the QGV but also connect the change in the quark propagator's analytic 
properties to modifications in the solution for the dynamically generated quark mass function for Euclidean momenta upon taking the full QGV into account. Hereby, including the full kinematical dependence of the 
QGV is essential. The surprisingly large effect of a seemingly weak angular dependence will be
demonstrated clearly herein.

Keeping the QGV's full kinematical dependencies requires a substantial numerical effort. In addition,
very accurate input in form of the gluon propagator and the three-gluon vertex function is needed. 
Therefore, the herein presented study restricts to the quenched approximation, and the respective
input for the Yang-Mills sector is taken from s\cite{MQHgithub}
which is based on ref.\ \cite{Huber:2020keu}, see also \cite{Huber:2018ned}. 
The main reason for choosing this source as input is given by the fact that these results for the Yang-Mills 
correlation functions are based on the solutions for the most sophisticated truncation of DSEs available so 
far, and that they lead to a very precise determination of the glueball spectrum
 \cite{Huber:2020ngt,Huber:2021yfy,Huber:2025kwy}. 

An interesting point to note hereby relates to the fact that there is not only one unique solution to the Yang-Mills DSEs but one obtains several solutions which differ by the behaviour of the correlation functions in the 
deep infrared \cite{Fischer:2008uz,Alkofer:2008jy}. There is one specific solution in which the correlation functions behave power-like in the infrared, the so-called scaling solution, as found and discussed in refs.\ 
\cite{vonSmekal:1997ohs,vonSmekal:1997ern,Zwanziger:2001kw,Lerche:2002ep,Fischer:2002hna,Fischer:2005ui,Huber:2007kc}. In addition, there is an one-parameter family of solutions characterised by infrared 
finite values for the gluon propagator and the ghost renormalisation function, see, e.g.,  \cite{Aguilar:2008xm,Aguilar:2015bud,Eichmann:2021zuv}. 
Following \cite{Fischer:2008uz} we will denote them here as decoupling solutions. 
A potential explanation for the appearance of several solutions might be that in this context some further 
non-perturbative gauge fixing is required \cite{Maas:2009se}.
This hypothesis is now further substantiated by the fact that the glueball spectrum is independent of the type
of solution for the Yang-Mills correlation functions as long as a consistent set is taken into account
 \cite{Huber:2020ngt,Huber:2021yfy,Huber:2025kwy}. Of course, such an independence should apply to 
all gauge-independent physical states, cf.\ the discussion in \cite{Blank:2010pa}.

As the Yang-Mills sector and the matter sector are independently gauge invariant one may even speculate 
whether the quark $n$-point functions are independent of the scaling vs.\ decoupling solution dichotomy. 
As a first step investigating this question we will present and compare results for the quark propagator 
obtained from the Yang-Mills scaling and two different decoupling solutions.

This paper is organised as follows: In Sec.\ \ref{sec:dses} we introduce and discuss the system of coupled 
DSEs for the quark propagator and the QGV to be solved. Furthermore, we present the key features of 
the three sets of input for the Yang-Mills correlation functions taken from ref.\ \cite{Huber:2020keu}. 
In Sec.\ \ref{sec:kin} we define the kinematical variables to be used as arguments of the QGV and provide a 
rationale for choosing exactly these variables. As this is very important for the further analysis we 
discuss in detail the employed tensor basis including its properties w.r.t.\ D$\chi$SB, tensorial 
classification and dimension of the respective form factors. In Sec.\ \ref{sec:numres} we present our numerical 
results as well as some of their consequential facts about the quark propagator and the QGV. In Sec.\ \ref{sec:Conclusion} we close with a summary and some concluding remarks. We also offer an outlook 
on some potential applications of the here presented features of the QGV.

\section{System of Dyson-Schwinger equations}
\label{sec:dses}

All expressions in the following are understood to be formulated in
Euclidean momentum space, {\it i.e.}, after a Wick rotation.

\begin{figure}[b]
\centering
\begin{equation*}
\begin{split}
        \begin{tikzpicture}[baseline=(a)]
        \begin{feynman}
            \vertex [large, dot, fill=black] (a) at (0.2,0) {};
            \vertex (b)  at (-1,0);
            \vertex (c) at (1,0);
            \diagram* {
                (b) -- [fermion] (c),
            };
            \vertex[large, dot, label={\small $p$}] at (0, -0.6);
        \end{feynman}
    \end{tikzpicture}^{-1}
        \ = \  \hspace{40mm} ~
\end{split}
\end{equation*}
\begin{equation*}
\begin{split}
        \begin{tikzpicture}[baseline=(a)]
        \begin{feynman}
            \vertex (b)  at (-1,0);
            \vertex (c) at (1,0);
            \diagram* {
                (b) -- [fermion] (c),
            };
            \vertex[large, dot, label={\small $p$}] at (0, -0.6);
        \end{feynman}
    \end{tikzpicture}^{-1}
    \ - \
    \begin{tikzpicture}[baseline=(a)]
        \begin{feynman}
            \vertex [small, dot, fill=black] (a1) at (0,0) {};
            \vertex [large, dot, fill=black] (a2) at (2,0) {};
            \vertex [large, dot, fill=black] at (1.0,1.04){};
            \vertex [large, dot, fill=black] at (1.2,0) {};
            \vertex [large, dot] (a22) at (2,0);
            \vertex (b)  at (-1,0);
            \vertex (c) at (3,0);
            \diagram* {
                (b) -- [fermion] (a1), (a1) -- [fermion] (a2), (a22) -- [fermion] (c),
            };
            \draw[gluon] (a22) arc [start angle=0, end angle=180, radius=0.99cm];
            \vertex[large, dot, label={\small $p$}] at (-0.55, -0.6);
            \vertex[large, dot, label={\small $p$}] at (2.55, -0.6);
            \vertex[large, dot, label={\small $q$}] at (1, -0.6);
            \vertex[large, dot, label={\small $k$}] at (1, 1.1);
        \end{feynman}
    \end{tikzpicture}  \\ \\
\end{split}
\end{equation*}
\caption{DSE for the quark propagator including one bare (small black dot) and one proper, i.e., fully dressed (large black dot), quark-gluon vertex.
The large black dots on the lines denote fully dressed propagators. 
\label{quarkdse0}}
\end{figure}
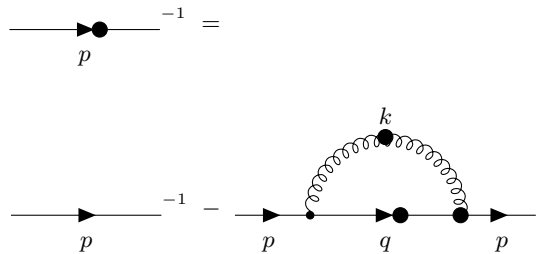

With given input from the Yang-Mills sector we will solve a set of coupled equations for the quark propagator
and the QGV. 
The DSE for the quark propagator,
\begin{align}
S^{-1}(p) & = \nonumber  \\
S_0^{-1}(p) &+ Z_{1F} C_F g^2 \int_q D^{\mu\nu}(k) \gamma^\mu S(q) \Gamma^\nu(k;q,-p) \, ,
\label{eq:quarkDSE}
\end{align}
is displayed in fig.~\ref{quarkdse0}, it involves one bare and one dressed QGV. 
Herein, $S(p)$ denotes the quark propagator for quark momentum $p$ and $D_{\mu\nu}(k=p-q)$ the gluon propagator. The abbreviation $\int_q$ stands for $\int \text{d}^4 q / (2\pi)^4$.
$S_0(p)$ is the bare quark propagator, $Z_{1F}$ the corresponding
multiplicative renormalisation constant, and $C_F$ the
Casimir of the fundamental representation taking the value $4/3$ for SU(3).


In eq.\ \eqref{eq:quarkDSE} the coupling and the generator of the fundamental colour representation,
$t^a_{ij}$, have been already factored out from the dressed quark-gluon vertex 
\begin{equation} \label{ qqgdressed}
\Gamma^{a, ij}_\mu (k; q,-p) = i g t^a_{ij} \Gamma_\mu(k;q,-p) \, .
\end{equation}
~

The dressed quark propagator is parametrised by two scalar functions which are conventionally chosen
as $A(p^2)$ and $B(p^2)$ or as $Z_f(p^2)$ and $M(p^2)$:
\begin{equation}        
        S(p) = \frac{-iA(p^2)\slashed{p} + B(p^2)}{p^2 A^2(p^2) + B^2(p^2)} = Z_f(p^2) \frac{-i\slashed{p} +M(p^2)}{p^2 + M^2(p^2)}.       
\end{equation}
Hereby,  $Z_f(p^2) = 1 / A(p^2)$ is the quark renormalisation function and 
$M(p^2) = B(p^2)/A(p^2)$ represents the quark mass function. 
In the chiral limit, i.e., for vanishing current masses $m=0$, the quark mass function is a 
renormalisation group (RG) invariant quantity.
The corresponding tree-level quantities are proportional to renormalisation constants,
 $A^{(tl)}(p^2)=Z_2$ and $B^{(tl)}(p^2) = Z_2 m_{\text{bare}} = Z_2 Z_m m_r$
where  $m_{\text{bare}}$ and $m_r$ denote the bare and the renormalised mass, respectively.
For further details w.r.t. the definition of the renormalisation constants we refer to, e.g., 
\cite{Alkofer:2000wg,Fischer:2003rp,Fischer:2003zc}.

In fig.\  \ref{fig:QGV_DSE} we depict the functional equation for the QGV in 3PI -- 3-loop truncation.
The rationale for choosing this form of the equation for the QGV can be found in refs.\ \cite{Alkofer:2008tt,Windisch:2014lce,Hopfer:2014szm,Sanchis-Alepuz:2015tha}, see also, e.g., refs.~\cite{Williams:2015cvx,Eichmann:2016yit,MQHunpublished} for the use of the 3PI -- 3-loop truncation in bound state studies.
The first term on the r.h.s\ of the QGV equation is given by the tree-level contribution to this vertex, $ {\Gamma}^{(0), a, ij}_\mu = Z_{1F} \,  ig t^{a,ij}\gamma_\mu$. 
Following the usual nomenclature we will denote the first one-loop diagram, i.e., the one with the 
three-gluon vertex, as non-Abelian diagram, and the one with three dressed QGVs Abelian diagram. Note that the pre-factors of
these two diagrams originating from the colour algebra come with opposite sign, and the Abelian diagram
is suppressed by a factor $1/N_c^2=1/9$ as compared to the non-Abelian one. Additionally, the Abelian diagram is in the non-perturbative region further suppressed dynamically 
\cite{Alkofer:2008tt,Windisch:2014lce,Hopfer:2014szm,Aguilar:2024ciu}.

\begin{widetext}

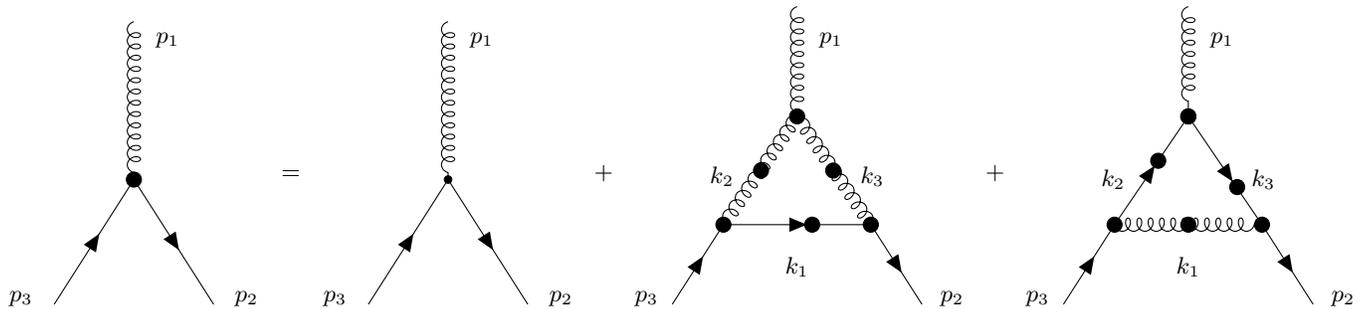
\begin{figure}[h!]
\centering
\begin{equation*}
\begin{split}
        \begin{tikzpicture}[baseline=(a)]
        \begin{feynman}
            \vertex [large, dot, black] (a) at (0,0);
            \vertex [large, dot, fill=black] (a1) at (0,0) {};
            \vertex (b)  at (0,2.1);
            \vertex (c) at (1.06,-1.66);
            \vertex (d) at (-1.06,-1.66);
            \diagram* {
                (b) -- [gluon] (a), (a) -- [fermion] (c), (d) -- [fermion] (a),
            };
            \vertex[large, dot, label={\small $p_1$}] at (0.45, 1.64);
            \vertex[large, dot, label={\small $p_3$}] at (-1.5, -1.8);
            \vertex[large, dot, label={\small $p_2$}] at (1.5, -1.8);
        \end{feynman}
    \end{tikzpicture}
    \ = \
    \begin{tikzpicture}[baseline=(a)]
        \begin{feynman}
            \vertex [small, dot, black] (a) at (0,0);
            \vertex [small, dot, fill=black] (a1) at (0,0) {};
            \vertex (b)  at (0,2.1);
            \vertex (c) at (1.06,-1.66);
            \vertex (d) at (-1.06,-1.66);
            \diagram* {
                (b) -- [gluon] (a), (a) -- [fermion] (c), (d) -- [fermion] (a),
            };
            \vertex[large, dot, label={\small $p_1$}] at (0.45, 1.64);
            \vertex[large, dot, label={\small $p_3$}] at (-1.5, -1.8);
            \vertex[large, dot, label={\small $p_2$}] at (1.5, -1.8);
        \end{feynman}
    \end{tikzpicture}
    \ + \
    \begin{tikzpicture}[baseline=(a)]
        \begin{feynman}
            \vertex [small, dot] (a) at (0,0);
            \vertex [large, dot, black] (a1) at (0,0.84);
            \vertex [large, dot, black] (a2) at (0.98,-0.6);
            \vertex [large, dot, black] (a3) at (-0.98,-0.6);
            \vertex [large, dot, fill=black] (a11) at (0,0.84) {};
            \vertex [large, dot, fill=black] (a21) at (0.98,-0.6) {};
            \vertex [large, dot, fill=black] (a31) at (-0.98,-0.6) {};
            \vertex [large, dot, fill=black] (z) at (0.48,0.12) {};
            \vertex [large, dot, fill=black] (zz) at (-0.48,0.12) {};
            \vertex [large, dot, fill=black] (zzz) at (0.2,-0.6) {};
            \vertex (b) at (0,2.3);
            \vertex (c) at (1.66,-1.66);
            \vertex (d) at (-1.66,-1.66);
            \diagram* {
                (b) -- [gluon] (a1), (a2) -- [fermion] (c),
                (a3) -- [fermion] (a2), (d) -- [fermion] (a3), (a2) -- [gluon] (a1), (a1) -- [gluon] (a3),
            };
            \vertex[large, dot, label={\small $p_1$}] at (0.45, 1.64);
            \vertex[large, dot, label={\small $p_3$}] at (-2, -1.8);
            \vertex[large, dot, label={\small $p_2$}] at (2.06, -1.8);
            \vertex[large, dot, label={\small $k_1$}] at (0, -1.4);
            \vertex[large, dot, label={\small $k_3$}] at (1, -0.2);
            \vertex[large, dot, label={\small $k_2$}] at (-1, -0.2);
        \end{feynman}
    \end{tikzpicture}
    \ + \
    \begin{tikzpicture}[baseline=(a)]
        \begin{feynman}
            \vertex [small, dot] (a) at (0,0);
            \vertex [large, dot, black] (a1) at (0,0.84);
            \vertex [large, dot, black] (a2) at (0.94,-0.6);
            \vertex [large, dot, black] (a3) at (-0.94,-0.6);
            \vertex [large, dot, fill=black] (a11) at (0,0.84) {};
            \vertex [large, dot, black] (a21) at (0.98,-0.6);
            \vertex [small, dot, black] (a31) at (-0.98,-0.6);
            \vertex [large, dot, fill=black] (a1) at (0,0.84) {};
            \vertex [large, dot, fill=black] (a2) at (0.98,-0.6) {};
            \vertex [large, dot, fill=black] (a3) at (-0.98,-0.6) {};
            \vertex [large, dot, fill=black] (z) at (0.65,-0.1) {};
            \vertex [large, dot, fill=black] (zz) at (-0.4,0.25) {};
            \vertex [large, dot, fill=black] (zzz) at (0,-0.6) {};
            \vertex (b) at (0,2.3);
            \vertex (c) at (1.66,-1.66);
            \vertex (d) at (-1.66,-1.66);
            \diagram* {
                (b) -- [gluon] (a11), (a21) -- [fermion] (c),
                (a31) -- [gluon] (a21), (d) -- [fermion] (a31), (a11) -- [fermion] (a21), (a31) -- [fermion] (a11),
            };
            \vertex[large, dot, label={\small $p_1$}] at (0.45, 1.64);
            \vertex[large, dot, label={\small $p_3$}] at (-2, -1.8);
            \vertex[large, dot, label={\small $p_2$}] at (2.06, -1.8);
            \vertex[large, dot, label={\small$k_1$}] at (0, -1.4);
            \vertex[large, dot, label={\small $k_3$}] at (1, -0.24);
            \vertex[large, dot, label={\small $k_2$}] at (-1, -0.24);
        \end{feynman}
    \end{tikzpicture} 
\end{split}
\end{equation*}
\caption{Diagrammatic representation of the equation for the
 quark-gluon vertex including the non-Abelian and the Abelian diagram.}
\label{fig:QGV_DSE}
\end{figure}
\end{widetext}

Due to this, the results presented below will either take only the non-Abelian diagram into account, 
and the quark propagator DSE is solved consistently together with the QGV equation with non-Abelian diagram only, or we add to the QGV obtained in this way the result for the Abelian diagram. Phrased
otherwise, in the second version the QGV appearing three times in the Abelian diagram is the one determined from the non-Abelian diagram.\footnote{As we will argue below it is 
necessary to keep the full momentum dependence of the QGV. This makes the numerical 
evaluation of the Abelian diagram extremely expensive. As our second method provides further evidence
that the Abelian diagram provides a small contribution we leave a self-consistent and CPU-intensive 
implementation of the Abelian diagram for further work.}

As already stated, for the gluon propagator and the three-gluon vertex as well as for the respective
values of the renormalisation constants the results of ref.\ \cite{Huber:2020keu} are 
employed.\footnote{The employed input has been cross-checked with our own solution of the coupled DSEs in the 
Yang-Mills sector which, however, has been obtained in a less sophisticated truncation.  }
These renormalisation constants are then also used to determine the value of $Z_{1F}$ via the relations 
$Z_{1F}= Z_2 / \widetilde Z_3 = Z_2Z_1/Z_3$. 

As has been done in ref.\ \cite{Huber:2020keu} scale setting is performed by fixing the position of the maximum of the gluon 
propagator to lattice data.

For the three-gluon vertex only the tree-level tensor structure is taken into account, and we furthermore
exploit the planar degeneracy of the three-gluon vertex by approximating the corresponding dressing function to depend only on the symmetric variable $\mathcal{S}_0 = (p_1^2+p_2^2+p_3^2)/6$:
\begin{equation}
        \Gamma^{abc}_{\mu \nu \rho} (p_1,p_2,p_3) \to 
i g f^{abc} \, \eta \, \Gamma^{\mathrm{eff}}(\mathcal{S}_0) \, T^{\mathrm{tl}}_{\mu \nu \rho}(p_1,p_2,p_3) \, .
\end{equation}
It turned out to be necessary to enhance the three-gluon vertex by a factor $\eta$ in order to obtain within
the used quenched approximation and further simplifications
D$\chi$SB. For the results to be presented we used $\eta=1.5$ which
 provided a quark mass function of the expected numerical values. As we are here mostly interested 
in the characteristic features of the quark propagator and the QGV we consider this an acceptable 
procedure.\footnote{It is left to a future investigation whether a more accurate representation of the 
three-gluon vertex or going beyond the quenched approximation or both will remove the need for this
work-around. It should be noted that the results obtained here provide additional guidance in how to implement either of these further steps.}

\section{Quark-gluon vertex: Kinematical variables and tensor basis}
\label{sec:kin}

\subsection{Momentum dependence}

\begin{figure}
\begin{center}
$\Gamma^{a, ij}_{\mu , \alpha \beta} (k; q,-p) =$
 \begin{tikzpicture}[baseline=(a)]
        \begin{feynman}
            \vertex [large, dot, black] (a) at (0,0);
            \vertex [large, dot, fill=black] (a1) at (0,0) {};
            \vertex (b)  at (0,2.1);
            \vertex (c) at (1.06,-1.66);
            \vertex (d) at (-1.06,-1.66);
            \diagram* {
                (b) -- [gluon] (a), (a) -- [fermion] (c), (d) -- [fermion] (a),
            };
            \vertex[large, dot, label={\scriptsize $k$}] at (0.9, 1);
            \vertex[large, dot, label={\scriptsize $\mu, a$}] at (0.86, 0.56);
            \vertex[large, dot, label={\scriptsize $q$}] at (-1, -1.1);
            \vertex[large, dot, label={\scriptsize $p$}] at (1, -1.1);
            \vertex[large, dot, label={\scriptsize $\alpha$}] at (-1.4, -1.96);
            \vertex[large, dot, label={\scriptsize $\beta$}] at (1.4, -1.96);
                        \vertex[large, dot, label={\scriptsize $i$}] at (-1.4, -2.56);      
            \vertex[large, dot, label={\scriptsize $j$}] at (1.4, -2.56);
        \end{feynman}
    \end{tikzpicture}
\end{center}
\caption{\label{fig:QGV}
Pictorial representation of the quark-gluon vertex. The momentum dependence as well as 
the discrete  indices (colour and Lorentz/Dirac) are ordered such
that one starts at the top and goes from there anti-clockwise (mathematically positive). 
All momenta, $k$, $q$ and $-p$,  are counted as incoming.}
\end{figure}
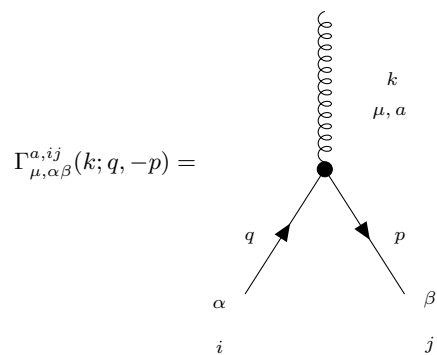

The quark-gluon vertex for a gluon (with colour index $a$ and Lorentz index $\mu$) and for 
quarks (with colour indices $i$ and $j$ as well as Dirac indices $\alpha$ and $\beta$), ~
${\Gamma}^{a, ij}_{\mu, \, \alpha \beta} (k;q, -p) = i g t^a_{ij} \Gamma_{\mu, \, \alpha \beta} (k;q,-p) $,
is depicted in Fig.~\ref{fig:QGV}. Momentum conservation relates the
gluon momentum $k$ to the quark momenta $q$ and $p$, $k=p-q$. Correspondingly, this three-point
function depends on three Lorentz invariant combinations of the momenta. As stated in the introduction
we want to investigate whether the quark-gluon vertex displays a planar degeneracy similar to the one
found for the full three-gluon vertex.

In order to minimise the dependence on an angular variable we want to exploit the symmetry 
between the two quark lines and introduce first the average quark momentum $\bar p = \frac 1 2 (q+p)$
and then  $w = \cos \left( \angle (k, \bar{p})\right)$, the cosine of the angle between $k$ and $\bar{p}$.
The quark-gluon vertex is in the following treated as function of the Lorentz invariant quantities 
$k^2$, $\bar p^2$ and $w= k \cdot \bar p / \sqrt{k^2 \bar p^2}$. This choice can be better understood 
by looking at the charge conjugation properties of the vertex. From the transformation rules of the 
tree-level vertex one can deduce the effect of charge conjugation on the fully dressed vertex,
\begin{equation}
        C^{-1} \Gamma_\mu (k;q,-p)C = - \Bigl( \Gamma_\mu (k;-q,p) \Bigr)^T \, ,
\end{equation}
where $C$ is the conventional charge conjugation matrix.\footnote{We use Hermitean Dirac matrices, 
$\left(\gamma_\mu\right)^\dagger = \gamma_\mu$, which fulfil the Clifford algebra,
$        \{ \gamma_\mu, \gamma_\nu \} = 2 \delta_{\mu \nu} $. The charge conjugation matrix is then given 
by $C=\gamma^4 \gamma^2$.
Hence, transposition and complex  conjugation coincide.
}
Therefore, under charge conjugation the sign of $\bar p$ flips, i.e., $\bar p \to - \bar p$,
which, under the condition that the tensor basis transforms correctly under charge conjugation,
 implies for the dressing functions
\begin{equation}
        \Gamma^{(i)} (k^2, \bar p^2, w) =  \Gamma^{(i)} (k^2,\bar p^2, -w)  \, ,
\label{ChConj}
\end{equation}
see below.

\subsection{Constructing the basis}

The quark-gluon vertex is a Dirac-matrix valued four-vector with a total of 64 components. 
As not all of these are independent it is useful to span this vertex over an appropriate basis. 
Of course, this basis is ambiguous, and its precise choice is guided on the one hand by practicality 
and on the other hand by amenability to physical interpretation, see, e.g., \cite{Eichmann:2026ttr} and references therein. 
Furthermore, it enables the representation of the vertex in terms of a set of dressing functions, resp.,
form factors, which then in turn describe, as a function of three Lorentz-invariant kinematical variables,
the weight of the related tensor structures. In order to facilitate comparisons in between different calculations general accepted conventions for the QGV's basis are desirable, and corresponding efforts have been undertaken in the past. Herein we want to proceed on this task and construct systematically, i.e., based on some ordering principles, a basis which is as unique as possible.\footnote{This systematic ordering of basis elements has been devised together with Markus Q.\ Huber to allow for a simpler comparison of our results
to the ones reported in \cite{MQHunpublished}.
Furthermore, in Appendix \ref{app:basis} we provide a translation table of our resulting basis to the respective tensors described in the recent review \cite{Eichmann:2026ttr}.}

The fully dressed quark-gluon vertex can be decomposed into twelve tensor structures, eight purely
transverse tensors ${R}_\mu^{(i)}$ and four so-called ``longitudinal'' tensors 
 that 
are constrained by corresponding Slavnov-Taylor identities. As we are working in Landau gauge, in the
respective decomposition
only the eight transverse tensor structures are needed. Correspondingly, 
the transversely projected QGV  is of the form
\begin{equation}
 \Gamma^{T}_\mu (k; q,-p) = \sum_{i=1}^8 {\Gamma}^{(i)} (k^2, \bar p^2, w)  \, {R}^{(i)}_\mu (k; \bar{p}) 
\, .
\end{equation}

To define a suitable basis, i.e., to choose the tensors ${R}^{(i)}_\mu (k; \bar{p})$, $i=1,\ldots 8$, and order them according to some principles, we will exploit two different properties of them. 

First, let us recall the
properties of these tensors w.r.t\ to chiral symmetry. As a chiral transformation of the QGV 
with parameter $\alpha$ is given by acting with 
$e^{-i\alpha \gamma^5}  $
from the left and the right a chirally symmetric QGV fulfils the relation $\{\gamma^5,\Gamma^\mu\}=0$
analogously to a chirally symmetric  quark propagator for which one also has $ \{\gamma^5,S\}=0$.
 D$\chi$SB will lead to a violation of the vanishing of these anti-commutators, and for the quark
propagator, as discussed in the previous section, a non-vanishing scalar part signalled by $B(p^2)\not= 0$
emerges. 

As for the QGV, all tensor structures with an \emph{odd} number of Dirac matrices will have 
a vanishing anti-commutator with $\gamma^5$ and are thus chirally symmetric.  
Those tensor structures with an \emph{even} number of 
gamma matrices can only be present when chiral symmetry is broken.

It is important to note that the chirally symmetric ($\chi$S) and the chirality violating ($\chi$V)
parts of the QGV act in a 
different way in the quark propagator DSE when solving the equation for the quark propagator and
the QGV self-consistently. As it is evident from the structure of the quark propagator DSE, and as also demonstrated below on the basis of the obtained numerical results, the QGV's $\chi$S parts contribute
the major part to the $\chi$S quark propagator dressing function of $A(p^2)$. On the other 
hand, the scalar dressing function $B(p^2)$ originates mostly from the QGV's $\chi$V parts. Phrased 
otherwise: The QGV's $\chi$V tensor structures are driving in a highly self-consistent manner D$\chi$SB.
This motivates our choice to group the $\chi$S and $\chi$V structures in the ordering of the basis.

The basis can be constructed from the following elements:
\begin{equation}
        \begin{pmatrix}
                \mathds{1} \\ \slashed{k} \\ \slashed{\bar{p}} \\ \slashed{k}\slashed{\bar{p}}
        \end{pmatrix}
        \otimes
        \begin{pmatrix}
                \gamma_\mu \\ k_\mu \\ \bar{p}_\mu
        \end{pmatrix} .
\end{equation}
This leads to twelve tensor structures, eight of which are transverse to the gluon momentum. Six are 
$\chi$S and six $\chi$V, resp., out of the transverse tensors four are $\chi$S and four $\chi$V. 
Additionally, we require good charge conjugation properties such that the corresponding dressing 
functions fulfil \eqref{ChConj}.

For the presentation of the transverse tensor structures it is helpful to define the transverse projector
\begin{equation}
\mathcal{T}^{\mu\nu}_k = \delta^{\mu \nu} - k^\mu k^\nu / k^2 \, .
\end{equation}
Furthermore, besides the the conventional commutator we define $\left[ a,b,c \right] = \left[ a, b \right] c + \left[ b, c \right] a + \left[ c, a \right] b$. 

The four $\chi$S tensors we are using are given by
\begin{align}
\label{T1}
{R}^{(1),\mu } (k; \bar{p}) &= \mathcal{T}^{\mu\nu}_k \gamma_\nu \,
,\\
\label{T2}
{R}^{(2),\mu } (k; \bar{p}) &= \frac 1 6 \left[ \gamma^\mu, \slashed{\bar p}, \slashed{k} \right]
 \, , \\
\label{T3}
{R}^{(3),\mu } (k; \bar{p}) &= \mathcal{T}^{\mu\nu}_k\bar p_\nu  \slashed{\bar p} \, , \\
\label{T4}
{R}^{(4),\mu } (k; \bar{p}) &= (\bar p \cdot k) \bigl( (\bar p \cdot k) \delta^{\mu\nu} - \bar p^\mu k^\nu \bigr) \gamma_\nu \, .
\end{align}
The $\chi$V tensors are then chosen as
\begin{align}
\label{T5}
{R}^{(5),\mu } (k; \bar{p}) &= i \mathcal{T}^{\mu\nu}_k \bar p_\nu \, ,\\
\label{T6}
{R}^{(6),\mu } (k; \bar{p}) &= \frac i 2 \,  \left[ \gamma^\mu, \slashed{k} \right] \, \\
\label{T7}
{R}^{(7),\mu } (k; \bar{p}) &= \frac i 2  (\bar p \cdot k)  \mathcal{T}^{\mu\nu}_k  
\left[ \gamma_\nu, \slashed{\bar p} \right]\, , \\
\label{T8}
{R}^{(8),\mu } (k; \bar{p}) &= \frac i 2  \bigl( (\bar p \cdot k) \delta^{\mu\nu} - \bar p^\mu k^\nu \bigr) \left[ \gamma_\nu, \slashed{\bar p} \right] \, .
\end{align}

Some properties of the tensors $R^{(i)}$ are summarised in table~\ref{tab:tensors}. The relation to the tensors as described in the recent review \cite{Eichmann:2026ttr} is given in eq.~\eqref{eq:transl} in
Appendix~\ref{app:basis}.

The tensor $R ^{(1)}$ \eqref{T1} is the transverse projection of the tree-level structure. Therefore, it has been labeled as number one. The following three $\chi$S tensor structure are then ordered according to dimension and the importance as determined {\it {a posteriori}} from the numerical solution (see below).
As for the first ordering 
criterion we note that the QGV is dimensionless. Therefore, the (energy) dimension of the tensors $R^{(i)}$, as given by the number of momenta appearing in them, has to be balanced by the dressing functions. Thus, a large negative dimension implies less weight. 

The $\chi$V tensors $R^{(i)}$, $i=5,6,7,8$, are ordered w.r.t. the same principles. It is important to note 
here that $R^{(5)}$ \eqref{T5} is a scalar in the Clifford algebra of Dirac matrices, the other three are
tensors, they contain a commutator of Dirac matrices and are thus proportional to 
$\sigma_{\alpha \beta}$.\footnote{ 
The quark's chromomagnetic moment is given by the infrared value of $\Gamma^{(6)}$, the dressing function of $R^{(6)}$ \eqref{T6}.}

Note also that $R^{(2)}$ contains three Dirac matrices and is thus a higher-rank tensor in the Clifford algebra of Dirac matrices. The other three $\chi$S tensors contain one Dirac matrix and relate thus
to a vectorial coupling of the quarks to gluons.

\begin{table}[t]
\begin{tabular}{||c|c|c|c||}
\hline
Tensor& \# $\gamma$'s &chirality & dim($\Gamma^{(i)}$)  \\
\hline
\hline
$R^{(1)}$ & 1 & $\chi$S & ~0 \\
\hline
$R^{(2)}$ & 3 & $\chi$S & -2 \\
\hline
$R^{(3)}$ & 1 & $\chi$S & -2 \\
\hline
$R^{(4)}$ & 1 & $\chi$S & -4 \\
\hline
\hline
$R^{(5)}$ & 0 & $\chi$V & -1 \\
\hline
$R^{(6)}$ & 2 & $\chi$V & -1 \\
\hline
$R^{(7)}$ & 2 & $\chi$V & -3 \\
\hline
$R^{(8)}$ & 2 & $\chi$V & -3 \\
\hline
\end{tabular}
\caption{\label{tab:tensors} For the eight transverse tensors chosen as basis for the QGV the number
of Dirac matrices is displayed (2nd column), the behaviour
under chiral transformations is indicated (3rd column), and the energy dimension of the corresponding 
dressing function is given (4th column).
}
\end{table}

From all eight dressing functions only $\Gamma^{(1)}$ is subject to renormalisation. Using the 
renormalisation constant $Z_{1F}$ as determined from Slavnov-Taylor identities, see the previous 
section, is then sufficient for producing finite and stable results. Correspondingly,
for large momenta $\Gamma^{(1)}$ decreases logarithmically according to its anomalous dimension,
which is in leading order in the Landau gauge and  in quenched approximation 
$\gamma_{A\bar qq}= -9/22$.
The other seven dressing functions behave power-like for large momenta, they are only sizeable in the non-perturbative domain. As a matter of fact, as will be seen from the numerical results, with the exception of $\Gamma^{(2)}$, the non-tree-level form factors $\Gamma^{(3\ldots 8)}$ are very much restricted to
the sub-GeV region.

\section{Numerical results}
\label{sec:numres}

\subsection{Numerical method and error estimate}
\label{subsec:nummet}

The equations shown in figs.~\ref{quarkdse0} and~\ref{fig:QGV_DSE} form a system of $2+8$ coupled integral equations: two equations for the dressing functions of the quark propagator and eight for the QGV. As input, the gluon propagator, the three-gluon vertex, the corresponding renormalisation constants, the coupling $\alpha(\mu^2)$, the UV cutoff $\Lambda^2$, and a rescaling factor relating internal and physical units are taken from~\cite{Huber:2020keu}.

The DSE for the quark propagator is given in eq.~\eqref{eq:quarkDSE}. Projecting onto appropriate Dirac structures and taking colour and Dirac traces yields coupled equations for the dressing functions $A(p^2)$ and $B(p^2)$. These equations are renormalised using a momentum subtraction scheme by imposing $A(p^2=\mu_f^2) = 1$ and $B(p^2=\mu_f^2) = m$, where $\mu_f^2$ denotes the subtraction scale and $m$ the renormalised current quark mass. An additional equation for the renormalisation constant $Z_2$ is solved simultaneously. After renormalisation, the quark propagator equations take the form
\begin{align}
A(p^2) & = 1 + \Pi_{A}(p^2) - \Pi_{A}(\mu_f^2), \\
B(p^2) & = m + \Pi_{B}(p^2) - \Pi_{B}(\mu_f^2), \\
Z_2 & = 1 - \Pi_{A}(\mu_f^2),
\end{align}
where $\Pi_{A}$ and $\Pi_{B}$ denote the corresponding self-energies.

To obtain equations for the individual vertex dressing functions, the QGV on the left-hand side of fig.~\ref{fig:QGV_DSE} is expanded in the tensor basis~\eqref{T1}-\eqref{T8}. Multiplication with the conjugate basis tensors $\bar{R}^{(j)}_\mu$ from the left, followed by Dirac and colour traces, leads to a matrix equation for the eight vertex components,
\begin{equation}
\begin{split}
\label{eq:GammaNum}
	\Gamma^{(i)} = Z_{1F} \delta^{(1)(i)} & + \sum_{(j)} \left(M^{-1}\right)^{(i)(j)} V^{(j)}_{\neg \text{abel}} \\
	& + \sum_{(j)} \left(M^{-1}\right)^{(i)(j)} V^{(j)}_{\text{abel}}
\end{split}
\end{equation}
with 
\begin{align}
	M^{(j)(i)} &= \frac{1}{4} \text{tr} \left[\bar{R}^{(j)}_\mu R^{(i),\mu}\right], \\
    V^{(j)}_{\neg \text{abel},  \,   \text{abel}} &= 
\frac{1}{4} \text{tr} \left[\bar{R}^{(j)}_\mu \triangle_{\neg \text{abel},  \,   \text{abel}}^\mu \right].
\end{align}
Including only the non-Abelian triangle diagram, denoted here by $\triangle^\mu_{\neg \text{abel}}$, one 
takes into account only the first line in \eqref{eq:GammaNum}. The herein employed version of 
including the Abelian triangle diagram, denoted by $\triangle^\mu_{ \text{abel}}$, consists 
of taking additionally the second line of \eqref{eq:GammaNum} into account, however, without iterating the quark propagator and the vertex DSEs further.  Details like, e.g., the precise form of the kernels 
$V^{(j)}_{\neg \text{abel},  \,   \text{abel}} $ and the expressions derived from the triangle diagrams,
$\triangle_{\neg \text{abel},  \,   \text{abel}}^\mu $, 
can be found in our code release~\cite{code_release}. 

The vertex renormalisation constant calculated via the relation $Z_{1F}=Z_2/\tilde Z_3$ is updated
with every full iteration of the quark propagator. Whereas $\tilde{Z}_3 = Z_3 / Z_1$ is given as part
of the input data and is therefore constant $Z_2$ gets evaluated newly every time when solving for the quark propagator.

For numerical stability, the tensor structures defined in eqs.~\eqref{T1} to~\eqref{T8} are constructed from normalised momenta. After convergence, the dressing functions are rescaled by the appropriate powers of momenta to recover the original basis.

The full system of $3+8$ coupled equations is solved using a fixed-point iteration scheme. At each iteration step, the vertex equations are updated twice, followed by iterations of the propagator equations until a convergent solution for $A(p^2)$ and $B(p^2)$ is found. Convergence is monitored using the Euclidean norm of the difference between successive iteration steps for both propagator and vertex dressing functions. The iterative procedure continues until the discrepancy between successive iterations falls below a prescribed tolerance, indicating that a sufficiently accurate solution has been obtained.

To improve numerical stability, a relaxation step for the first few vertex updates is added by forming a weighted average of the current and previous solutions. This procedure is implemented to prevent updates that may lead to a runaway solution and to facilitate smoother convergence.

The quark propagator dressings are initialised as $A(p^2)=1$ and $B(p^2)=const.$ (typically chosen
as one in internal units), 
while the vertex is initialised with $\Gamma^{(1)}=Z_{1F}$ and $\Gamma^{(i)}=0$ for $i>1$. 
Particularly,  in the chiral limit, a non-vanishing initial quark mass function to start the iterative process is very helpful, since the trivial solution $M(p^2)=0$ also satisfies the equations. Initialising $\chi$V vertex components as non-zero instead is possible but induces an incorrect UV behaviour for the non-leading tensor structures, which can hinder correct convergence.

The evaluation of the dressing functions at intermediate momenta requires multidimensional interpolation. Cubic spline interpolation is employed for all functions. Given the structure of the form factors, constant extrapolation is sufficient in the infrared (IR). In the ultraviolet (UV), the propagator dressing 
$A(p^2)$ is extrapolated using $A(p^2) = a_{A} (p^2)^{f_{A}}$ 
with parameters $a_A$ and $f_A$ determined from least squares fits.
In the chiral limit, $B(p^2)$ is neglected in the UV extrapolation, i.e., we set $B(p^2)=0$ above the 
UV cutoff. For non-vanishing current masses least-square fits of the form 
$B(p^2) = a_{B} \text{ln}(p^2)^{f_{B}}$ are used.
 For the non-leading vertex dressings, constant UV extrapolation is adequate, while linear extrapolation is used for the tree-level component.

After introducing spherical coordinates for the loop momenta, all integrations are performed using Gaussian quadrature. For the radial integration and parts of the angular integration, Gauss-Legendre quadrature is employed, while the remaining angular integral is evaluated using Gauss-Chebyshev quadrature of the second kind. The radial integration is split at lower and upper endpoints, enabling the inclusion of additional integration points in the IR and UV. Further splits are implemented at additional kinematically sensitive points. Typically, for $N$ external momenta grid points, $3N$ to $4N$ radial integration points and $N$ angular integration points are used. 

Further technical details, including a comprehensive description of the implemented equations and the explicit form of all integral kernels, the Mathematica code used to 
generate the kernels, and the Julia scripts to solve the equations as described above, are provided in our code release~\cite{code_release}. All input parameters used in the present study are documented in the corresponding data release~\cite{data_release}. 

The DSEs employed in this work are solved within a specific truncation scheme. As a consequence, systematic uncertainties associated with truncation errors and artefacts are intrinsically unknown and cannot be quantified on the basis of one specific study alone.\footnote{Indeed, our 
here presented investigation indicates that to obtain apparent convergence for the quark-gluon vertex
quite sophisticated truncation schemes will be needed. } 
In the following, we therefore restrict ourselves to an assessment of the numerical error of our solution, while disregarding both systematic truncation errors and  artefacts as well as potential uncertainties in the input data. To this end, we repeat our calculations including only the non-Abelian diagram with increased external and internal grid points. This procedure allows us to test the numerical accuracy of our implementation.

The numerical errors for the dressing functions of the quark propagator, 
$A(p^2)$ and $B(p^2)$, resp., $M(p^2)$, 
and the QGV functions $\Gamma^{(i)} (k^2, \bar p^2, w) $ 
are estimated by computing the relative deviation of the two solutions with a different number 
of grid points for momenta 
$p^2,k^2,\bar{p}^2 \in \left[10^{-3} \text{ GeV}^2, 10^1 \text{ GeV}^2\right]$ and taking the absolute maximum of the resulting values thereby leading to a conservative estimate of the resulting errors. 
The UV region is excluded from this analysis because either the respective dressing functions attain negligibly small absolute values or their behaviour is fixed by the renormalisation condition. Furthermore, momentum regions in which the vertex dressings fall below $10^{-3}$ are omitted, as their contribution to the quark propagator DSE is numerically insignificant. Nevertheless, relatively large relative errors occur only
where the
functions assume small values. Due to this we consider our error estimate as quite robust and reliable. 

The error estimates have been performed based on input for a decoupling and the
scaling solution.
The here and in the next subsections used decoupling solution is selected, if not stated explicitly otherwise,
 via a normalisation condition for the IR value of the ghost renormalisation function, $G(0)=5$.

The external momentum grids for the QGV dressing functions
 in $k^2$, $\bar p^2$ and $w$ are in both cases
$64 \times 64 \times 48$ vs.\ $72 \times 72 \times 48$, the number of  internal Gau\ss ian integration 
points for one radial and two angular integration points are 
$256 \times 64 \times 64$ vs.\ $288 \times 72 \times 72$. For the quark propagator functions the
external grid has been raised from 64 to 72 points, the number of integration points did rise
from $256 \times 64$ to $288 \times 72$. In view of the discussion in the next section it is important 
to note that the number of external points in the angular variable $w$ has been chosen large enough 
such that a variation of this number has practically no influence on the result any more. 
The resulting relative errors for these data sets are summarised in table~\ref{tab:errors}.

From this analysis we infer that, in the relevant momentum domain, the numerical uncertainty of the QGV dressing functions remains at a level of approximately 5~\% or slightly above.\footnote{The
relative error for the dressing function $\Gamma^{(2)}$ based on the scaling solution seems with 
7\% to be significantly larger than the others. To this end we note that  $\Gamma^{(2)}$  as obtained
from the decoupling solution assumes a relatively large IR value whereas $\Gamma^{(2)}$  as obtained
from the scaling solution displays a zero crossing in the IR and a small value in the deep IR, cf.\ Sect.\ \ref{subsec:scadec}. Thus, this error
estimate is distorted by a ``small-basis-effect''.}
 Because the contributions of the vertex structures enter the propagator DSE with individual kernels, these uncertainties are further suppressed at the level of the propagator. We estimate the numerical error of the 
function $A(p^2)$ to be of the order of $0.5$~\% and the one of the function $B(p^2)$ of the order of 
$2$~\%. This leaves for the quark mass function as a conservative error estimate  $2.5$~\%.

Finally, in order to assess the numerical impact of the Abelian diagram, we include its contribution as described above and perform an  error analysis analogous to the one described above.
The effect on the QGV dressing functions will be discussed in the next subsections. The consequences 
for the quark propagator functions of adding the Abelian diagram confirm the sub-leading nature of this
diagram. E.g., in the chiral limit, for the quark mass function $M(p^2)$ the approximate inclusion of the Abelian diagram lowers its infrared value by  slightly more than $5$~\%, see the next subsection. This then also justifies that the very expensive iteration with full inclusion of the Abelian diagram is left for future investigations. 

\begin{table}[t]
\begin{tabular}{||c|c|c||}
\hline
Function & rel.\ error (DC1) [\%] & rel.\ error (SC) [\%] \\
\hline
\hline
$A$ & $0.1$ &  $0.1$ \\
\hline
$M$ & $1.4$  & $1.4$ \\
\hline
\hline
$\Gamma^{(1)}$ & $0.3$ & $0.3$ \\
\hline
$\Gamma^{(2)}$ & $0.3$ & $7.2$ \\
\hline
$\Gamma^{(3)}$ & $2.8$ & $2.9$ \\
\hline
$\Gamma^{(4)}$ & $1.9$ & $2.1$ \\
\hline
\hline
$\Gamma^{(5)}$ & $4.0$ & $4.0$ \\
\hline
$\Gamma^{(6)}$ & $2.0$ & $2.0$ \\
\hline
$\Gamma^{(7)}$ & $5.7$  & $5.8$ \\
\hline
$\Gamma^{(8)}$ & $5.2$ & $4.9$ \\
\hline
\end{tabular}
\caption{\label{tab:errors} Estimation of the relative numerical error for dressing functions of the quark propagator and the quark-gluon vertex for a decoupling-type solution (DC1) and scaling solution (SC) input as described in the main text. }
\end{table}

\subsection{Overview over  and importance of the  vertex form factors}
\label{subsec:ove}

In the following we will present only results in the chiral limit, the results with non-vanishing current quark masses will be published elsewhere. We only want to state here that the results for small current quark masses deviate in the more or less expected manner from the results in the chiral limit.

First, the values in the deep infrared are provided in table \ref{tab:sizes}. In case the maximal value of a function does not coincide with its IR value, respectively, for negative functions the minimal value of a function does not coincide with its IR value, this maximal / minimal value is given after a slash.The dependence on $w$ is at the maximal / minimal value so weak that it does not appear at the provided 3-digit accuracy. 

All the vertex form factors assume a constant value in the deep IR. As, with the exception of $\Gamma^{(2)}$ based on the scaling solution, 
also these values are typically of order one, see   table \ref{tab:sizes}, it is herewith {\it {a posteriori}} justified to order the functions according to their energy dimension.

In table \ref{tab:sizesA} we provide the same data but now for results with the Abelian diagram added as described above. Although some details change the overall picture remains the same. As for the quark
propagator, the function $A(p^2)$ is hardly influenced by the Abelian diagram, the IR value of the quark mass function, $M(0)$, is lowered by 17 MeV in both cases, for the decoupling and the scaling input.
The effect of the Abelian diagram on the QGV vertex form factors is also quite limited with the exception of $\Gamma_5$. We will return to this issue in subsection \ref{subsec:Gordon}.

\begin{table}[t]
\begin{tabular}{||l|c|c||}
\hline
Function & IR / maximal (DC1) & IR / maximal  (SC) \\
\hline
\hline
$A$ & ~ 1.364 &  ~ 1.366  \\
\hline
$M$ ~~ [GeV] & ~ 0.321  &  ~ 0.329 \\
\hline
\hline
$\Gamma^{(1)}$ & ~ 3.654 &  ~ 3.388 \\
\hline
$\Gamma^{(2)}$ [GeV$^{-2}$]& - 3.064 & $\sim$ 0.1 / - 2.493 \\
\hline
$\Gamma^{(3)}$ [GeV$^{-2}$] & - 2.423 &  - 1.512\\
\hline
$\Gamma^{(4)}$ [GeV$^{-4}$] &  0.556 & 0.251 \\
\hline
\hline
$\Gamma^{(5)}$ [GeV$^{-1}$]&  ~ 3.338 & ~ 2.560 \\
\hline
$\Gamma^{(6)}$ [GeV$^{-1}$]&  - 2.382 &  ~ - 0.938 / -1.979 \\
\hline
$\Gamma^{(7)}$ [GeV$^{-3}$]&  ~  1.317 & ~ 0.857 / 1.237 \\
\hline
$\Gamma^{(8)}$ [GeV$^{-3}$]&  - 0.511 &  ~ 0.746  \\
\hline
\end{tabular}
\caption{\label{tab:sizes} Given are the IR values, i.e., the ones for $p^2=0$, resp., $k^2=\bar p^2 =0$.
If the maximal absolute value of a dressing function does not agree with the IR one it is additionally 
provided. 
}
\end{table}

\begin{table}[t]
\begin{tabular}{||l|c|c||}
\hline
Function & IR / maximal (DC1) & IR / maximal  (SC) \\
\hline
\hline
$A$ & ~ 1.358 &  ~ 1.360  \\
\hline
$M$ ~~ [GeV] & ~ 0.305  &  ~ 0.312 \\
\hline
\hline
$\Gamma^{(1)}$ & ~ 3.231 / 3.617 &  ~ 3.002 / 3.687 \\
\hline
$\Gamma^{(2)}$ [GeV$^{-2}$]& - 3.514 & - 0.354 / - 2.782 \\
\hline
$\Gamma^{(3)}$ [GeV$^{-2}$] & - 2.132 &  - 1.435\\
\hline
$\Gamma^{(4)}$ [GeV$^{-4}$] &  ~ 0.373 & ~ 0.101 \\
\hline
\hline
$\Gamma^{(5)}$ [GeV$^{-1}$]&  ~ 1.980 & ~ 1.434 \\
\hline
$\Gamma^{(6)}$ [GeV$^{-1}$]&  - 2.167 &  ~ - 0.747 / -1.833 \\
\hline
$\Gamma^{(7)}$ [GeV$^{-3}$]&  ~  1.073 / 1.357 & ~ 0.614 / 1.193 \\
\hline
$\Gamma^{(8)}$ [GeV$^{-3}$]&  0.001 / -0.199  &  ~ 1.392  \\
\hline
\end{tabular}
\caption{\label{tab:sizesA} Same as table \ref{tab:sizes} but now with the Abelian diagram included as described in the text. 
}
\end{table}

\begin{figure}
\includegraphics[width=0.49\textwidth]{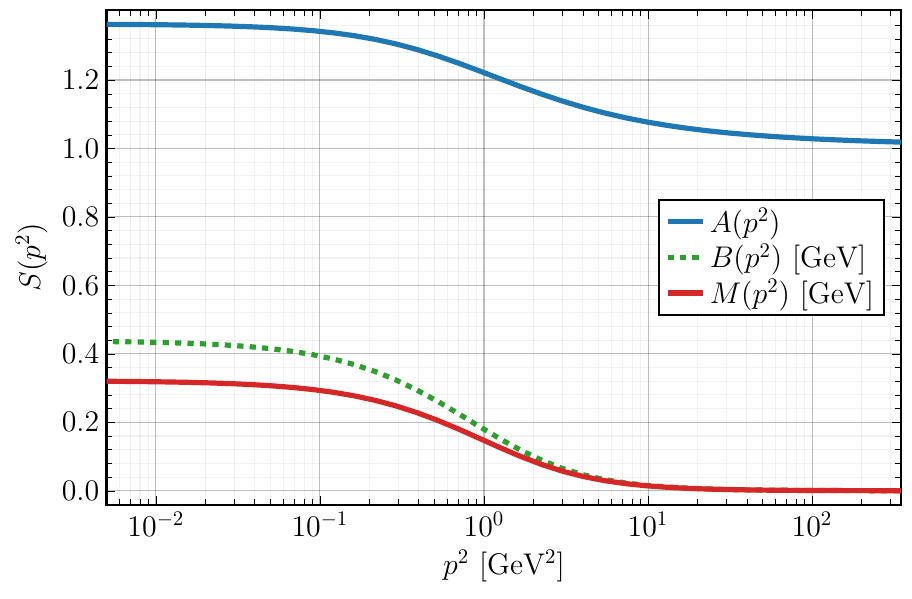}
\caption{Displayed are the quark propagator functions based on the Yang-Mills decoupling solution.
\label{fig:prop1}}
\end{figure}

\begin{figure}[h]
\includegraphics[width=0.49\textwidth]{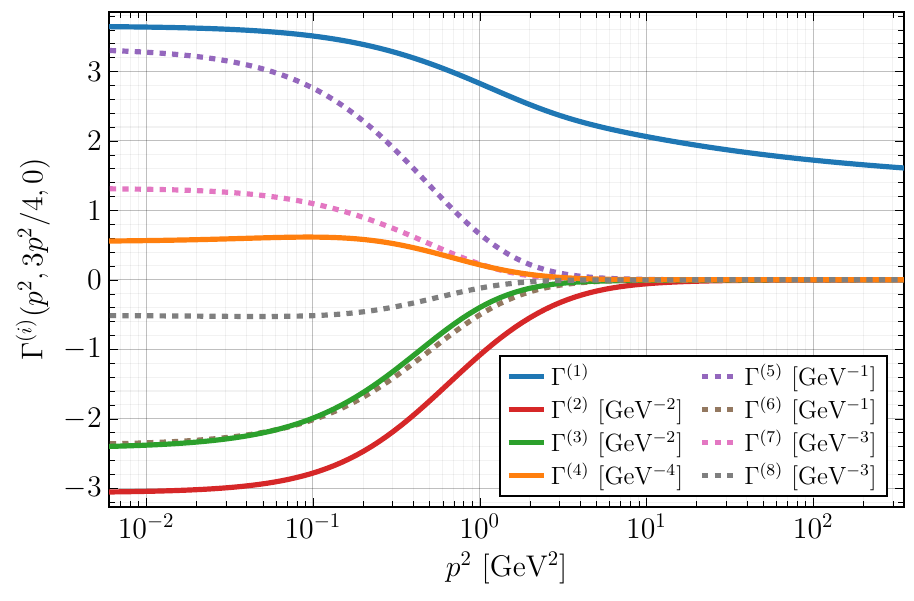}
\caption{Displayed are the QGV functions (units are GeV) at the symmetric point $p^2=q^2=k^2$,
resp., $\bar p^2 = 3k^2/4, \, w=0$, based on the YM decoupling solution. 
\label{fig:QGV1}}
\end{figure}

Returning to the comparison in between the solutions based on either decoupling or scaling in subsection \ref{subsec:scadec} some illustrations for the obtained functions based on the decoupling solution are presented in figs. \ref{fig:prop1} and \ref{fig:QGV1}. W.r.t.\ the discussion of the analytic structure below
we want to note that the function $A(p^2)$ extends to quite large momenta with a corresponding ``width scale'' (named width in the following) being 1.7~GeV$^2 \approx (1.3$ GeV$)^2$ whereas the width of $M(p^2)$ is in the sub-GeV region: 0.8 GeV$^2 \approx (0.9$ GeV$)^2$. At the symmetric point, as a function of $p^2$, the form factor $\Gamma^{(1)}$ displays a similar fall-off as the propagator function $A(p^2)$. The other $\chi S$ form factors $\Gamma^{(2,3,4)}$ display a width slightly below one GeV whereas the $\chi V$ ones $\Gamma^{(5,6,7,8)}$ are narrower, displaying a width 0.4 GeV$^2 \approx (0.65$ GeV$)^2$. Definitely, all non-tree-level form factors are negligibly small above momenta squared of the order of several GeV$^2$.

\begin{figure}
\includegraphics[width=0.49\textwidth]{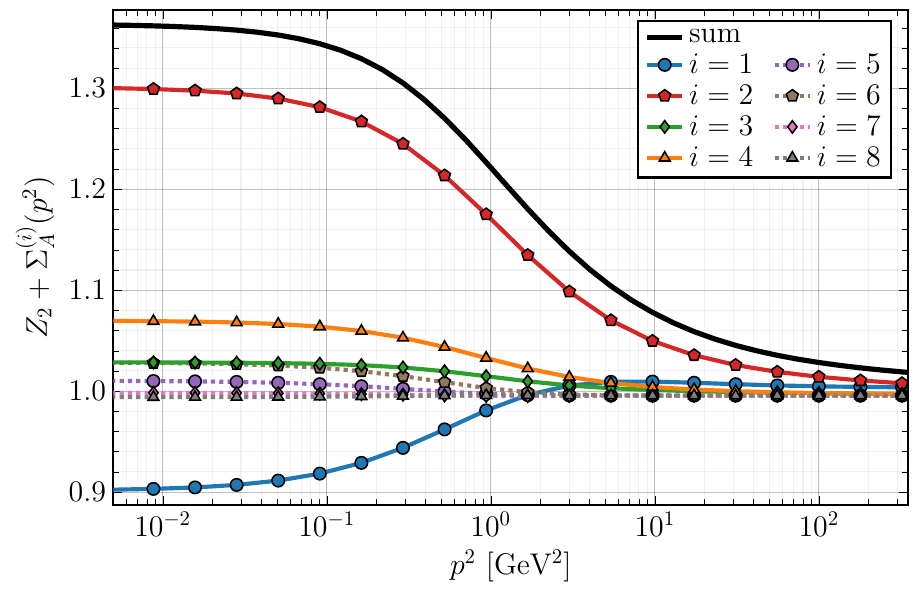}
\includegraphics[width=0.49\textwidth]{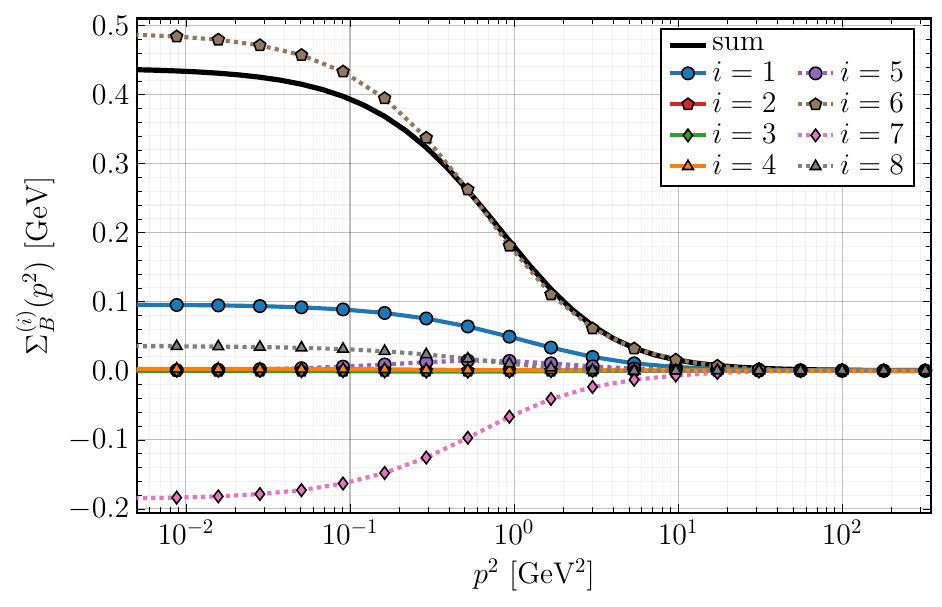}
\caption{Displayed are the contributions of the QGV functions to the propagator function $A(p^2)$ (upper panel) and to $B(p^2)$ (lower panel, units are GeV).
\label{fig:QpropCon}}
\end{figure}

In order to gain more insight we can split the contributions to the quark propagator functions according to the different QGV form factors,
\begin{equation}
A(p^2) = Z_2 + \sum _{i=1}^8 \Sigma_A^{(i)} \, , \quad \, B(p^2) =  \sum _{i=1}^8 \Sigma_B^{(i)} \, , 
\label{eq:QpropCon}
\end{equation}
where we also exploited the fact that in the chiral limit the function $B(p^2)$ does not get renormalised. In the upper panel of Fig.~\ref{fig:QpropCon} we display the function $A(p^2)$ (full black line) and the eight different contributions (more precisely, $Z_2 + \Sigma_A^{(i)}$). First of all, we note that all major contributions originate from the $\chi$S QGV form factors with $\Gamma^{(2)}$ providing the biggest one.
This behaviour is mirrored for the function $B(p^2)$, here the $\chi$V QGV form factors, in particular $\Gamma^{(6)}$, are providing the dominating contributions. 

A surprising fact is hereby that in the sub-GeV region the tree-level contribution, $\Sigma_B^{(1)}$ contributes even point-wise less than 1/4 to the function $B(p^2)$. Given the fact that $-\Sigma_A^{(1)} \lesssim 0.1$, see Fig.~\ref{fig:QpropCon}, we conclude that for $M(p^2) = B(p^2)/A(p^2)$ significantly less than 1/3 originates from the tree-level structure. As the momentum-dependence of the $\chi$V kernels is different from the $\chi$S ones the mass function possesses a different momentum dependence as if calculated in the rainbow approximation. This in turn has a big impact on the analytic structure of the quark
propagator, see the corresponding discussion in subsection \ref{subsec:qprop}.

At large $p^2$ the tree-level contributions $\Sigma_{A,B}^{(1)}$ are, of course, the dominant ones. However, for the function $B(p^2)$ the tree-level contribution dominates only for momenta above six GeV,
for the function $A(p^2)$ even momenta above 30 GeV are required before $\Sigma_{A}^{(1)}$ becomes the dominant contribution.\footnote{Such a behaviour has also be seen in ref.\ \cite{Williams:2014iea}.}

Some simple but quite accurate fits of the QGV functions displayed in Fig.\ \ref{fig:QGV1} but then (i) as a function of the gluon momentum $k^2$ at (almost) vanishing average quark momentum $\bar p^2$ and (ii) as a function of $\bar p^2$ at (almost) vanishing $k^2$ are given in the appendix \ref{App1}. As seen from these fits, when written as functions $\Gamma^{(i)} (k^2,\bar p^2,w)$
the behaviour w.r.t. the gluon momentum $k$ is quite different as w.r.t. to the averaged quark momentum $\bar p$. Partially, this is a trivial effect due to the pre-factor in the definition $\bar p = \frac 1 2 (p+q)$ but 
in essence it reflects also the difference from the scale of the gapped Yang-Mills system, respectively, the gluon mass scale, to the induced scale in the quark sector although the latter is induced by the former.
From the non-tree-level form factors $\Gamma^{(2)}$,  $\Gamma^{(5)}$ and  $\Gamma^{(6)}$  fall off quite slowly with the gluon momentum, they reach half their infrared value at $k^2\approx 2$GeV$^2$. The other four non-tree-level form factors decrease somewhat more rapidly with $k^2$, reaching the half-value at approximately one GeV$^2$. In distinction to this, the drop-off with the averaged quark momentum goes fast, for all the non-tree-level form factors the half-value is reached around $\bar p^2 \approx 0.4$GeV$^2$,
see the appendix for more details. 

We refer also to the appendix for a discussion of the QGV functions' behaviour in the UV, in particular, the difference as seen as a function of the gluon momentum or the (averaged) quark momentum.

\subsection{On the angular dependence of the vertex form factors}
\label{subsec:angdep}

From the tables \ref{tab:sizes} and \ref{tab:sizesA} one can infer already that the IR values of the QGV form factors are to a high degree independent of the angular variable $w$. In the UV, if either the square of the gluon momentum $k^2$ or of  the averaged quark momentum $\bar p^2$ is very large,  the tree-level function $\Gamma^{(1)}$ is to high degree independent of the angular variable $w$. The other, non-tree-level, form factors are in the UV also only weakly dependent on $w$, in addition to being tiny in the UV anyhow.

\begin{figure}[b]
\includegraphics[width=0.49\textwidth]{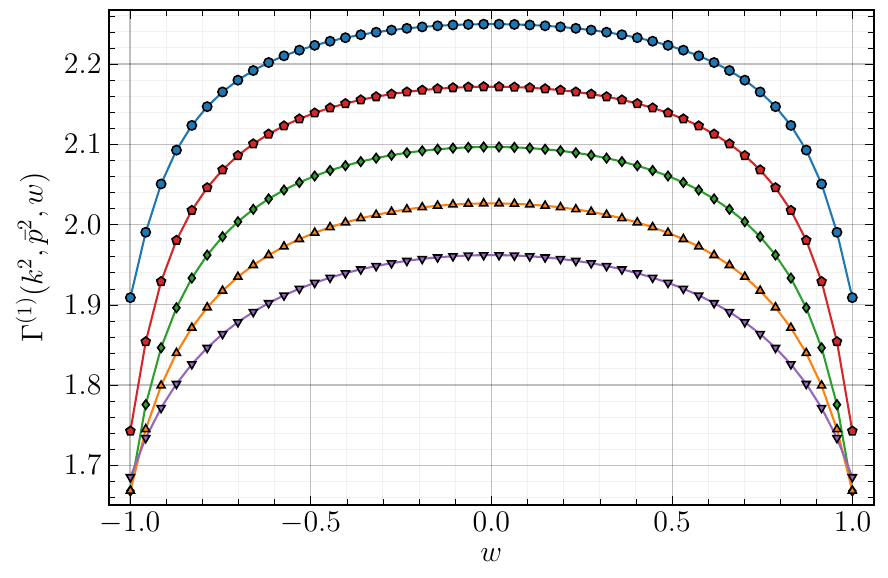}
\caption{Displayed are the $w$ - dependencies of $\Gamma^{(1)}$ for $k^2 = 17.3$ GeV$^2$ and $\bar p^2 = 4.0, 5.4, 7.2, 9.7,13.0$
GeV$^2$.
\label{fig:Gamma1w}}
\end{figure}

In the multi-GeV region one needs clearly to distinguish between the tree-level form factor  $\Gamma^{(1)}$  and the non-tree-level ones. For $\Gamma^{(1)}$ the strongest dependence on the angular variable $w$ occurs at $k^2\approx 20$ and $\bar p^2 \approx 10$ GeV$^2$. In Fig.\ \ref{fig:Gamma1w} this is exemplified for one value of $k^2$ and several values $\bar p^2$. For a reasonable polynomial fit of the $w$-dependence one needs to take into account terms proportional $w^2$ , $w^4$ and $w^6$. On the other hand, the non-tree-level form factors are for such values of momenta already quite small and additionally display a much weaker $w$-dependence.

\begin{figure}
\includegraphics[width=0.49\textwidth]{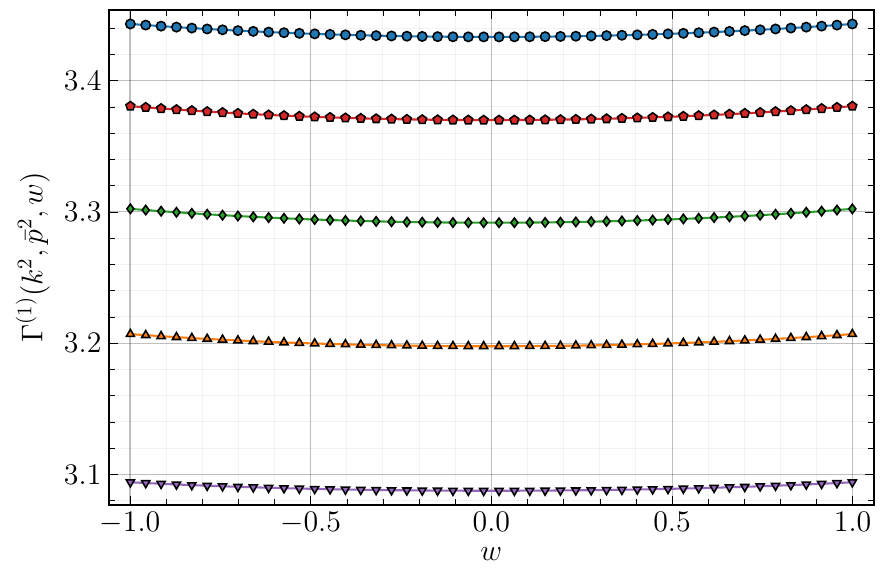}
\includegraphics[width=0.49\textwidth]{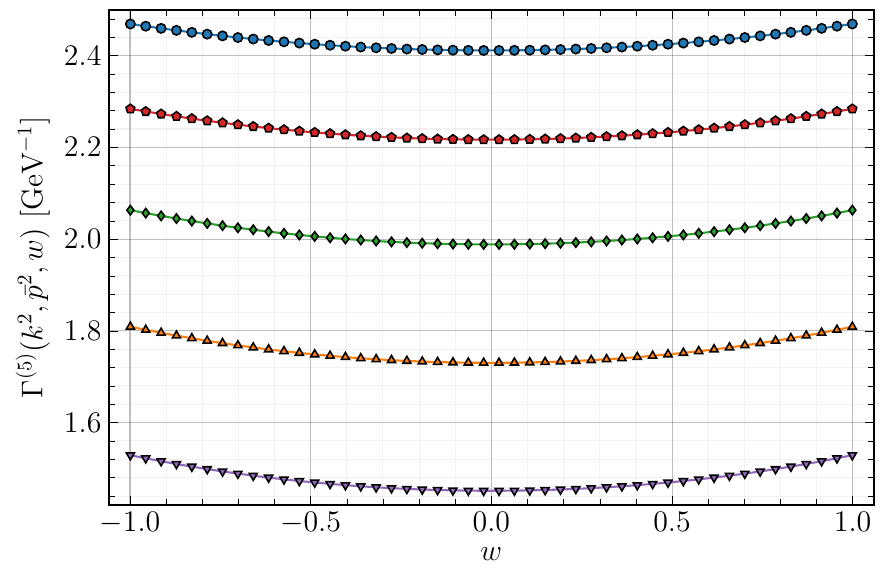}
\caption{Displayed are the $w$ - dependencies of $\Gamma^{(1)}$ (upper panel) and $\Gamma^{(5)}$
(lower panel) for $k^2 = 0.21767$ GeV$^2$ and $\bar p^2 = 0.12142, 0.16257, 0.21767, 0.29145, 0.39023$ 
GeV$^2$.
\label{fig:Gammaw}}
\end{figure}

Considering the dependence on the angular variable $w$ at intermediate momenta but still in the sub-GeV region, the tree-level form factor $\Gamma^{(1)}$ shows only some mild variation. The non-tree-level form factors possess in this kinematical region their strongest dependence on $w$ which is, however, still relatively mild. E.g., 
in Fig.~\ref{fig:Gammaw} the form factors $\Gamma^{(1)}$ (upper panel) and $\Gamma^{(5)}$ (lower panel) are displayed as function of $w$ for $k^2 = 0.21767$  GeV$^2$ and $ \bar p^2 = 0.12142, 0.16257, 0.21767, 0.29145, 0.39023$ GeV$^2$. One clearly notices the quite weak dependence
on $w$. Generically, one infers from the solutions in sub-GeV region
\begin{equation}
\Gamma^{(i)} = a^{(i)} (k^2,\bar p^2) \bigl( 1+ b^{(i)} (k^2, \bar p^2) w^2 \bigr) + ``{\mathrm{tiny}}`` \, , 
\end{equation}
where the coefficient of the quadratic term is quite strictly bounded. In this region one has with only very few exceptions $-0.5 \le b(k^2,\bar p^2) \le 0.5$, and for most momenta even $|b| < 0.1$.

\begin{figure}[h]
\includegraphics[width=0.49\textwidth]{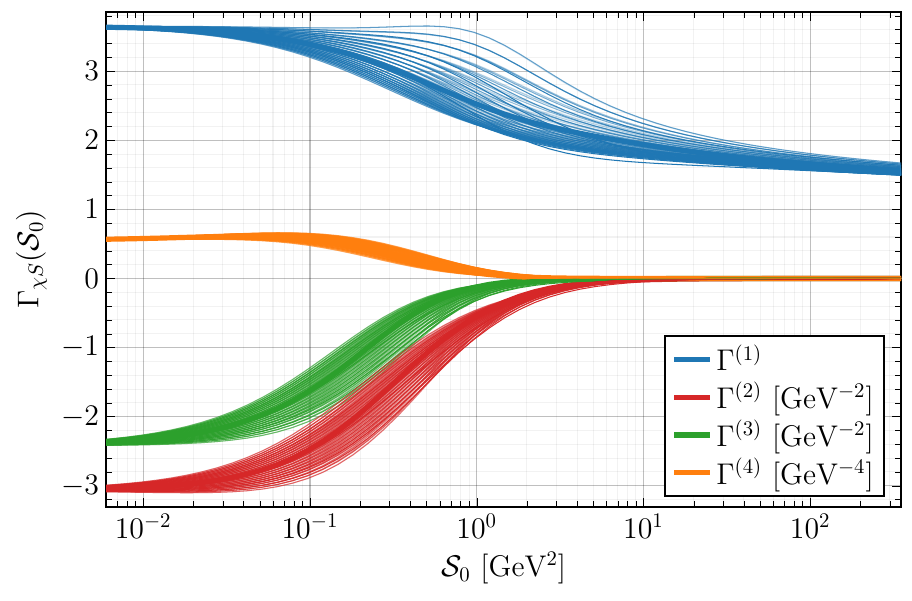}
\includegraphics[width=0.49\textwidth]{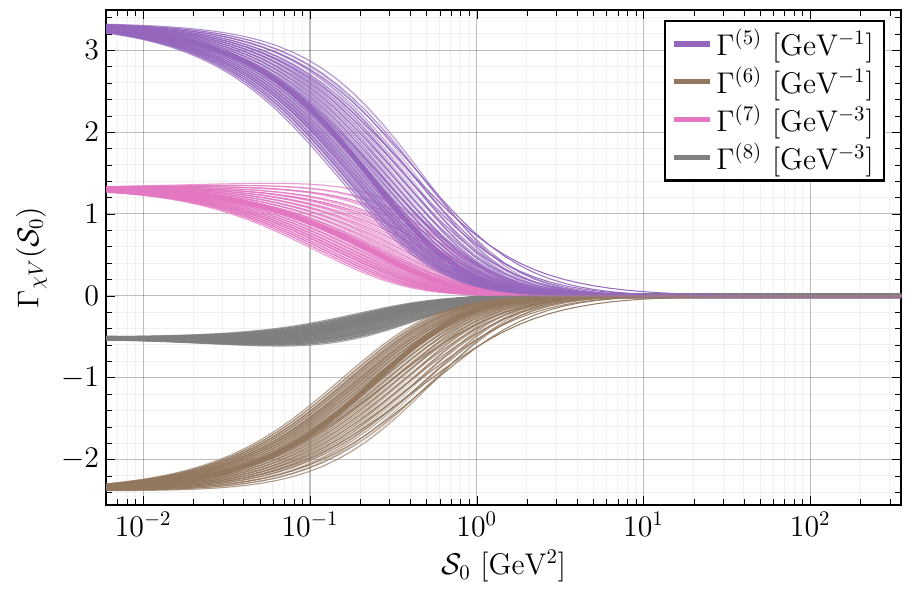}
\caption{Displayed are the QGV form factors as function of $S_3$ permutation group symmetric Lorentz invariant $\mathcal{S}_0$ where the dependence on the variables $a$ and $s$ are displayed as bands. 
\label{fig:GammaS0}}
\end{figure}

This corroborates that choosing this kinematical variables, based on charge conjugation properties and symmetry w.r.t. the quark and antiquark, leads to a generally weak and for most kinematical regions a very weak angular dependence. If one chooses, e.g., Lorentz invariant kinematical variables based on 
the permutation group $S_3$ one obtains a larger spread\footnote{These variables are given by 
$\mathcal{S}_0=(k^2+p^2+q^2)/6$,
$a=\sqrt{3}(p^2-q^2)/(k^2+p^2+q^2)$ and $s= (p^2+q^2-2k^2)/(k^2+p^2+q^2)$ and related to here
used kinematical variables via $\mathcal{S}_0 = k^2 / 4 + \bar{p}^2 / 3$, 
$a = k \cdot \bar{p} / (\sqrt{3} \mathcal{S}_0) = \sqrt{k^2 \bar{p}^2} w / (\sqrt{3} \mathcal{S}_0)$
and $s = 1 - k^2 / (2 \mathcal{S}_0)$. Note that $a$ and $s$ are compact, they are restricted to  $a^2 + s^2 \leq 1 $. }. In Fig.\ \ref{fig:GammaS0} the QGV form factors are plotted as function of the $S_3$-symmetric variable $\mathcal{S}_0$ with the dependencies on the compact kinematical variables $a$ and $s$ shown as bands. Clearly, for most functions the spread is significant.

The weak dependence on the angular variable $w$ is certainly helpful in understanding the momentum dependence of the QGV form factors. Can it be also used for introducing a quite accurate approximation scheme? This question particularly arises because solving for the $\Gamma^{(i)} (k^2, \bar p^2, w)$ for a fixed value of $w$ decreases, given a pre-required accuracy, the numerical effort by significantly more than one order of magnitude. This is quite plain from the fact that grid sizes are strongly reduced if instead of a three-dimensional grid a two-dimensional one is used. 

\begin{figure}
\includegraphics[width=0.49\textwidth]{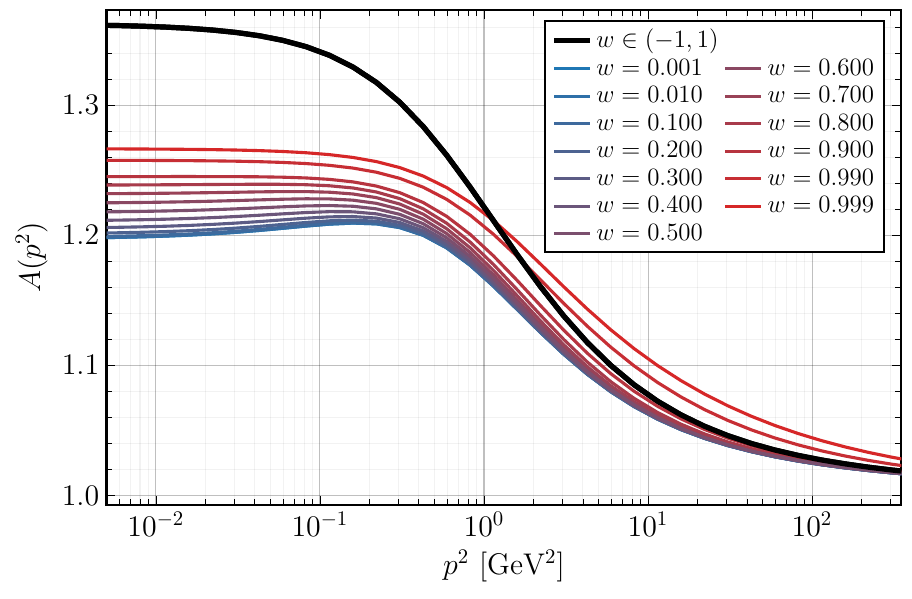}
\includegraphics[width=0.49\textwidth]{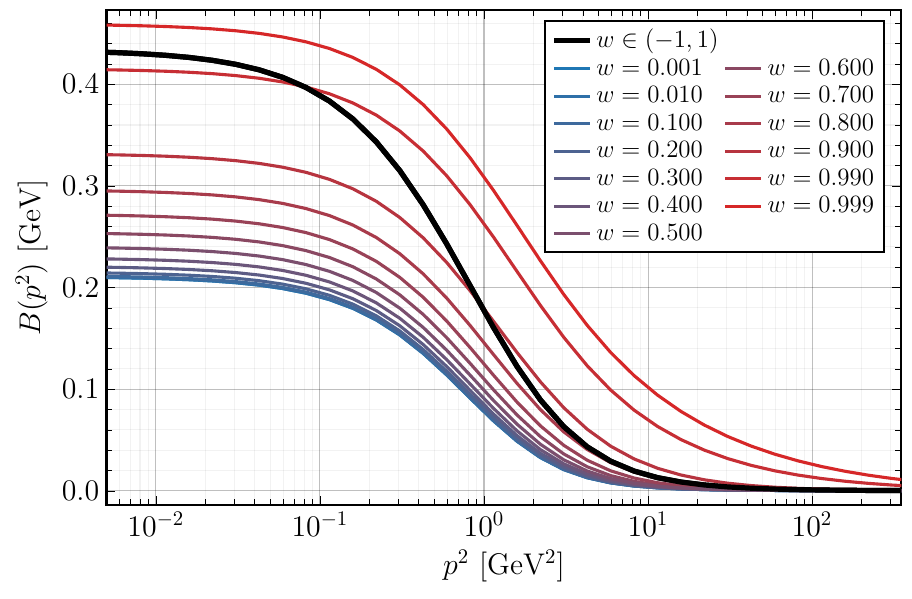}
\caption{Displayed are the quark propagator functions $A(p^2)$ (upper panel) and $B(p^2)$ (lower panel) as resulting from the full calculation labeled $w\in (-1,1)$ versus thirteen calculations with a fixed value of $w$.
\label{fig:propDepw}}
\end{figure}

Unfortunately, we find that the $w$-dependence, although it is weak, has a quite remarkable impact on the numerical solution. As can be seen on Fig.\ \ref{fig:propDepw} the impact on the quark propagator of such a reduced numerical scheme is surprisingly large, in particular, either the quark propagator functions reach only a fraction of the value obtained in the full calculation, or the decrease in the UV is much too weak, or both.

\begin{figure}
\includegraphics[width=0.49\textwidth]{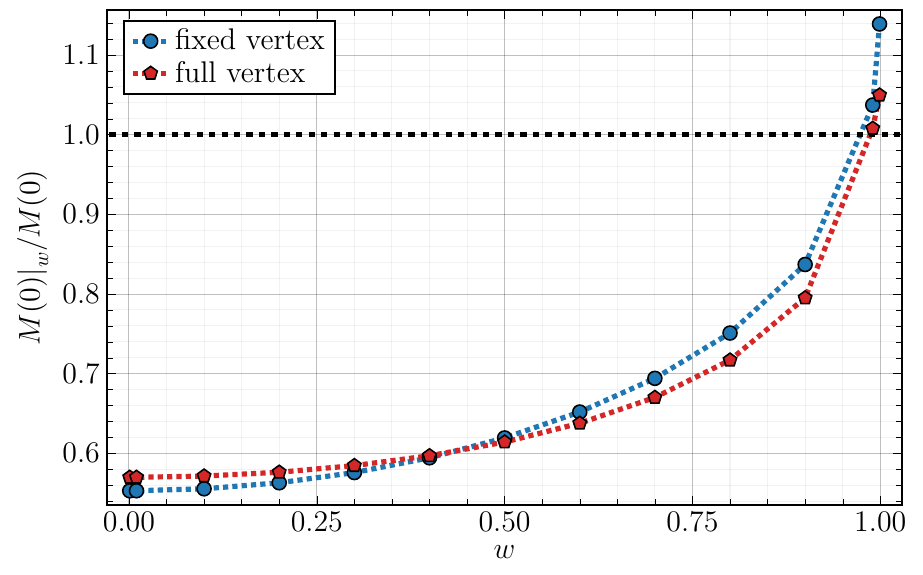}
\includegraphics[width=0.49\textwidth]{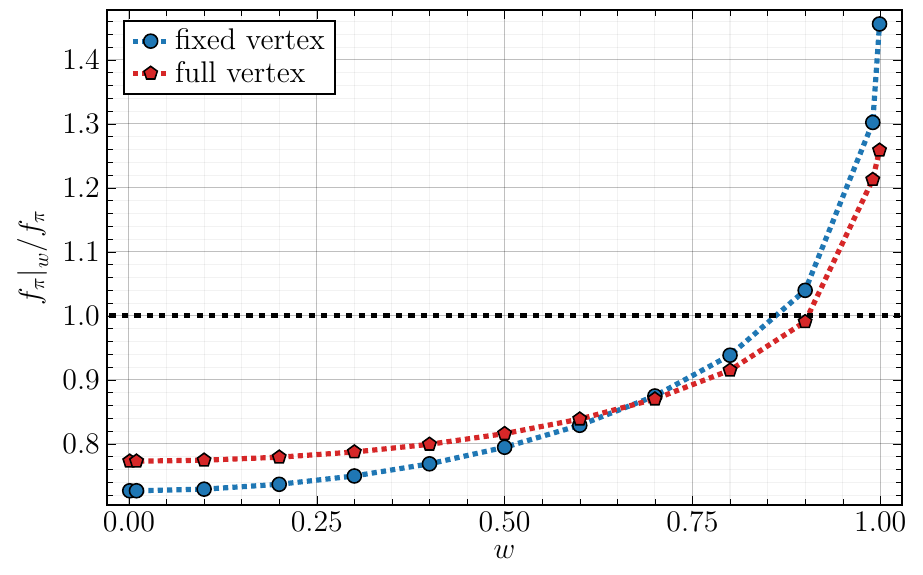}
\includegraphics[width=0.49\textwidth]{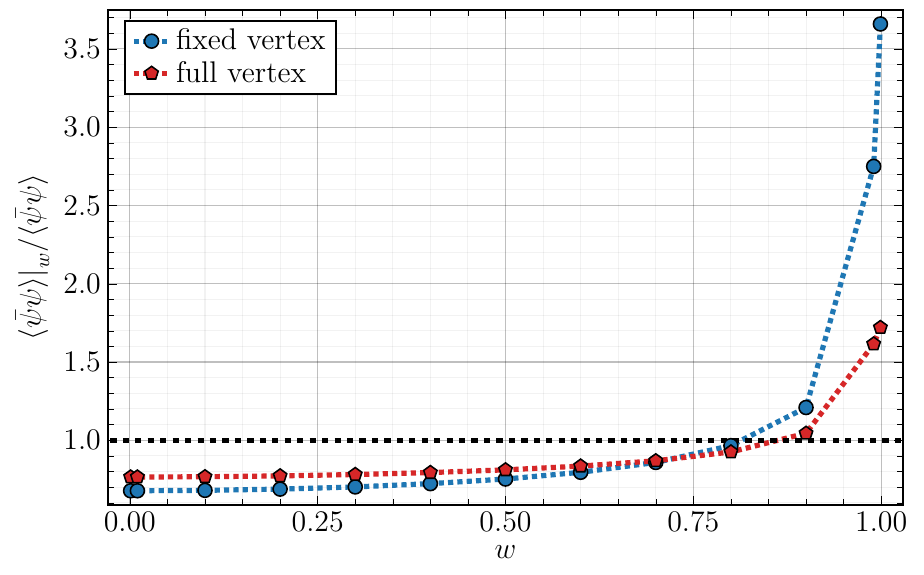}
\caption{Displayed are the ratios of (i) the IR value of the quark mass function $M(0)$ evaluated with a fixed value of $w$ to the value from the corresponding full solution (upper panel), (ii) the analogous ratio for the pion decay constant $f_\pi$ in Pagels-Stokar approximation (middle panel), and (iii) the analogous ratio for the quark condensate (lower panel) as a function of the selected value for $w$. The blue curves labeled ``fixed vertex'' are obtained by choosing the value of $w$ already during the iterative process, the red curves labeled ``full vertex'' are based on the solution for the QGV form factors with all three kinematical variables but after being calculated are then restricted to one value of $w$. 
\label{fig:MfpiCondDepw}}
\end{figure}

One might now ask whether this is an effect of the iterative procedure, or whether also the fully calculated vertex but then used only at some value of $w$, will lead to such deviations. To clarify this question ratios of quantities calculated with a fixed value of $w$ w.r.t.\ the full value have been calculated for two cases:
The restriction to a given value of $w$ is used from the start of the iterative procedure, or only used for the full QGV when calculating the respective quantity. In the upper panel of Fig.\ \ref{fig:MfpiCondDepw} the ratios for the IR value of the quark mass function, $M(0)$, are plotted. In the middle panel these ratios for the pion decay constant in Pagels-Stokar approximation \cite{Pagels:1979hd} (see also, e.g., \cite{Alkofer:2000wg}),
\begin{equation}
f_\pi^2 = Z_2 \frac N {\pi^2} \int dq^2 q^2 
\frac {M(q^2)\left( M(q^2) + \frac {q^2} 2 \frac{dM(q^2)}{dq^2}\right)}{A(q^2) \left( q^2 + M(q^2) \right)^2}
\end{equation}
 are displayed, and in the lower panel the ones for the quark condensate. These plots lead to the conclusion that even restricting the $w$-dependence of an already fully iterated solution to only one value of $w$ is already sufficient for quite some sizeable differences. Furthermore, we can infer that most of the contribution to the kernels constituting the integrands originates from regions close to the endpoints $w=\pm 1$.

Although the QGV vertex form factors have, in the chosen kinematical variables, a seemingly weak angular dependence already the effect of this angular dependence on the quark propagator and quantities derived from it makes it evident that {\it {no planar degeneracy of the quark-gluon vertex}} can be extracted on the basis of the presented results. Phrased otherwise, we have not been able to deduce on the basis of the observed weak dependence on the angular variable $w$ any simply implementable efficacious two-dimensional scheme for calculating the QGV in an even approximately valid way. 

\goodbreak

\subsection{A relation between vertex form factors}
\label{subsec:Gordon}

As a direct consequence of the Dirac equation 
\begin{equation}
(i\slashed{p} +m) \, u(p) =0
\end{equation}
 on-shell Dirac fermions spinors $u(p)$ fulfil the Gordon identity
\begin{equation}
2m \bar u(p) \gamma^\mu u(q) +  \bar u(p) \left(  2 i \bar p ^\mu - \frac i 2 [\slashed{k},\gamma^\mu] \right) u(q)  =0 \, ,
\label{eq:Gordon}
\end{equation}
where $k=p-q$ and $\bar p = \frac 1 2(p+q)$ in analogy to the definitions above. 
This implies that ``on-shell'' there should exist a relation between the vector coupling as in eq.\ \eqref{T1}, the  scalar coupling as in eq.\ \eqref{T5} and the tensorial coupling as in eq.\ \eqref{T6} 

As a matter of fact, at a fixed $k^2$ and $w$ we find for large $\bar p^2$ where the corresponding quark mass function is tiny and thus the first term in eq.~\eqref{eq:Gordon} is negligible that $\Gamma^{(5)} \approx
\Gamma^{(6)}$, however, both form factors are for these momenta already quite small.


Somewhat surprisingly,   within numerical accuracy for almost the whole kinematical range up to multi-GeV momenta  we observe
\begin{equation}
\Gamma^{(5)}(k^2,\bar p^2,w) = - 1.40 \,\, \Gamma^{(6)}(k^2,\bar p^2,w) \, .
\end{equation}
Phrased otherwise, these two terms of the QGV can be simplified to
\begin{align}
\Gamma^{(5)}(k^2,\bar p^2,w) R^{(5)} (k;\bar p) + \Gamma^{(6)}(k^2,\bar p^2,w) R^{(6)} (k;\bar p) 
\nonumber \\
= - \Gamma^{(6)}(k^2,\bar p^2,w)  \left( 1.40 R^{(5)} (k;\bar p) - R^{(6)} (k;\bar p)\right) \, .
\end{align}

One might now speculate whether such relations exist in between the other form factors of equal dimension. 
To this end we note that for the functions $\Gamma^{(7)}$ and $\Gamma^{(8)}$ we observe, however with less accuracy then in the previous case, approximate equality for large $\bar p^2$ and a proportionality in the sub-GeV region such that $\Gamma^{(7)}\approx -2.58\, \Gamma^{(8)}$.
However, the momentum dependencies of  $\Gamma^{(2)}$ and $\Gamma^{(3)}$ are distinctively different.

\subsection{Comparing results for scaling and decoupling inputs}
\label{subsec:scadec}

Suppose the difference between the decoupling solutions (with a finite gluon curvature mass scale) and the scaling (power-law) solutions are a non-perturbative gauge-fixing issue, cf.\ \cite{Maas:2009se}.
Then, because the Yang-Mills and matter sectors are independently gauge invariant the matter sector Green functions obtained from different types of solutions should be Landau-Khalatnikov transforms of each other.

As demonstrated below one observes even that,
taking from the Yang-Mills sector consistent input for decoupling and the scaling solutions, respectively, and including the full kinematical dependence for the QGV, the results for the quark propagator coincide 
for the different solutions  within the numerical precision.

In order to keep here the presentation tractable
 the discussion here restricts to results obtained with the non-Abelian diagram only, for a corresponding discussion when taking the Abelian diagram into account as described above, see also Appendix \ref{App2}.

\begin{figure}
\includegraphics[width=0.49\textwidth]{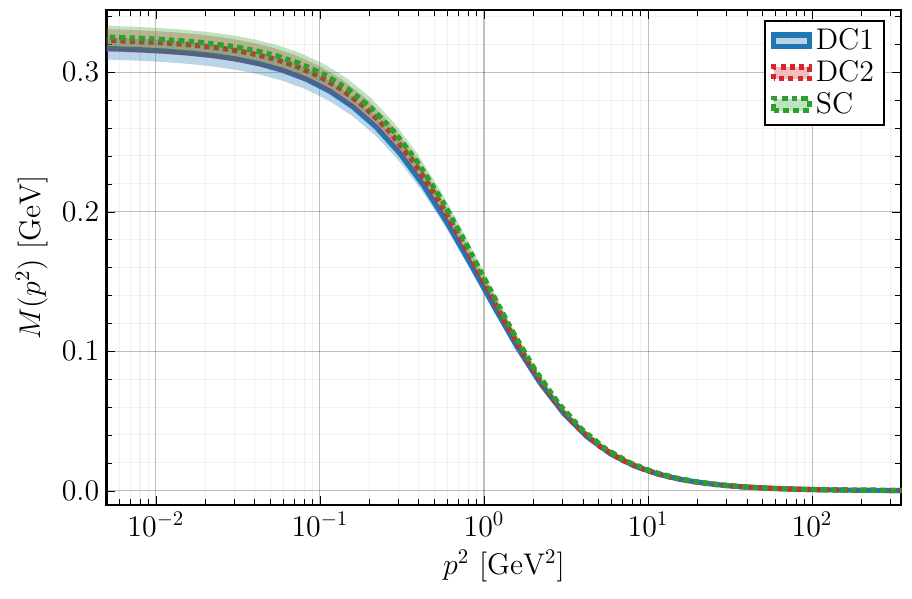}
\caption{Displayed are the quark mass functions as obtained from the decoupling solutions DC1 (blue full line) and DC2 (dotted red line) as well as the scaling solution (dotted green line).The respective numerical errors are shown as bands in the corresponding colour.
\label{fig:CompM}}
\end{figure}

In Fig.\ \ref{fig:CompM} the quark mass functions are compared for the above used decoupling solution ($G(0)=5$) labeled DC1, another decoupling solution ($G(0)=30$) labeled DC2 and the scaling solution labeled SC.\footnote{For this comparison the calculations of the QGV with 256 $\times$ 64 $\times$ 64 integration points and 64 $\times$ 64 $\times$ 48 external grid points have been used.} In order to highlight the agreement differences of the quark mass functions are shown
in Fig.\ \ref{fig:CompMdiff}. (NB: For the ratios see Fig.\ \ref{fig:CompMratio} in Appendix \ref{App2}.) Hereby, the bands display an error determined by addition of the respective two individual errors in order to provide a conservative estimate. Clearly, the deviations stay well inside the 
error bands, and one concludes equivalence within numerical errors. 

\begin{figure}
\includegraphics[width=0.49\textwidth]{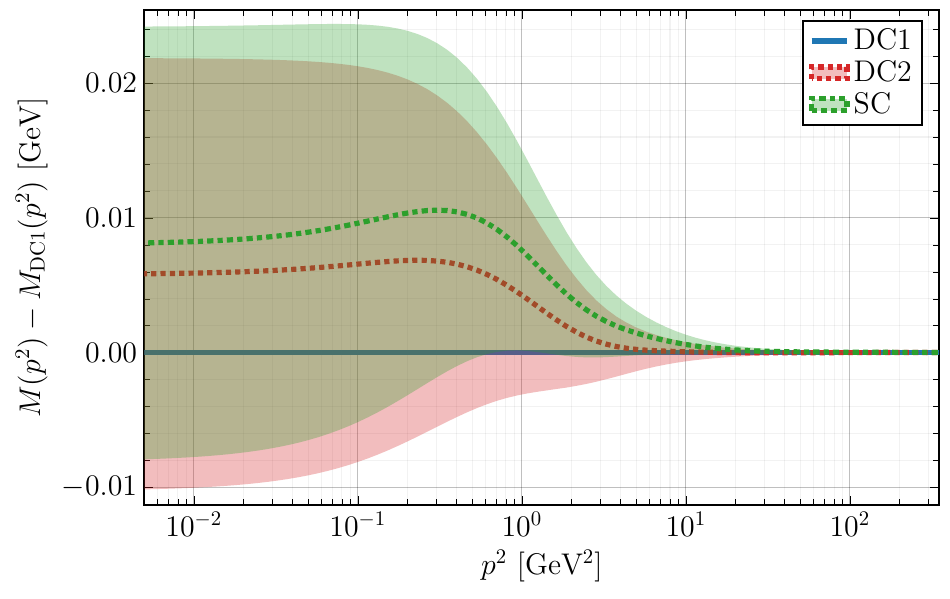}
\caption{Displayed are the differences of quark mass functions as obtained form the decoupling solution DC2 (dotted red line) and the scaling solution (dotted green line) to the one obtained from the solution DC1. The respective numerical errors are shown as bands in the corresponding colour.
\label{fig:CompMdiff}}
\end{figure}

In Table~\ref{tab:comp} the level of agreement is demonstrated for infrared value of the quark mass function $M(0)$, the pion decay constant in Pagels-Stokar approximation $f_\pi$, and the quark condensate as calculated from the DC1, DC2 and SC Yang-Mills solutions. Bearing in mind that the numerical accuracy is approximately 2.5 \% the deviations might very well be only due to the limitations in accuracy.

\begin{table}
\begin{tabular}{||c|c|c|c||}
\hline
Input & ~$M(0)$ [MeV]~ & ~$f_\pi$ [MeV]~ & $ (- \langle \bar{\psi} \psi \rangle )^{1/3}$ [MeV] \\
\hline
\hline
DC1 & $318$ & ~$73.8$~ & ~$331$~ \\
\hline
DC2 & $324$ & $74.5$ & $332$ \\
\hline
SC & $326$ & $75.3$ & $335$ \\
\hline
\end{tabular}
\caption{\label{tab:comp} The infrared value of the quark mass function $M(0)$, the pion decay constant in Pagels-Stokar approximation, and the quark condensate calculated from the three employed Yang-Mills inputs.}
\end{table}

\begin{figure}
\begin{center}
\includegraphics[width=0.46\textwidth]{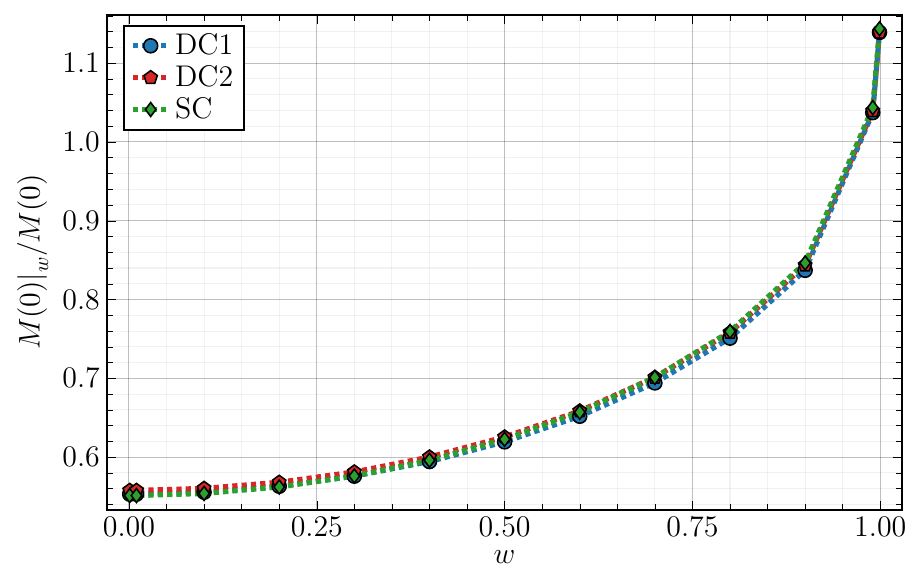}
\includegraphics[width=0.46\textwidth]{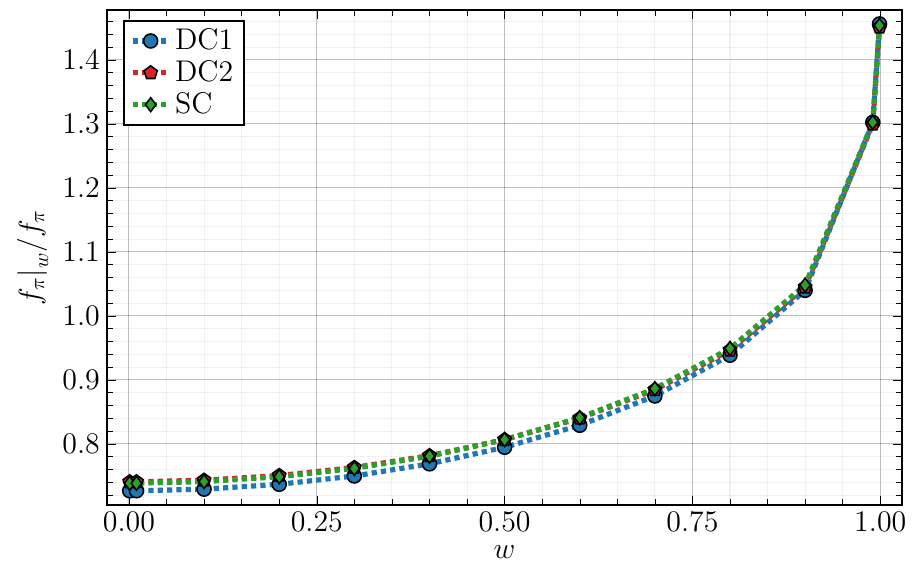}
\includegraphics[width=0.46\textwidth]{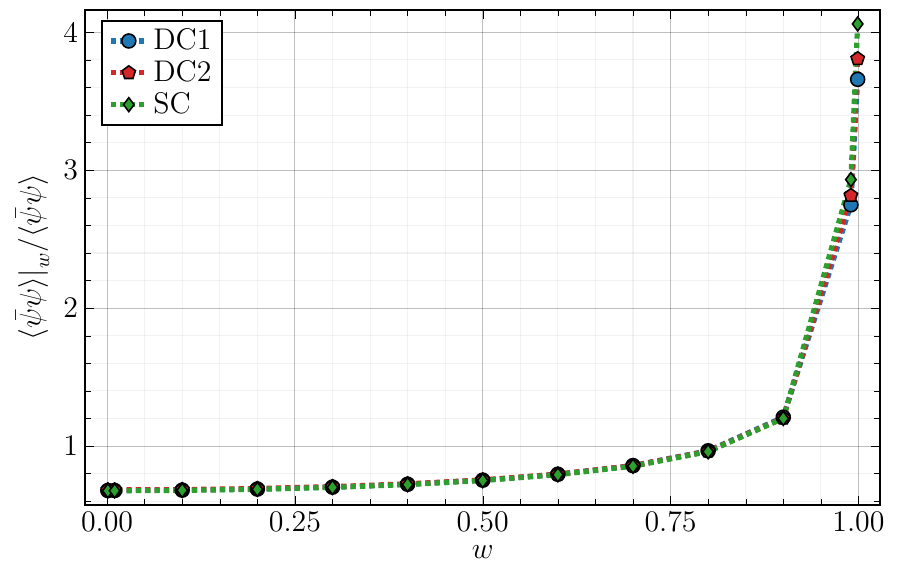}
\end{center}
\caption{Displayed are the ratios to the respective full three-dimensional solution of the IR value of the quark mass function $M(0)$  (upper panel),  the analogous ratios for the pion decay constant $f_\pi$ in Pagels-Stokar approximation (middle panel), and the analogous ratios for the quark condensate (lower panel) as a function of the selected value for $w$ in a two-dimensional calculation. 
The blue curves with balls labeled ``DC1'', the red curves with pentagons labeled ``DC2'' and green curves with diamonds labeled "SC" show the solutions with the respective Yang-Mills input.
\label{fig:MfpiCondCompw}}
\end{figure}

As discussed in subsection \ref{subsec:angdep} calculations at a fixed value of the angular variable $w$ lead to significant deviations for the quark propagator as compared to the full calculation. Somewhat surprisingly, the quark propagators calculated from the three different Yang-Mills solutions show quantitatively the same behaviour under this truncation. The level of agreement  for infrared value of the quark mass function $M(0)$, the pion decay constant in Pagels-Stokar approximation $f_\pi$, and the quark condensateas can be inferred from Fig.~\ref{fig:MfpiCondCompw}.

\begin{figure}
\includegraphics[width=0.49\textwidth]{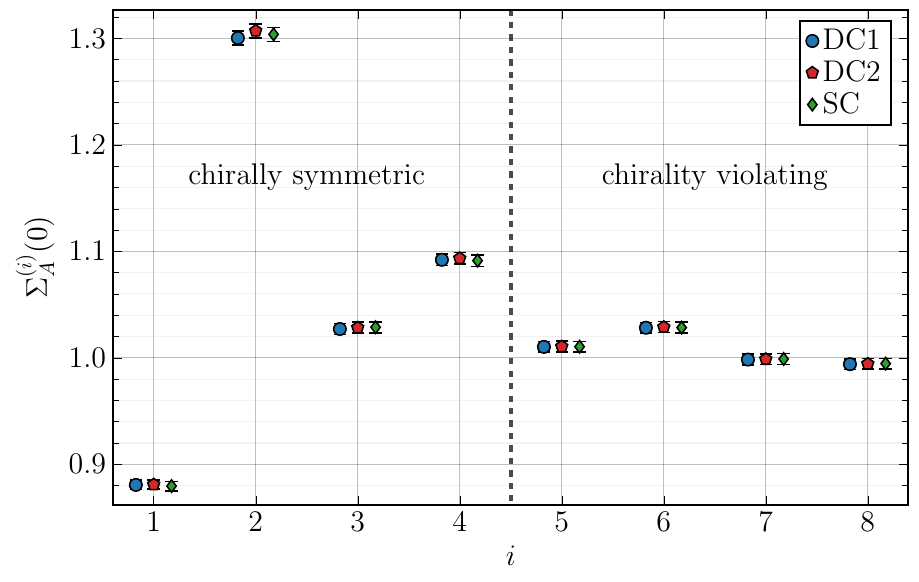}
\includegraphics[width=0.49\textwidth]{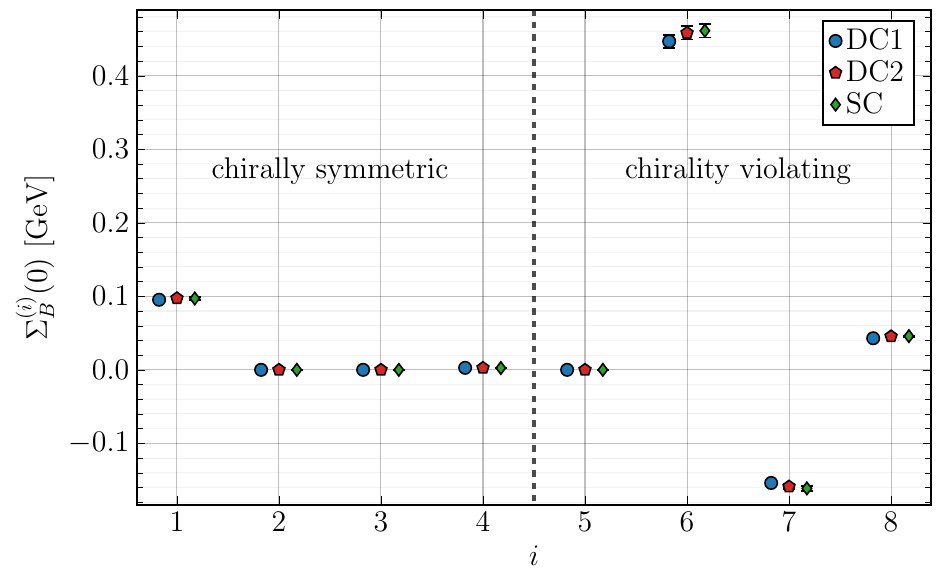}
\caption{Displayed are  the contributions from the different QGV tensor structures to the quark propagator function $A(p^2)$ (upper panel) and $B(p^2)$ (lower panel) for the three types of Yang-Mills inputs DC1, DC2 and SC.
\label{fig:SigmaComp}}
\end{figure}

Even the contributions from the different tensor structures are identical within the numerical precision.  The corresponding infrared values,  $\Sigma^{(i)}_A(0)$ and $\Sigma^{(i)}_B(0)$ are shown in 
Fig.\ \ref{fig:SigmaComp} for DC1, DC2 and SC input from the Yang-Mills sector.

\begin{figure}
\includegraphics[width=0.49\textwidth]{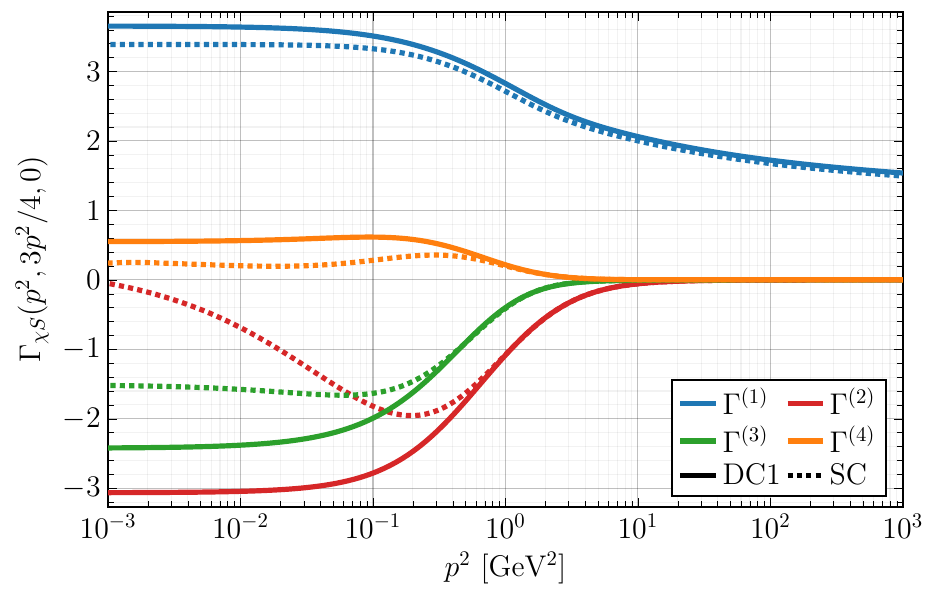}
\includegraphics[width=0.49\textwidth]{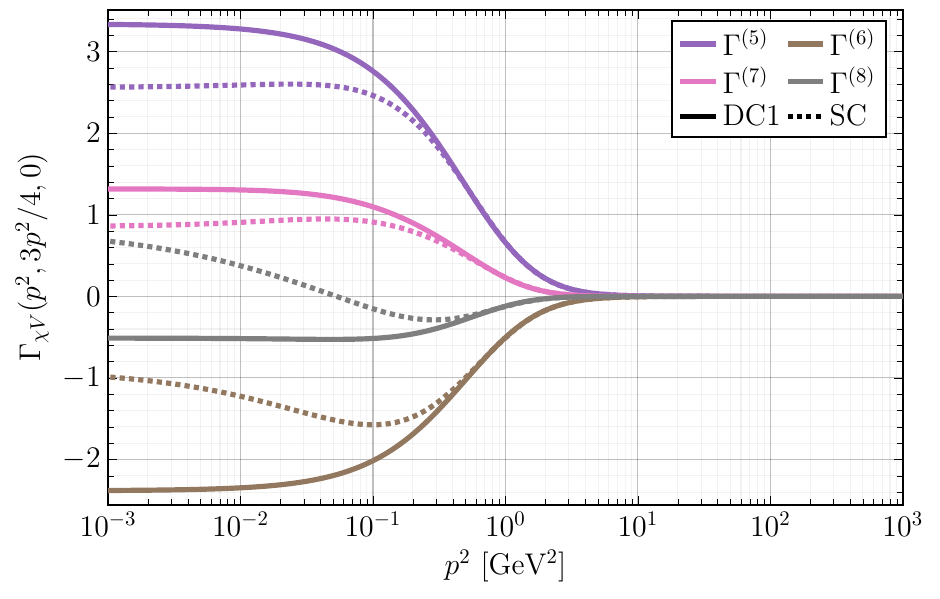}
\caption{Displayed are the QGV's form factors  for symmetric momenta, $\chi$S ones in the upper and $\chi$V ones in the lower panel, for the Yang-Mills inputs DC1 (full lines) and SC (dashed lines). 
\label{fig:CompScDc1}}
\end{figure}

\begin{figure}
\includegraphics[width=0.49\textwidth]{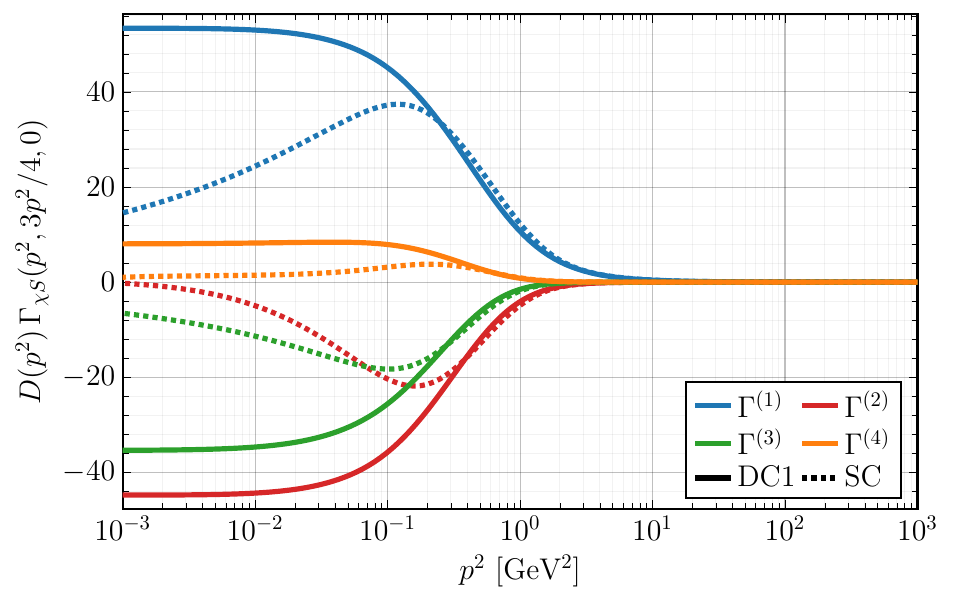}
\includegraphics[width=0.49\textwidth]{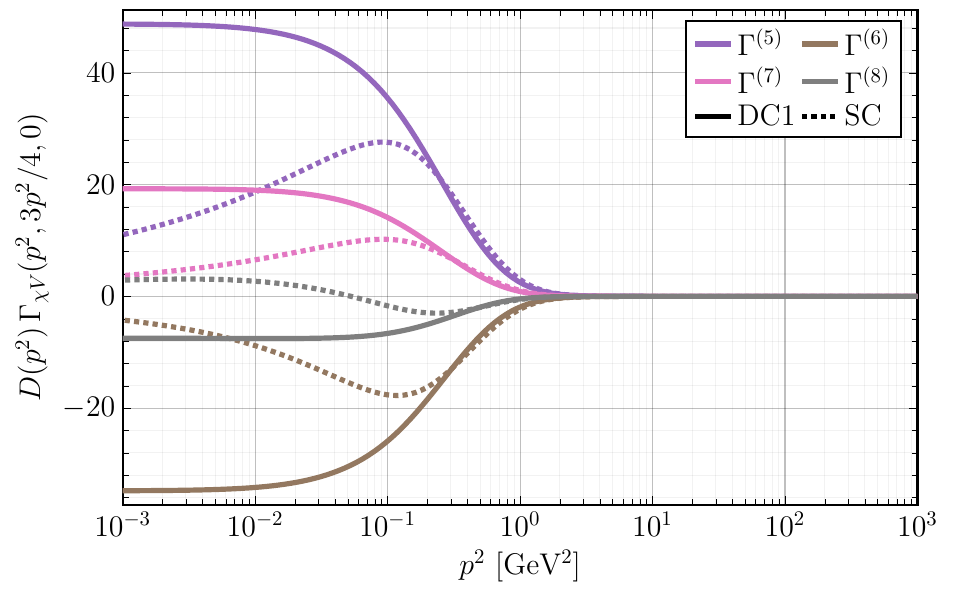}
\caption{The product of the gluon propagator with QGV's form factors  for symmetric momenta, $\chi$S ones in the upper and $\chi$V ones in the lower panel, for the Yang-Mills inputs DC1 (full lines) and SC (dashed lines). 
\label{fig:CompScDc1D}}
\end{figure}

This is the case although one notices quite some differences in between the QGV form factors, 
see Fig.\ \ref{fig:CompScDc1}.

In the quark DSE kernel the product of the gluon propagator times the QGV enters. All other elements in the quark DSE do not show any difference between the decoupling and the scaling solution. Therefore, the 
question arises whether the product of gluon propagator function $D(p^2)=Z(p^2)/p^2$ times a QGV form factor is invariant when changing the type of YM solution. This is clearly not the case, see Fig.\ \ref{fig:CompScDc1D}. However, within the quark DSE kernels and after integration the contributions 
$\Sigma_{A,B}^{(i)}$ become, at least to an astonishing degree, identical, see Fig.~\ref{fig:SigmaComp}.
This behaviour certainly deserves a further study which, however, is beyond the scope here.

\subsection{On the analytic structure of the quark propagator}
\label{subsec:qprop}

As truncations of the QGV in the quark DSE have a direct impact on the resulting analytic structure of the quark propagator, see, e.g., refs.~\cite{Alkofer:2003jj,Pawlowski:2024kxc}, it is interesting to study the corresponding impact of the here presented results for the QGV with the complete kinematics and all form factors kept.

Therefore, in this subsection results for the analytic structure of the quark propagator, or more precisely of
\begin{equation}
\sigma_S(p^2) = \frac 1 {A(p^2)} \frac {M(p^2)} {p^2+M^2(p^2)}
\end{equation}
and
\begin{equation}
\sigma_V(p^2) = \frac 1 {A(p^2)} \frac 1 {p^2+M^2(p^2)}
\end{equation}
will be presented. This structure will be extracted in two steps:  The corresponding numerical results for Euclidean $p^2$ are fitted to a high precision, and then these fits are employed to determine the poles of 
the quantities $\sigma_S(p^2)$ and $\sigma_V(p^2)$.

As we elaborated above results based on the decoupling and the scaling solutions are within numerical errors identical. Therefore also the extracted analytic structure will be the same modulo some small deviations in the pole position due to the numerical precision. We will also remark on the analytic structure for the case the Abelian diagram is added in the way described above. 

Also in this subsection we will focus exclusively on results in the chiral limit, $m_0=0$.
As the determination of pole positions from Euclidean data is a notoriously hard problem several fit functions for $M(p^2)$ and $A(p^2)$ have been employed to test and ensure the robustness of the procedure. 
For all the corresponding results for the quark mass function fits of the form 
\begin{equation}
M(p^2) =\frac{a_1}{(1+p^2/b_1) (\log(e+p^2/c_1))^{7/11}} +\frac{a_2}{(1+p^2/b_2)^2}
\label{eq:fullfit}
\end{equation}
have proven to be highly precise. The first term possesses a pole and a cut, the second term a double pole.

Not surprisingly the parameter $c_1$ is least constrained by the fit. Note that the exponent of the logarithmic term $1-d_m=7/11$ is related to the one-loop value of anomalous dimension of the quark mass $d_m$, and it is quite reassuring that taking this weak logarithmic dependence into account is improving the fit for large Euclidean $p^2$. As $c_1$ is determined to be in the range from (1.4  GeV)$^2$ to (2 GeV)$^2$ the influence of the logarithmic term is negligible for $0< p^2 < 10 $ GeV$^2$. We also checked explicitly that neglecting the logarithmic term has no influence when determining the position of the two lowest-lying poles of $\sigma_S(p^2)$ and $\sigma_V(p^2)$.

Therefore for the reported results fits of the form 
\begin{equation}
M(p^2) =\frac{a_1}{1+p^2/b_1} +\frac{a_2}{(1+p^2/b_2)^2}
\label{eq:mainfit}
\end{equation}
have been used. (NB: Also results from respective lattice calculations can be fitted quite accurately by this form, see, e.g., \cite{Oliveira:2025boh}). As a fit function of the form $\frac{a_1}{1+p^2/b_1}$ (or $\frac{a_1}{(1+p^2/b_1) (\log(e+p^2/c_1))^{7/11}}$) performs very badly we have cross-checked the extracted pole positions and 
residues by a simpler and by an extended fit function,
\begin{equation}
M(p^2) =\frac{a_2}{(1+p^2/b_2)^2}
\label{eq:dimfit}
\end{equation}
and
\begin{equation}
M(p^2) =\frac{a_1}{1+p^2/b_1} +\frac{a_2}{(1+p^2/b_2)^2}+\frac{a_3}{(1+p^2/b_3)^3} \, ,
\label{eq:extfit}
\end{equation}
respectively. Here we only report on poles of the quark propagator which have proven to be stable under these variations. 

The function $A(p^2)$ varies only slowly, nevertheless its curvature in the region of a few GeV requires 
a relatively flexible fit function. We use
\begin{equation}
A(p^2) =c_0 + \frac{c_1}{1+p^2/d_1} +\frac{c_2}{(1+p^2/d_2)^2}
\label{eq:Afit}
\end{equation}
to be quite precise. The parameter $c_0$ coincides, of course, with the asymptotic value of $A(p^2)$. 

As $d_1$ turns out to be of the order of slightly below 2 GeV$^2$, and the function $1/A(p^2)$ thus develops a pole at approximately $-d_1(1+c_1) > d_1$ the low-lying poles of $\sigma_{S,V} (p^2)$  
are generated by the factor $1/(p^2+M^2(p^2))$. Phrased otherwise, for determining low-lying poles the zeros of the expression $p^2+M^2(p^2)$ are the main quantities of interest. 

For the further analysis it is interesting to note that $a_1$ is small, $a_1\ll a_2$, and $b_1$ is a quite large scale. Thus, the by far largest term in the fit \eqref{eq:mainfit} is the double-pole term. This also then explains why the fit \eqref{eq:dimfit} provides qualitatively and semi-quantitively the same pole terms as the two other fits. 

We want to note here that the dominance of the double-pole term is directly related to the form of the kernels which are generated by the QGV's $\chi$V structures  in the quark propagator DSE. This is due to the fact that the $\chi$S tensors contain an odd number of Dirac matrices whereas the $\chi$V tensors an even number. Furthermore, when projecting on equations for the quark propagator functions $A(p^2)$ and $B(p^2)$ one performs Dirac traces. This leads now to crucial differences because Dirac traces with an odd number of Dirac matrices vanish. Consequently,  an additional function $A(q^2)$ in the numerator of the quark propagator's DSE kernel with a $\chi$S structure in the QGV appear and one has an additional $B(q^2)$ with a $\chi$V structure, and vice versa. One consequence of this can be nicely seen in Fig.\ \ref{fig:QpropCon} where the contributions of the QGV functions to the propagator functions $A(p^2)$ and $B(p^2)$ is shown: $\chi$S structures back-feed on $A(p^2)$ and $\chi$V ones on $B(p^2)$. Another consequence is the changed momentum dependence due to different back-coupling mechanisms which then results in the observed  dominance of the double-pole term.

The scale in the double-pole term, $b_2$, is of order 2~GeV$^2$, see below. Furthermore, we note 
that in the fit  \eqref{eq:dimfit} the parameter $a_2$ is given by $a_2=M(0)$.
Plugging the form \eqref{eq:dimfit} into the expression $p^2+M^2(p^2)$ and determining its zeros is 
equivalent to finding the roots of the fifth order polynomial $P(x)=x(1+x/b_2)^4+a_2^2$. For the parameters $a_2=M(0)$ and $b_2\approx$  2 GeV$^2$ one finds five roots, three real ones and one complex-conjugated pair. The appearance of a complex-conjugated pair of poles in the quark propagator might or might not be an artefact of the various employed approximations and truncations. 

The lowest-lying pole can now easily be approximated to occur at
\begin{equation}
P^2_{llp} \approx - \frac {M^2(0)}{(1-M^2(0)/b_2)^4} \approx - 1.2 M^2(0) .
\label{eq:llp}
\end{equation}
The residues at the pole are also easily determined to be 
\begin{align}
\frac 1 {A(0)} \left( 1-\frac{M^2(0)}{b_2}\right)^4 & \mathrm{ for} \,\, \sigma_V \,\,  \mathrm{and} 
\nonumber \\
\frac {M(0)} {A(0)} \left( 1-\frac{M^2(0)}{b_2}\right)^4 & \mathrm{ for} \,\, \sigma_S 
\end{align}
and clearly positive.

We now turn to the numerical analysis based on the fit \eqref{eq:mainfit}. Hereby we quote only poles which stay stable when employing the extended fit \eqref{eq:extfit}.
For the quark propagator based on the decoupling DC1 or the scaling SC solution the fit parameters for the form \eqref{eq:mainfit} are (units are GeV for $a_i$ and GeV$^2$ for $b_i$)
\begin{align}
a_1 & = & ~0.0092 &~~~\mathrm{(DC1)},  &\qquad a_1  &= & ~0.0049  &~~~\mathrm{(SC)} , \nonumber \\
a_2  & = & ~0.312  &~~~\mathrm{(DC1)},  &\qquad a_2  &= & ~0.325  &~~~\mathrm{(SC)}  , \nonumber \\
b_1 & =  & ~11.8   &~~~\mathrm{(DC1)},    &\qquad b_1  &= & ~28.8  &~~~\mathrm{(SC )}  , \nonumber \\
b_2  &= & ~2.01 &~~~\mathrm{(DC1)},     &\qquad b_2  &= & ~2.14  &~~~\mathrm{(SC)} .
\end{align}
Clearly, the double pole term is the by far dominating one in the fit. 

This leads to zeros of the expression $p^2+M^2(p^2)$  and thus poles in the $\sigma$'s at
\begin{align}
p^2 &= - 0.136 {\mathrm {GeV}}^2  & = & - (0.369 {\mathrm {GeV}})^2   ~~~\mathrm{(DC1)}  \nonumber \\
   &= - 0.143 {\mathrm {GeV}}^2 & = &  - (0.378 {\mathrm {GeV}})^2  ~~~\mathrm{(SC)}  \, .
\end{align}
and
\begin{align}
p^2 & = - 0.823 {\mathrm {GeV}}^2 & = & - (0.907 {\mathrm {GeV}})^2  ~~~\mathrm{(DC1)}  \nonumber \\
 & =- 0.875 {\mathrm {GeV}}^2 & = & - (0.935 {\mathrm {GeV}})^2 ~~~\mathrm{(SC)}  \, .
\end{align}
These two lowest-lying poles are extremely stable under variations of the parameterisation and are always on the real axis. The first pole possesses a positive residue\footnote{The lowest-lying pole influences very directly the low-lying meson spectrum, see, e.g., \cite{Alkofer:2026vux}.} (as one would expect for a physical excitation) and is in quantitative agreement with the analysis above. Also, the position of the lowest pole can be fairly well approximated by the expression \eqref{eq:llp}. Additionally, we verified the location of this pole by applying Schlessinger's point method (see, e.g., \cite{Tripolt:2018xeo}) to $\sigma_S(p^2)$ and $\sigma_V(p^2)$.

The second pole possesses a negative residue and thus corresponds to a ``ghost'' excitation.

There is also always a third real pole on the real axis well beyond one GeV$^2$ which then possesses a positive residue again. However, its location depends significantly on the parameterisation.

\begin{figure}
\includegraphics[width=0.49\textwidth]{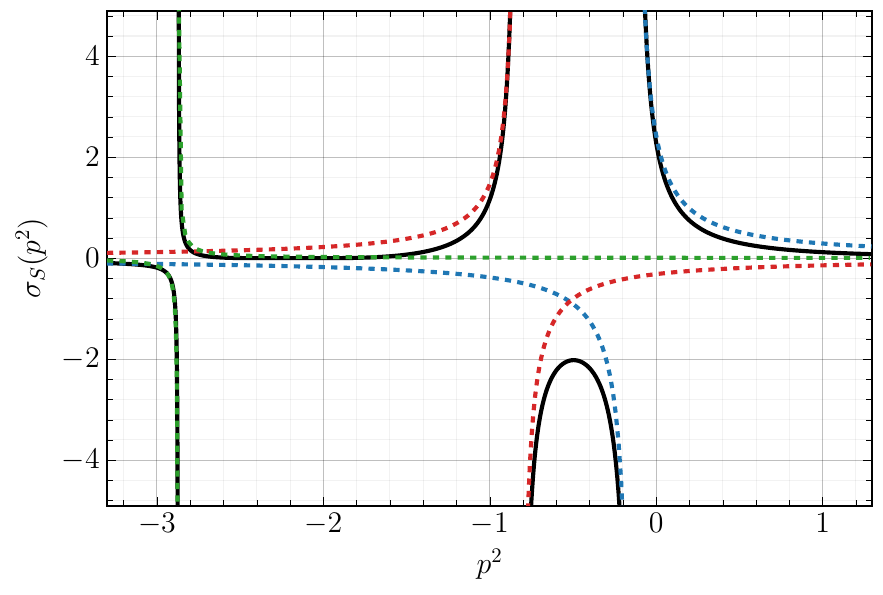}
\caption{$\sigma_S(p^2)$ (full black line) in the chiral limit based on the decoupling solution DC1. The different dashed lines are based on simple pole expressions and there to guide the eye: Blue dashed line $0.33/(x+0.136)$; red dashed line $-0.26/(x+0.823)$;
green dashed line $0.02/(x+2.872)$.
\label{fig:poles}}
\end{figure}

In Fig.~\ref{fig:poles},  based on the fits \eqref{eq:mainfit} for $M(p^2)$ and 
\eqref{eq:Afit} for $A(p^2)$, the function $\sigma_S(p^2)$
in the chiral limit based on the scaling solution DC1 and without the Abelian diagram is shown. One can very nicely identify the features as described above, in particular, the alternating signs for the residues. In particular, single-pole expressions shown as dashed lines allow to extract the pole positions and the residues.

Here a remark is in order: One could na\"ively expect that in the functions $\sigma_S(p^2)$ the factor $M(p^2)$ leads to a double pole at 2.01 GeV$^2$. However, this pole is canceled by the factor $1/A(p^2)$, i.e., the expression $M(p^2)/A(p^2)$ does not display a singularity at the location of double-pole of the fit \eqref{eq:mainfit} for $M(p^2)$.

Adding the Abelian diagram as described above the number, positions and residues of the poles are qualitatively and almost quantitatively the same. Noting, however, that taking into account the Abelian diagram results in a decrease of $M(p^2)$ it is no surprise that the lowest lying pole is somewhat closer to the origin. The numbers for the quark propagator\footnote{Here we employ the solution based on the scaling input.}
are
\begin{align}
a_1 & = 0.0073 \, , \, \, & a_2 &= 0.305 \, \, ,  \nonumber \\
 b_1 & = 20.7\, , \, \, & b_2 &= 2.04 \, ,
\end{align}
which leads to poles at 
\begin{equation}
p^2 = - 0.125 {\mathrm {GeV}}^2 = - (0.354 {\mathrm {GeV}})^2 
\end{equation}
and
\begin{equation}
p^2 = - 0.864 {\mathrm {GeV}}^2 = - (0.930 {\mathrm {GeV}})^2 \, .
\end{equation}

From the above discussion we conclude that taking the full QGV self-consistently into account lead to poles on the real time-like half-axis for the quark propagator. The lowest-lying pole is related to the infrared value of the quark mass function, cf.\ eq.\ \eqref{eq:llp}. The second pole occurs around 900 MeV, and it possesses a negative residue.

To investigate the analytic structure of the QGV's form factors is beyond the scope of the present work, in particular, as one anticipates that they possess at least the Landau singularities, cf.\ ref.\ \cite{Huber:2023uzd}. Nevertheless, the obtained results indicate that the $\chi$V form factors might possess singularities (poles or cuts) at $\bar p^2 \approx -M^2(0)$.

\section{Conclusions}
\label{sec:Conclusion}

In summary, we have studied the QGV in quenched approximation keeping its full kinematical dependence.  We reported on results in the chiral limit for a
self-consistently coupled systems of 3PI Dyson-Schwinger equations for the quark propagator and
the QGV. 

We analysed the resulting QGV's form factors in a basis ordered according to chiral symmetry and dimension. In particular, we also identified the different behaviour and the very distinct impact of the chirally symmetric and the chirality-violating  form factors. The interplay in between the dynamically generated chirality-violating tensor coupling and the scalar quark propagator function is identified to be driving dynamical chiral symmetry breaking, verifying hereby the highly self-consistent nature of chiral symmetry breaking in QCD. 

The numerical results suggest that the non-perturbative scalar and tensor couplings of the gluons to the quarks are related to each other. It would be certainly worthwhile to understand whether some type of off-shell Gordon identity is fulfilled by these form factors. 

We noted that keeping the QGV's full kinematical dependence in the quark propagator's equation is of uttermost importance for the results. In particular, the angular dependence of the QGV's form factors is essential for the resulting quark propagator. Clearly, the QGV does not show any kind of planar degeneracy.  Furthermore, the dependencies on the gluon and the quark momentum differ significantly. Whereas for the non-tree-level form factors the support region in gluon momentum is typically in between one and 1.5 GeV the support region in (averaged) quark momentum is in between 0.6 and 0.7 GeV.

Quite surprisingly, the resulting quark propagator is identical within numerical errors when obtained
either from decoupling solutions or the scaling solution for the Yang-Mills propagators and vertex
functions. This further indicates that the multitude of solutions in the Yang-Mills sector is indeed a non-perturbative gauge-fixing issue. To clarify this further a corresponding study of the quark four-point function and its, based on the here presented results anticipated independence, is highly desirable. 

The interplay of the tensorial coupling and the scalar quark propagator function has a significant impact on the momentum dependence of the quark mass function: With the tensorial coupling taken into account the quark mass function is dominated by a double-pole term whereas such a term is sub-leading in truncations without chirality-violating QGV's form factors. This in turn then implies the existence of low-lying real poles of the quark propagator. The location of the lowest-lying pole is directly related to the quark mass function. As novel feature we identified that the second pole, appearing around 0.9 GeV, possesses a negative residue. 

A determination of the analytic structure of the QGV's form factors, e.g., by contour deformation, will likely further highlight the differences in between the chirally symmetric and chirality-violating structures and thus provide some further insights. Additionally, relations bet\-ween the poles and zeros of the propagators and the vertex functions, maybe indicating some systematic cancelations, are of particular interest.

In a forth coming publication we will report on results with non-vanishing current quark masses and discuss the flavour dependence of the QGV in quenched approximation. Back-coupling the quarks into the 
Yang-Mills sector at this level of truncation will be a demanding task but certainly worth to be studied. Besides a comparison to real QCD a determination of the onset of the conformal window, extending the results of refs.\  \cite{Hopfer:2014zna,Zierler:2023qvz}, will be very interesting.

Last but not least, we want to mention that the phenomenologically very successful 3P0 model for meson decays might have its explanation in the chirality violating QGV form factors  \cite{Alkofer:2023syz}.  Extending such studies to the decays of exotic mesons, aiming to identify observable consequences of the chirality-violating quark-gluon couplings, is currently under investigation.

\section*{Acknowledgements}
We are in particular grateful to Markus Q.\ Huber for the exchange on the importance of the different 
structures in the quark-gluon vertex and the many related discussions. 

We furthermore thank Gernot Eichmann, Christian S.\ Fischer, Eduardo Ferreira, Niko Heinemann, 
Felipe Llanes-Estrada, Axel Maas, Jan M.\ Pawlowski, Alexandre Salas-Bernardez and Fabian Zierler for helpful discussions.

We are grateful to Gernot Eichmann, Markus Q.\ Huber and Jan M.\ Pawlowski for a critical reading of the manuscript.

GW acknowledges financial support by the Austrian Science Fund (FWF) [\href{https://www.fwf.ac.at/en/research-radar/10.55776/PAT6443923}{10.55776/PAT6443923}].

The numerical results presented here have been obtained using the facilities of the HPC center at the University of Graz.

\bigskip 

{\bf Open Access Statement}---For the purpose of open access, the authors have applied a Creative Commons Attribution (CC BY) licence  to any Author Accepted Manuscript version arising.

{\bf Research Data Access Statement}---The data generated for this manuscript can be downloaded from ref.~\cite{data_release}, and the software used to generate it is similarly available from ref.~\cite{code_release}.

\bigskip

\appendix

\section{Translation table for the QGV's tensor basis}\label{app:basis}

In  this Appendix a translation table of the basis given in eqs.\ \eqref{T1} \ldots \eqref{T8} to the tensors $G_i$ and $T_j$ given in table II of ref.\ \cite{Eichmann:2018ytt} as well as fig.\ 8 of ref.\ \cite{Eichmann:2026ttr} is provided.\footnote{To the best of our knowledge this basis has been suggested for the first time in ref.\ \cite{Kizilersu:1995iz}.} As both basis systems are based on the same principles with the exception that in the basis used here the $\chi$S and $\chi$V parts are separated this translation reduces to a re-ordering with some transverse projections:
\begin{align}
R^{(1),\mu} &= \mathcal{T}_k^{\mu \nu} G_{1, \nu} \, , \quad &R^{(5),\mu} &= \mathcal{T}_k^{\mu \nu} G_{3, \nu} \, , \nonumber \\
R^{(2),\mu} &= T_4^\mu \, , \quad &R^{(6),\mu} &= T_3^\mu \, , \nonumber \\
R^{(3),\mu} &= \mathcal{T}_k^{\mu \nu} G_{2, \nu} \, , \quad &R^{(7),\mu} &= \mathcal{T}_k^{\mu \nu} G_{4, \nu} \, , \nonumber \\
R^{(4),\mu} &= T_7^\mu \, , \quad &R^{(8),\mu} &= T_8^\mu \, .
\label{eq:transl}
\end{align}

\section{Fits of the QGV functions} \label{App1}

Here fits of the QGV form factors based on the DC1 input will be provided. First, they are considered as a function of the gluon momentum, $k^2$, for a fixed very small  value of the averaged quark momentum, $\bar p^2_{\mathrm min} = 0.822 \cdot 10^{-4}$ GeV$^2$. Note that for this quark momentum all QGV  form factors are independent of the angular variable $w$.

Allowing for a large enough number of parameters to achieve point-wise good fits one obtains for the $\chi$S ones:
\begin{align}
\Gamma^{(1)} (k^2,\bar p^2_{\mathrm min}, w) & = 1.47 +  \\
& +  \frac{2.18 \left( 1+ y/7.26 +(y/17.0)^2 \right)} {1+y/12.6+(y/8.05)^2 + (y/41.3)^3} \, , \nonumber  \\
\Gamma^{(2)} (k^2,\bar p^2_{\mathrm min}, w) & = \frac {-3.07 }{1+(y/2.08)^2} \, , \\
\Gamma^{(3)} (k^2,\bar p^2_{\mathrm min}, w) & = \frac {-2.40 }{1+(y/1.35)^2} \, , \\
\Gamma^{(4)} (k^2,\bar p^2_{\mathrm min}, w) & = \frac {0.550 \left( 1+(y/0.0430)^2\right) }{1+(y/0.0471)^2+(y/0.360)^{17/4}} \, ,
\end{align}
where $y=k^2/$ GeV$^2$. For the $\chi$V form factors the corresponding fits are:
\begin{align}
\Gamma^{(5)} (k^2,\bar p^2_{\mathrm min}, w) & = \frac{3.35}{1+y/2.70 +(y/2.72)^{5/2}}  \\
\Gamma^{(6)} (k^2,\bar p^2_{\mathrm min}, w) & = \frac {-2.40 }{1+y/3.74+ (y/2.84)^{5/2}} \, , \\
\Gamma^{(7)} (k^2,\bar p^2_{\mathrm min}, w) & = \frac {1.32 \left(1+y/0.0658\right) }{1+y/0.0713+(y/0.98)^{7/2}} \, , \\
\Gamma^{(8)} (k^2,\bar p^2_{\mathrm min}, w) & = \frac {-0.511\left( 1+y/0.0500\right) }{1+y/0.0616+(y/0.668)^{7/2}} \, .
\end{align}

Likewise, for a fixed very small  value of the gluon momentum, $k^2_{\mathrm min} = 0.822 \cdot 10^{-4}$ GeV$^2$, the following fits for the $\chi$S and $\chi$V form factors as function of $\bar p^2$ are point-wise precise:
\begin{align}
\Gamma^{(1)} (k^2_{\mathrm min},\bar p^2, w) & = 1.32 + \\
& + \frac{2.33 \left(1+ x/3.92 +(x/29.4)^2 \right)} {1+x/0.855+(x/8.25)^2 + (x/71.6)^3} \, , \nonumber  \\
\Gamma^{(2)} (k^2_{\mathrm min},\bar p^2, w) & = \frac {-3.06 }{1+x/0.573} \, , \\
\Gamma^{(3)} (k^2_{\mathrm min},\bar p^2, w) & =  \frac {-2.42 \left( 1+x/16.4\right) }{1+x/0.408+(x/0.635)^2} \, , \\
\Gamma^{(4)} (k^2_{\mathrm min},\bar p^2, w) & = \frac {0.550\left(1+x/0.00408\right)}{1+x/0.00414+(x/0.130)^3} \, ,
\end{align}
and
\begin{align}
\Gamma^{(5)} (k^2_{\mathrm min},\bar p^2, w) & = \frac{3.34}{1+x/0.426+(x/0.731)^{9/4}} \, , \nonumber  \\
\Gamma^{(6)} (k^2_{\mathrm min},\bar p^2, w) & = \frac {-2.38 }{1+x/0.444+(x/0.714)^{9/4}} \, , \\
\Gamma^{(7)} (k^2_{\mathrm min},\bar p^2, w) & = \frac {1.32}{1+x/0.300+(x/0.488)^3} \, , \\
\Gamma^{(8)} (k^2_{\mathrm min},\bar p^2, w) & = \frac {-0.511}{1+x/0.598+(x/0.759)^3} \, ,
\end{align}
  where $x=\bar p^2$/GeV$^2$.

In particular, with exception of $\Gamma^{(1)}$, these fits also allow to extract the numerically observed UV behaviour of the QGV's form factors.\footnote{The UV behaviour of $\Gamma^{(1)}$ is known from perturbation theory and determined by the QGV's anomalous dimension which is -9/22 in the Landau gauge in the quenched approximation.} In table~\ref{tab:uv} the numerically extracted exponents $\xi$ of the power laws $(k^2)^{-\xi}$, $(\bar p^2)^{-\xi}$ and $(\mathcal{S}_0)^{-\xi}$ in the UV  are listed. 
In all cases, except for $\Gamma^{(8)}$ vs.\ $\mathcal{S}_0$, power laws are clearly visible in log-log-plots of the numerical results starting at a few GeV$^2$ up to the numerical cutoff. The exponents are extracted from the data with a precision $\pm 1/4$. In particular, one notes that the decrease vs.\ $k^2$ is faster than vs.\ $\bar p^2$. 

\begin{table}
\begin{tabular}{|c|c|c|c|}
\hline
  & vs.\ $k^2$ & vs.\ $\bar p^2$ & vs.\ $\mathcal{S}_0$ \\
\hline
$\Gamma^{(2)}$ & 2 & 1 & 3/2 \\
$\Gamma^{(3)}$ & 2 & 1 & 5/4 \\
$\Gamma^{(4)}$ & 9/4 & 2 & 5/2 \\
\hline
$\Gamma^{(5)}$ & 5/2 & 9/4 & 9/4 \\
$\Gamma^{(6)}$ & 5/2 & 9/4 & 9/4 \\
$\Gamma^{(7)}$ & 7/2 & 3 & 7/2 \\
$\Gamma^{(8)}$ & 7/2 & 3 & --- \\ 
\hline
\end{tabular}
\caption{\label{tab:uv} 
The numerically extracted exponents of the UV power laws for the non-tree-level form factors as function of squared gluon momentum $k^2$, squared averaged quark momentum $\bar p^2$, and the permutation symmetric variable $\mathcal{S}_0$. }
\end{table}

\section{More on the comparison in between scaling and decoupling solutions }\label{App2}

\begin{figure}[t]
\includegraphics[width=0.49\textwidth]{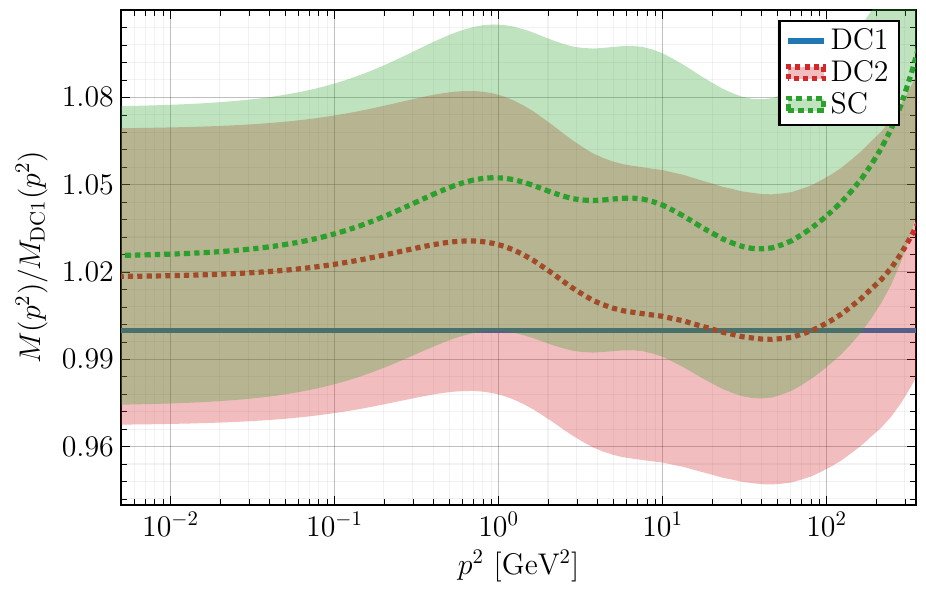}
\caption{Displayed are the ratios of quark mass functions as obtained from the decoupling solution DC2 (dotted red line) and the scaling soution (dotted green line) to the one obtained from the solution DC1. The respective numerical errors are shown as bands in the corresponding colour.
\label{fig:CompMratio}}
\end{figure}

\begin{figure}
\includegraphics[width=0.49\textwidth]{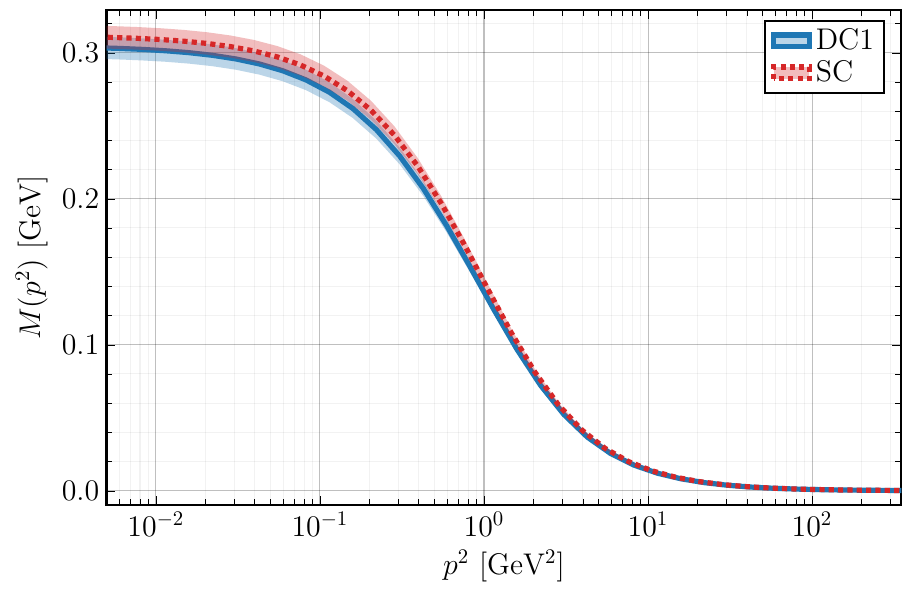}
\caption{Shown are the chiral-limit quark mass functions based on DC1 and SC input with Abelian diagram taken into account and including the respective error bars.
\label{fig:CompMabel}}
\end{figure}

\begin{figure}
\includegraphics[width=0.49\textwidth]{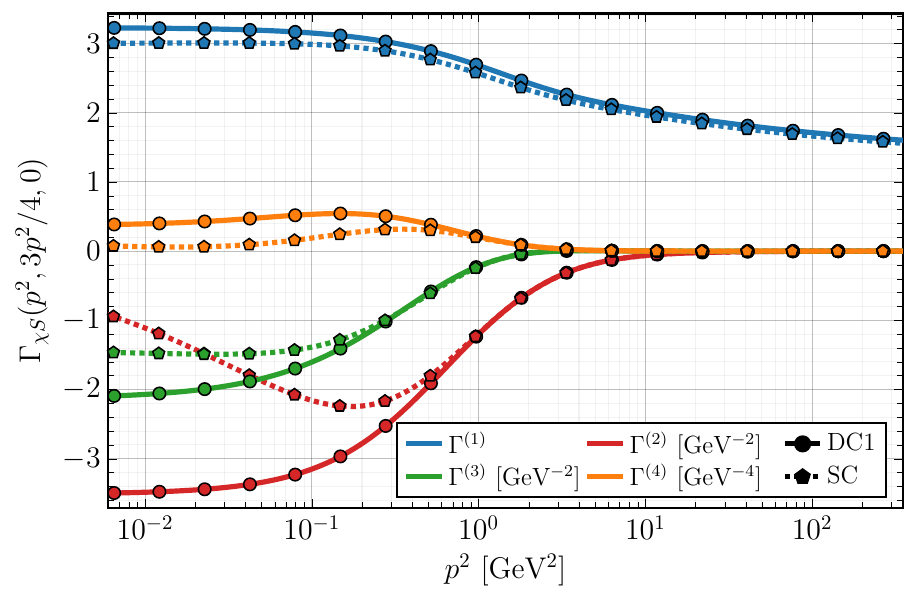}
\includegraphics[width=0.49\textwidth]{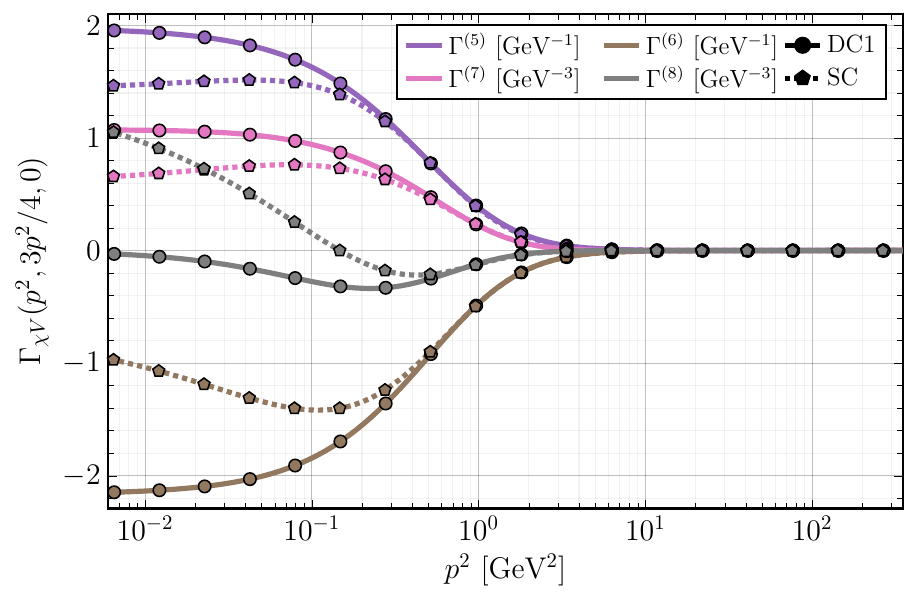}
\caption{Shown are the QGV form factors at symmetric momenta based on DC1 and SC input with Abelian diagram taken into account. The upper panel displays the $\chi$S ones, and the lower panel the $\chi$V ones.
\label{fig:CompQGVabel}}
\end{figure}

\begin{figure}
\includegraphics[width=0.49\textwidth]{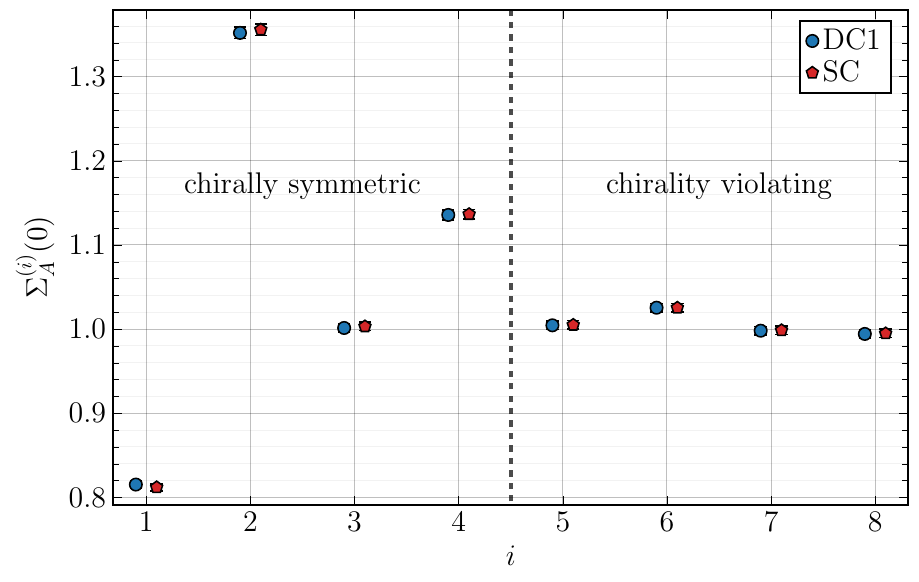}
\includegraphics[width=0.49\textwidth]{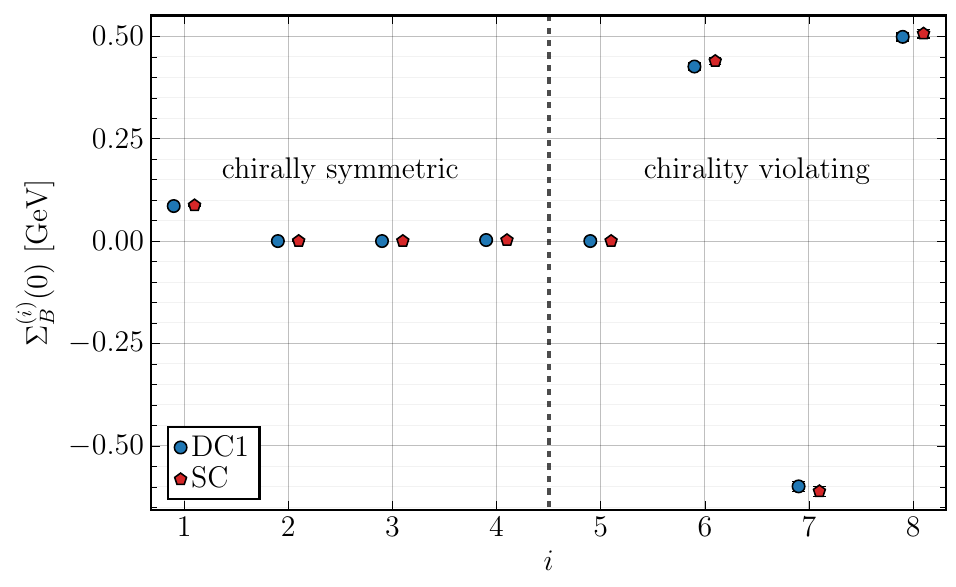}
\caption{Shown are the contributions to the infrared value of the quark propagator functions $A(p^2)$ (upper panel) and $B(p^2)$ (lower panel) based on DC1 and SC input with Abelian diagram taken into account.
\label{fig:CompContABabel}}
\end{figure}

First, to complement the presentation in Fig.\ \ref{fig:CompMdiff},
 in Fig.\ \ref{fig:CompMratio} instead of the differences of the chiral-limit quark mass functions
as obtained from the decoupling solution DC2 and the scaling solution to the one
obtained from the solution DC1 the corresponding ratios are displayed allowing to read off the relative error. Taking into account that the chiral-limit quark mass functions decrease rapidly for large momenta, and thus the ratios above a 100 GeV$^2$ suffer from a ``small-basis-effect'', again the agreement within error bounds are verified.

As stated in Sect.\ \ref{sec:dses} we estimate the importance of the Abelian diagram by adding  to the QGV obtained self-consistently with the non-Abelian diagram the result for the Abelian diagram. This amounts to using for the QGV appearing three times in the Abelian diagram the one determined from the non-Abelian diagram. Due to the smallness of this correction and the then enormous required cost in computing time we refrain from iterating this system until self-consistency. Furthermore, for this calculation only a comparison between DC1 and SC input   is performed.

From Fig.\ \ref{fig:CompMabel} one can infer nicely that  also for the case with the Abelian diagram included the difference between the solutions based on DC1 and SC input is smaller than the numerical errors.

As seen from Fig.\ \ref{fig:CompQGVabel} the change of the QGV form factors in between the DC1 and SC solution are very comparable to the ones obtained without the Abelian diagram, and this despite the approximate way the Abelian diagram is taken into account. As in the case with the non-Abelian diagram only, the different contributions to the quark propagator functions are within error bounds identical. In Fig.\ \ref{fig:CompContABabel} the corresponding infrared values are displayed demonstrating the level of agreement\footnote{Note that with Abelian diagram added $\Sigma^{(7)}$ and $\Sigma^{(8)}$ are individually sizeable but cancel each other pretty precisely thus matching the picture without Abelian diagram. We attribute this as a peculiarity due to the tensor basis, respectively, to the specific form of $R^{(7)}$ \eqref{T7} and $R^{(8)}$ \eqref{T8}. 
It will be interesting to see whether the individual sizes of  $\Sigma^{(7)}$ and $\Sigma^{(8)}$ remain if the Abelian diagram is also iterated.}.

\bibliographystyle{elsarticle-num} 
\bibliography{refs}

\begin{thebibliography}{100}
\expandafter\ifx\csname url\endcsname\relax
  \def\url#1{\texttt{#1}}\fi
\expandafter\ifx\csname urlprefix\endcsname\relax\def\urlprefix{URL }\fi
\expandafter\ifx\csname href\endcsname\relax
  \def\href#1#2{#2} \def\path#1{#1}\fi

\bibitem{Gross:2022hyw}
F.~Gross, et~al., {50 Years of Quantum Chromodynamics}, Eur. Phys. J. C 83
  (2023) 1125.
\newblock \href {http://arxiv.org/abs/2212.11107} {\path{arXiv:2212.11107}},
  \href {http://dx.doi.org/10.1140/epjc/s10052-023-11949-2}
  {\path{doi:10.1140/epjc/s10052-023-11949-2}}.

\bibitem{Pagels:1977xv}
H.~Pagels, {A Nonperturbative Approach to Quantum Chromodynamics}, Phys. Rev. D
  15 (1977) 2991.
\newblock \href {http://dx.doi.org/10.1103/PhysRevD.15.2991}
  {\path{doi:10.1103/PhysRevD.15.2991}}.

\bibitem{Marciano:1977su}
W.~J. Marciano, H.~Pagels, {Quantum Chromodynamics: A Review}, Phys. Rept. 36
  (1978) 137.
\newblock \href {http://dx.doi.org/10.1016/0370-1573(78)90208-9}
  {\path{doi:10.1016/0370-1573(78)90208-9}}.

\bibitem{Pagels:1978ba}
H.~Pagels, {Dynamical Chiral Symmetry Breaking in Quantum Chromodynamics},
  Phys. Rev. D 19 (1979) 3080.
\newblock \href {http://dx.doi.org/10.1103/PhysRevD.19.3080}
  {\path{doi:10.1103/PhysRevD.19.3080}}.

\bibitem{Gusynin:1978tr}
V.~P. Gusynin, V.~A. Miransky, {On the Vacuum Rearrangement in Massless
  Chromodynamics}, Phys. Lett. B 76 (1978) 585--588.
\newblock \href {http://dx.doi.org/10.1016/0370-2693(78)90860-2}
  {\path{doi:10.1016/0370-2693(78)90860-2}}.

\bibitem{Eichmann:2025wgs}
G.~Eichmann, {Hadron physics with functional methods,~}\href
  {http://arxiv.org/abs/2503.10397} {\path{arXiv:2503.10397}}.

\bibitem{Huber:2025cbd}
M.~Q. Huber, {A beginner's guide to functional methods in particle
  physics,~}\href {http://arxiv.org/abs/2510.18960} {\path{arXiv:2510.18960}}.

\bibitem{Braun:2025gvq}
J.~Braun, A.~Gei{\ss}el, J.~M. Pawlowski, F.~R. Sattler, N.~Wink, {Juggling
  with tensor bases in functional approaches}, Annals Phys. 484 (2026) 170250.
\newblock \href {http://arxiv.org/abs/2503.05580} {\path{arXiv:2503.05580}},
  \href {http://dx.doi.org/10.1016/j.aop.2025.170250}
  {\path{doi:10.1016/j.aop.2025.170250}}.

\bibitem{Eichmann:2026ttr}
G.~Eichmann, {Getting a handle on correlation functions,~}\href
  {http://arxiv.org/abs/2603.00804} {\path{arXiv:2603.00804}}.

\bibitem{Skullerud:1995uk}
J.~I. Skullerud, {A Study of the quark - gluon vertex}, Nucl. Phys. B Proc.
  Suppl. 47 (1996) 398--401.
\newblock \href {http://arxiv.org/abs/hep-lat/9509068}
  {\path{arXiv:hep-lat/9509068}}, \href
  {http://dx.doi.org/10.1016/0920-5632(96)00082-5}
  {\path{doi:10.1016/0920-5632(96)00082-5}}.

\bibitem{Skullerud:1997wc}
J.~I. Skullerud, {The Running coupling from the quark gluon vertex}, Nucl.
  Phys. B Proc. Suppl. 63 (1998) 242--244.
\newblock \href {http://arxiv.org/abs/hep-lat/9710044}
  {\path{arXiv:hep-lat/9710044}}, \href
  {http://dx.doi.org/10.1016/S0920-5632(97)00733-0}
  {\path{doi:10.1016/S0920-5632(97)00733-0}}.

\bibitem{Skullerud:2001bu}
J.~Skullerud, A.~Kizilersu, A.~G. Williams, {Quark gluon vertex in a momentum
  subtraction scheme}, Nucl. Phys. B Proc. Suppl. 106 (2002) 841--843.
\newblock \href {http://arxiv.org/abs/hep-lat/0109027}
  {\path{arXiv:hep-lat/0109027}}, \href
  {http://dx.doi.org/10.1016/S0920-5632(01)01861-8}
  {\path{doi:10.1016/S0920-5632(01)01861-8}}.

\bibitem{Skullerud:2002sk}
J.~Skullerud, P.~O. Bowman, A.~Kizilersu, {The Nonperturbative quark gluon
  vertex}, in: {5th International Conference on Quark Confinement and the
  Hadron Spectrum}, 2002, pp. 270--272.
\newblock \href {http://arxiv.org/abs/hep-lat/0212011}
  {\path{arXiv:hep-lat/0212011}}, \href
  {http://dx.doi.org/10.1142/9789812704269_0033}
  {\path{doi:10.1142/9789812704269_0033}}.

\bibitem{Skullerud:2004pt}
J.~I. Skullerud, A.~Kizilersu, P.~O. Bowman, D.~B. Leinweber, A.~G. Williams,
  {Looking inside the quark-gluon vertex}, Nucl. Phys. B Proc. Suppl. 128
  (2004) 117--124.
\newblock \href {http://dx.doi.org/10.1016/S0920-5632(03)02467-8}
  {\path{doi:10.1016/S0920-5632(03)02467-8}}.

\bibitem{Lin:2005zd}
H.-W. Lin, {Quark-gluon vertex with an off-shell O(a)-improved chiral fermion
  action}, Phys. Rev. D 73 (2006) 094511.
\newblock \href {http://arxiv.org/abs/hep-lat/0510110}
  {\path{arXiv:hep-lat/0510110}}, \href
  {http://dx.doi.org/10.1103/PhysRevD.73.094511}
  {\path{doi:10.1103/PhysRevD.73.094511}}.

\bibitem{Kizilersu:2006et}
A.~Kizilersu, D.~B. Leinweber, J.-I. Skullerud, A.~G. Williams, {Quark-gluon
  vertex in general kinematics}, Eur. Phys. J. C 50 (2007) 871--875.
\newblock \href {http://arxiv.org/abs/hep-lat/0610078}
  {\path{arXiv:hep-lat/0610078}}, \href
  {http://dx.doi.org/10.1140/epjc/s10052-007-0250-6}
  {\path{doi:10.1140/epjc/s10052-007-0250-6}}.

\bibitem{Furui:2008zm}
S.~Furui, {A Study of quark-gluon vertices using the lattice Coulomb gauge
  domain wall fermion}, PoS LATTICE2008 (2008) 130.
\newblock \href {http://arxiv.org/abs/0808.1796} {\path{arXiv:0808.1796}},
  \href {http://dx.doi.org/10.22323/1.066.0130}
  {\path{doi:10.22323/1.066.0130}}.

\bibitem{Oliveira:2016muq}
O.~Oliveira, A.~K{\i}z{\i}lersu, P.~J. Silva, J.-I. Skullerud, A.~Sternbeck,
  A.~G. Williams, {Lattice Landau gauge quark propagator and the quark-gluon
  vertex}, Acta Phys. Polon. Supp. 9 (2016) 363--368.
\newblock \href {http://arxiv.org/abs/1605.09632} {\path{arXiv:1605.09632}},
  \href {http://dx.doi.org/10.5506/APhysPolBSupp.9.363}
  {\path{doi:10.5506/APhysPolBSupp.9.363}}.

\bibitem{Sternbeck:2017ntv}
A.~Sternbeck, P.-H. Balduf, A.~K\i{}z\i{}lersu, O.~Oliveira, P.~J. Silva, J.-I.
  Skullerud, A.~G. Williams, {Triple-gluon and quark-gluon vertex from lattice
  QCD in Landau gauge}, PoS LATTICE2016 (2017) 349.
\newblock \href {http://arxiv.org/abs/1702.00612} {\path{arXiv:1702.00612}},
  \href {http://dx.doi.org/10.22323/1.256.0349}
  {\path{doi:10.22323/1.256.0349}}.

\bibitem{Kizilersu:2021jen}
A.~K\i{}z\i{}lers\"u, O.~Oliveira, P.~J. Silva, J.-I. Skullerud, A.~Sternbeck,
  {Quark-gluon vertex from Nf=2 lattice QCD}, Phys. Rev. D 103~(11) (2021)
  114515.
\newblock \href {http://arxiv.org/abs/2103.02945} {\path{arXiv:2103.02945}},
  \href {http://dx.doi.org/10.1103/PhysRevD.103.114515}
  {\path{doi:10.1103/PhysRevD.103.114515}}.

\bibitem{Skullerud:2021pel}
J.-I. Skullerud, A.~K\i{}z\i{}lers\"u, O.~Oliveira, P.~Silva, A.~Sternbeck,
  {Quark-gluon vertex with 2 flavours of O(a) improved Wilson fermions}, PoS
  LATTICE2021 (2022) 305.
\newblock \href {http://arxiv.org/abs/2111.13455} {\path{arXiv:2111.13455}},
  \href {http://dx.doi.org/10.22323/1.396.0305}
  {\path{doi:10.22323/1.396.0305}}.

\bibitem{Marques:2023cmi}
J.~Marques, G.~Kalusche, T.~Mendes, P.~J. Silva, J.-I. Skullerud, O.~Oliveira,
  {The quark propagator and quark-gluon vertex from lattice QCD at finite
  temperature}, PoS LATTICE2022 (2023) 280.
\newblock \href {http://arxiv.org/abs/2301.10607} {\path{arXiv:2301.10607}},
  \href {http://dx.doi.org/10.22323/1.430.0280}
  {\path{doi:10.22323/1.430.0280}}.

\bibitem{Davydychev:2000rt}
A.~I. Davydychev, P.~Osland, L.~Saks, {Quark gluon vertex in arbitrary gauge
  and dimension}, Phys. Rev. D 63 (2001) 014022.
\newblock \href {http://arxiv.org/abs/hep-ph/0008171}
  {\path{arXiv:hep-ph/0008171}}, \href
  {http://dx.doi.org/10.1103/PhysRevD.63.014022}
  {\path{doi:10.1103/PhysRevD.63.014022}}.

\bibitem{Davydychev:2000yy}
A.~I. Davydychev, P.~Osland, L.~Saks, {One loop results for the quark gluon
  vertex in arbitrary dimension}, Nucl. Phys. B Proc. Suppl. 89 (2000)
  277--282.
\newblock \href {http://arxiv.org/abs/hep-ph/0008202}
  {\path{arXiv:hep-ph/0008202}}, \href
  {http://dx.doi.org/10.1016/S0920-5632(00)00856-2}
  {\path{doi:10.1016/S0920-5632(00)00856-2}}.

\bibitem{Gracey:2011vw}
J.~A. Gracey, {Two loop QCD vertices at the symmetric point}, Phys. Rev. D 84
  (2011) 085011.
\newblock \href {http://arxiv.org/abs/1108.4806} {\path{arXiv:1108.4806}},
  \href {http://dx.doi.org/10.1103/PhysRevD.84.085011}
  {\path{doi:10.1103/PhysRevD.84.085011}}.

\bibitem{Huang:2020fiu}
G.~Huang, P.~Zhuang, {Quark-quark-gluon vertex for heavy quarks up to order
  $1/m^5$}, Phys. Rev. D 102~(1) (2020) 014034.
\newblock \href {http://arxiv.org/abs/2004.10545} {\path{arXiv:2004.10545}},
  \href {http://dx.doi.org/10.1103/PhysRevD.102.014034}
  {\path{doi:10.1103/PhysRevD.102.014034}}.

\bibitem{vonSmekal:1990kgh}
L.~von Smekal, P.~A. Amundsen, R.~Alkofer, {A Covariant model for dynamical
  chiral symmetry breaking in QCD}, Nucl. Phys. A 529 (1991) 633--652.
\newblock \href {http://dx.doi.org/10.1016/0375-9474(91)90589-X}
  {\path{doi:10.1016/0375-9474(91)90589-X}}.

\bibitem{Bender:2002as}
A.~Bender, W.~Detmold, C.~D. Roberts, A.~W. Thomas, {Bethe-Salpeter equation
  and a nonperturbative quark gluon vertex}, Phys. Rev. C 65 (2002) 065203.
\newblock \href {http://arxiv.org/abs/nucl-th/0202082}
  {\path{arXiv:nucl-th/0202082}}, \href
  {http://dx.doi.org/10.1103/PhysRevC.65.065203}
  {\path{doi:10.1103/PhysRevC.65.065203}}.

\bibitem{Watson:2004kd}
P.~Watson, W.~Cassing, P.~C. Tandy, {Bethe-Salpeter meson masses beyond ladder
  approximation}, Few Body Syst. 35 (2004) 129--153.
\newblock \href {http://arxiv.org/abs/hep-ph/0406340}
  {\path{arXiv:hep-ph/0406340}}, \href
  {http://dx.doi.org/10.1007/s00601-004-0067-x}
  {\path{doi:10.1007/s00601-004-0067-x}}.

\bibitem{Bhagwat:2004kj}
M.~S. Bhagwat, P.~C. Tandy, {Quark-gluon vertex model and lattice-QCD data},
  Phys. Rev. D 70 (2004) 094039.
\newblock \href {http://arxiv.org/abs/hep-ph/0407163}
  {\path{arXiv:hep-ph/0407163}}, \href
  {http://dx.doi.org/10.1103/PhysRevD.70.094039}
  {\path{doi:10.1103/PhysRevD.70.094039}}.

\bibitem{Holl:2004qn}
A.~Holl, A.~Krassnigg, C.~D. Roberts, {Confinement, DCSB, bound states, and the
  quark-gluon vertex}, Nucl. Phys. B Proc. Suppl. 141 (2005) 47--52.
\newblock \href {http://arxiv.org/abs/nucl-th/0408015}
  {\path{arXiv:nucl-th/0408015}}, \href
  {http://dx.doi.org/10.1016/j.nuclphysbps.2004.12.009}
  {\path{doi:10.1016/j.nuclphysbps.2004.12.009}}.

\bibitem{Fischer:2004ym}
C.~S. Fischer, F.~J. Llanes-Estrada, R.~Alkofer, {Dynamical mass generation in
  Landau gauge QCD}, Nucl. Phys. B Proc. Suppl. 141 (2005) 128--133.
\newblock \href {http://arxiv.org/abs/hep-ph/0407294}
  {\path{arXiv:hep-ph/0407294}}, \href
  {http://dx.doi.org/10.1016/j.nuclphysbps.2004.12.020}
  {\path{doi:10.1016/j.nuclphysbps.2004.12.020}}.

\bibitem{Llanes-Estrada:2004hnb}
F.~J. Llanes-Estrada, C.~S. Fischer, R.~Alkofer, {Semiperturbative construction
  for the quark-gluon vertex}, Nucl. Phys. B Proc. Suppl. 152 (2006) 43--46.
\newblock \href {http://arxiv.org/abs/hep-ph/0407332}
  {\path{arXiv:hep-ph/0407332}}, \href
  {http://dx.doi.org/10.1016/j.nuclphysbps.2005.08.008}
  {\path{doi:10.1016/j.nuclphysbps.2005.08.008}}.

\bibitem{Alkofer:2006gz}
R.~Alkofer, C.~S. Fischer, F.~J. Llanes-Estrada, {Dynamically induced scalar
  quark confinement}, Mod. Phys. Lett. A 23 (2008) 1105--1113.
\newblock \href {http://arxiv.org/abs/hep-ph/0607293}
  {\path{arXiv:hep-ph/0607293}}, \href
  {http://dx.doi.org/10.1142/S021773230802700X}
  {\path{doi:10.1142/S021773230802700X}}.

\bibitem{Matevosyan:2006bk}
H.~H. Matevosyan, A.~W. Thomas, P.~C. Tandy, {Quark-gluon vertex dressing and
  meson masses beyond ladder-rainbow truncation}, Phys. Rev. C 75 (2007)
  045201.
\newblock \href {http://arxiv.org/abs/nucl-th/0605057}
  {\path{arXiv:nucl-th/0605057}}, \href
  {http://dx.doi.org/10.1103/PhysRevC.75.045201}
  {\path{doi:10.1103/PhysRevC.75.045201}}.

\bibitem{Matevosyan:2007cx}
H.~H. Matevosyan, A.~W. Thomas, P.~C. Tandy, {Consequences Of Fully Dressing
  Quark-Gluon Vertex Function With Two-Point Gluon Lines}, J. Phys. G 34 (2007)
  2153--2164.
\newblock \href {http://arxiv.org/abs/0706.2393} {\path{arXiv:0706.2393}},
  \href {http://dx.doi.org/10.1088/0954-3899/34/10/005}
  {\path{doi:10.1088/0954-3899/34/10/005}}.

\bibitem{Alkofer:2008et}
R.~Alkofer, C.~S. Fischer, R.~Williams, {U(A)(1) anomaly and eta-prime mass
  from an infrared singular quark-gluon vertex}, Eur. Phys. J. A 38 (2008)
  53--60.
\newblock \href {http://arxiv.org/abs/0804.3478} {\path{arXiv:0804.3478}},
  \href {http://dx.doi.org/10.1140/epja/i2008-10646-x}
  {\path{doi:10.1140/epja/i2008-10646-x}}.

\bibitem{Alkofer:2008tt}
R.~Alkofer, C.~S. Fischer, F.~J. Llanes-Estrada, K.~Schwenzer, {The Quark-gluon
  vertex in Landau gauge QCD: Its role in dynamical chiral symmetry breaking
  and quark confinement}, Annals Phys. 324 (2009) 106--172.
\newblock \href {http://arxiv.org/abs/0804.3042} {\path{arXiv:0804.3042}},
  \href {http://dx.doi.org/10.1016/j.aop.2008.07.001}
  {\path{doi:10.1016/j.aop.2008.07.001}}.

\bibitem{He:2009sj}
H.-x. He, {Transverse Symmetry Transformations and the Quark-Gluon Vertex
  Function in QCD}, Phys. Rev. D 80 (2009) 016004.
\newblock \href {http://arxiv.org/abs/0906.2834} {\path{arXiv:0906.2834}},
  \href {http://dx.doi.org/10.1103/PhysRevD.80.016004}
  {\path{doi:10.1103/PhysRevD.80.016004}}.

\bibitem{Fischer:2009jm}
C.~S. Fischer, R.~Williams, {Probing the gluon self-interaction in light
  mesons}, Phys. Rev. Lett. 103 (2009) 122001.
\newblock \href {http://arxiv.org/abs/0905.2291} {\path{arXiv:0905.2291}},
  \href {http://dx.doi.org/10.1103/PhysRevLett.103.122001}
  {\path{doi:10.1103/PhysRevLett.103.122001}}.

\bibitem{Windisch:2012de}
A.~Windisch, M.~Hopfer, R.~Alkofer, {Towards a self-consistent solution of the
  Landau gauge quark-gluon vertex Dyson-Schwinger equation}, Acta Phys. Polon.
  Supp. 6 (2013) 347--352.
\newblock \href {http://arxiv.org/abs/1210.8428} {\path{arXiv:1210.8428}},
  \href {http://dx.doi.org/10.5506/APhysPolBSupp.6.347}
  {\path{doi:10.5506/APhysPolBSupp.6.347}}.

\bibitem{Hopfer:2012cnq}
M.~Hopfer, A.~Windisch, R.~Alkofer, {The Quark-Gluon Vertex in Landau gauge
  QCD}, PoS ConfinementX (2012) 073.
\newblock \href {http://arxiv.org/abs/1301.3672} {\path{arXiv:1301.3672}},
  \href {http://dx.doi.org/10.22323/1.171.0073}
  {\path{doi:10.22323/1.171.0073}}.

\bibitem{Aguilar:2012pfj}
A.~C. Aguilar, D.~Binosi, J.~C. Cardona, J.~Papavassiliou, {Nonperturbative
  results on the quark-gluon vertex}, PoS ConfinementX (2012) 103.
\newblock \href {http://arxiv.org/abs/1301.4057} {\path{arXiv:1301.4057}},
  \href {http://dx.doi.org/10.22323/1.171.0103}
  {\path{doi:10.22323/1.171.0103}}.

\bibitem{Ayala:2012pb}
A.~Ayala, A.~Bashir, D.~Binosi, M.~Cristoforetti, J.~Rodriguez-Quintero, {Quark
  flavour effects on gluon and ghost propagators}, Phys. Rev. D 86 (2012)
  074512.
\newblock \href {http://arxiv.org/abs/1208.0795} {\path{arXiv:1208.0795}},
  \href {http://dx.doi.org/10.1103/PhysRevD.86.074512}
  {\path{doi:10.1103/PhysRevD.86.074512}}.

\bibitem{Rojas:2013tza}
E.~Rojas, J.~P. B.~C. de~Melo, B.~El-Bennich, O.~Oliveira, T.~Frederico, {On
  the Quark-Gluon Vertex and Quark-Ghost Kernel: combining Lattice Simulations
  with Dyson-Schwinger equations}, JHEP 10 (2013) 193.
\newblock \href {http://arxiv.org/abs/1306.3022} {\path{arXiv:1306.3022}},
  \href {http://dx.doi.org/10.1007/JHEP10(2013)193}
  {\path{doi:10.1007/JHEP10(2013)193}}.

\bibitem{Alkofer:2013qoc}
R.~Alkofer, G.~Eichmann, C.~S. Fischer, M.~Hopfer, M.~Vujinovic, R.~Williams,
  A.~Windisch, {On propagators and three-point functions in Landau gauge QCD
  and QCD-like theories}, PoS QCD-TNT-III (2013) 003.
\newblock \href {http://arxiv.org/abs/1405.7310} {\path{arXiv:1405.7310}},
  \href {http://dx.doi.org/10.22323/1.193.0003}
  {\path{doi:10.22323/1.193.0003}}.

\bibitem{Williams:2014iea}
R.~Williams, {The quark-gluon vertex in Landau gauge bound-state studies}, Eur.
  Phys. J. A 51~(5) (2015) 57.
\newblock \href {http://arxiv.org/abs/1404.2545} {\path{arXiv:1404.2545}},
  \href {http://dx.doi.org/10.1140/epja/i2015-15057-4}
  {\path{doi:10.1140/epja/i2015-15057-4}}.

\bibitem{Aguilar:2014lha}
A.~C. Aguilar, D.~Binosi, D.~Iba{\~n}ez, J.~Papavassiliou, {New method for
  determining the quark-gluon vertex}, Phys. Rev. D 90~(6) (2014) 065027.
\newblock \href {http://arxiv.org/abs/1405.3506} {\path{arXiv:1405.3506}},
  \href {http://dx.doi.org/10.1103/PhysRevD.90.065027}
  {\path{doi:10.1103/PhysRevD.90.065027}}.

\bibitem{Rojas:2014tya}
E.~Rojas, B.~El-Bennich, J.~P. B.~C. De~Melo, M.~A. Paracha, {Insights into the
  Quark{\textendash}Gluon Vertex from Lattice QCD and Meson Spectroscopy}, Few
  Body Syst. 56~(6-9) (2015) 639--644.
\newblock \href {http://arxiv.org/abs/1409.8620} {\path{arXiv:1409.8620}},
  \href {http://dx.doi.org/10.1007/s00601-015-1020-x}
  {\path{doi:10.1007/s00601-015-1020-x}}.

\bibitem{Ayala:2014uua}
A.~Ayala, J.~J. Cobos-Mart{\'\i}nez, M.~Loewe, M.~E. Tejeda-Yeomans, R.~Zamora,
  {Finite temperature quark-gluon vertex with a magnetic field in the Hard
  Thermal Loop approximation}, Phys. Rev. D 91~(1) (2015) 016007.
\newblock \href {http://arxiv.org/abs/1410.6388} {\path{arXiv:1410.6388}},
  \href {http://dx.doi.org/10.1103/PhysRevD.91.016007}
  {\path{doi:10.1103/PhysRevD.91.016007}}.

\bibitem{Mitter:2014wpa}
M.~Mitter, J.~M. Pawlowski, N.~Strodthoff, {Chiral symmetry breaking in
  continuum QCD}, Phys. Rev. D 91 (2015) 054035.
\newblock \href {http://arxiv.org/abs/1411.7978} {\path{arXiv:1411.7978}},
  \href {http://dx.doi.org/10.1103/PhysRevD.91.054035}
  {\path{doi:10.1103/PhysRevD.91.054035}}.

\bibitem{Chen:2015mda}
H.~Chen, J.~B. Wei, M.~Baldo, G.~F. Burgio, H.~J. Schulze, {Hybrid neutron
  stars with the Dyson-Schwinger quark model and various quark-gluon vertices},
  Phys. Rev. D 91~(10) (2015) 105002.
\newblock \href {http://arxiv.org/abs/1503.02795} {\path{arXiv:1503.02795}},
  \href {http://dx.doi.org/10.1103/PhysRevD.91.105002}
  {\path{doi:10.1103/PhysRevD.91.105002}}.

\bibitem{Pelaez:2015tba}
M.~Pel\'aez, M.~Tissier, N.~Wschebor, {Quark-gluon vertex from the Landau gauge
  Curci-Ferrari model}, Phys. Rev. D 92~(4) (2015) 045012.
\newblock \href {http://arxiv.org/abs/1504.05157} {\path{arXiv:1504.05157}},
  \href {http://dx.doi.org/10.1103/PhysRevD.92.045012}
  {\path{doi:10.1103/PhysRevD.92.045012}}.

\bibitem{Binosi:2016wcx}
D.~Binosi, L.~Chang, J.~Papavassiliou, S.-X. Qin, C.~D. Roberts, {Natural
  constraints on the gluon-quark vertex}, Phys. Rev. D 95~(3) (2017) 031501.
\newblock \href {http://arxiv.org/abs/1609.02568} {\path{arXiv:1609.02568}},
  \href {http://dx.doi.org/10.1103/PhysRevD.95.031501}
  {\path{doi:10.1103/PhysRevD.95.031501}}.

\bibitem{Williams:2016zpc}
R.~Williams, H.~Sanchis-Alepuz, {Influence of the nonperturbative quark-gluon
  vertex on the meson and baryon spectrum}, AIP Conf. Proc. 1701~(1) (2016)
  040021.
\newblock \href {http://dx.doi.org/10.1063/1.4938638}
  {\path{doi:10.1063/1.4938638}}.

\bibitem{Blum:2015lsa}
A.~L. Blum, R.~Alkofer, M.~Q. Huber, A.~Windisch, {Unquenching the three-gluon
  vertex: A status report}, Acta Phys. Polon. Supp. 8~(2) (2015) 321.
\newblock \href {http://arxiv.org/abs/1506.04275} {\path{arXiv:1506.04275}},
  \href {http://dx.doi.org/10.5506/APhysPolBSupp.8.321}
  {\path{doi:10.5506/APhysPolBSupp.8.321}}.

\bibitem{Blum:2016fib}
A.~L. Blum, R.~Alkofer, M.~Q. Huber, A.~Windisch, {Three-point vertex functions
  in Yang-Mills Theory and QCD in Landau gauge}, EPJ Web Conf. 137 (2017)
  03001.
\newblock \href {http://arxiv.org/abs/1611.04827} {\path{arXiv:1611.04827}},
  \href {http://dx.doi.org/10.1051/epjconf/201713703001}
  {\path{doi:10.1051/epjconf/201713703001}}.

\bibitem{Gomez-Rocha:2014vsa}
M.~G{\'o}mez-Rocha, T.~Hilger, A.~Krassnigg, {First Look at
  Heavy{\textendash}Light Mesons with a Dressed Quark{\textendash}Gluon
  Vertex}, Few Body Syst. 56~(6-9) (2015) 475--480.
\newblock \href {http://arxiv.org/abs/1408.1077} {\path{arXiv:1408.1077}},
  \href {http://dx.doi.org/10.1007/s00601-014-0938-8}
  {\path{doi:10.1007/s00601-014-0938-8}}.

\bibitem{Gomez-Rocha:2015qga}
M.~Gomez-Rocha, T.~Hilger, A.~Krassnigg, {Effects of a dressed quark-gluon
  vertex in pseudoscalar heavy-light mesons}, Phys. Rev. D 92~(5) (2015)
  054030.
\newblock \href {http://arxiv.org/abs/1506.03686} {\path{arXiv:1506.03686}},
  \href {http://dx.doi.org/10.1103/PhysRevD.92.054030}
  {\path{doi:10.1103/PhysRevD.92.054030}}.

\bibitem{Gomez-Rocha:2016cji}
M.~G{\'o}mez-Rocha, T.~Hilger, A.~Krassnigg, {Effects of a dressed quark-gluon
  vertex in vector heavy-light mesons and theory average of the $B_c^*$ meson
  mass}, Phys. Rev. D 93~(7) (2016) 074010.
\newblock \href {http://arxiv.org/abs/1602.05002} {\path{arXiv:1602.05002}},
  \href {http://dx.doi.org/10.1103/PhysRevD.93.074010}
  {\path{doi:10.1103/PhysRevD.93.074010}}.

\bibitem{Bermudez:2017bpx}
R.~Bermudez, L.~Albino, L.~X. Guti\'errez-Guerrero, M.~E. Tejeda-Yeomans,
  A.~Bashir, {Quark-gluon Vertex: A Perturbation Theory Primer and Beyond},
  Phys. Rev. D 95~(3) (2017) 034041.
\newblock \href {http://arxiv.org/abs/1702.04437} {\path{arXiv:1702.04437}},
  \href {http://dx.doi.org/10.1103/PhysRevD.95.034041}
  {\path{doi:10.1103/PhysRevD.95.034041}}.

\bibitem{Cyrol:2017ewj}
A.~K. Cyrol, M.~Mitter, J.~M. Pawlowski, N.~Strodthoff, {Nonperturbative quark,
  gluon, and meson correlators of unquenched QCD}, Phys. Rev. D 97~(5) (2018)
  054006.
\newblock \href {http://arxiv.org/abs/1706.06326} {\path{arXiv:1706.06326}},
  \href {http://dx.doi.org/10.1103/PhysRevD.97.054006}
  {\path{doi:10.1103/PhysRevD.97.054006}}.

\bibitem{Contant:2018zpi}
R.~Contant, M.~Q. Huber, C.~S. Fischer, C.~A. Welzbacher, R.~Williams, {On the
  quark-gluon vertex at non-vanishing temperature}, Acta Phys. Polon. Supp. 11
  (2018) 483.
\newblock \href {http://arxiv.org/abs/1805.05885} {\path{arXiv:1805.05885}},
  \href {http://dx.doi.org/10.5506/APhysPolBSupp.11.483}
  {\path{doi:10.5506/APhysPolBSupp.11.483}}.

\bibitem{Oliveira:2018fkj}
O.~Oliveira, T.~Frederico, W.~de~Paula, J.~P. B.~C. de~Melo, {Exploring the
  Quark-Gluon Vertex with Slavnov-Taylor Identities and Lattice Simulations},
  Eur. Phys. J. C 78~(7) (2018) 553.
\newblock \href {http://arxiv.org/abs/1807.00675} {\path{arXiv:1807.00675}},
  \href {http://dx.doi.org/10.1140/epjc/s10052-018-6037-0}
  {\path{doi:10.1140/epjc/s10052-018-6037-0}}.

\bibitem{Oliveira:2018ukh}
O.~Oliveira, W.~de~Paula, T.~Frederico, J.~P. B.~C. de~Melo, {The Quark-Gluon
  Vertex and the QCD Infrared Dynamics}, Eur. Phys. J. C 79~(2) (2019) 116.
\newblock \href {http://arxiv.org/abs/1807.10348} {\path{arXiv:1807.10348}},
  \href {http://dx.doi.org/10.1140/epjc/s10052-019-6617-7}
  {\path{doi:10.1140/epjc/s10052-019-6617-7}}.

\bibitem{Sultan:2018tet}
M.~A. Sultan, K.~Raya, F.~Akram, A.~Bashir, B.~Masud, {Effect of the
  quark-gluon vertex on dynamical chiral symmetry breaking}, Phys. Rev. D
  103~(5) (2021) 054036.
\newblock \href {http://arxiv.org/abs/1810.01396} {\path{arXiv:1810.01396}},
  \href {http://dx.doi.org/10.1103/PhysRevD.103.054036}
  {\path{doi:10.1103/PhysRevD.103.054036}}.

\bibitem{Vujinovic:2018nko}
M.~Vujinovic, R.~Alkofer, {Low-energy spectrum of an SU(2) gauge theory with
  dynamical fermions}, Phys. Rev. D 98~(9) (2018) 095030.
\newblock \href {http://arxiv.org/abs/1809.02650} {\path{arXiv:1809.02650}},
  \href {http://dx.doi.org/10.1103/PhysRevD.98.095030}
  {\path{doi:10.1103/PhysRevD.98.095030}}.

\bibitem{Oliveira:2020yac}
O.~Oliveira, T.~Frederico, W.~de~Paula, {The soft-gluon limit and the infrared
  enhancement of the quark-gluon vertex}, Eur. Phys. J. C 80~(5) (2020) 484.
\newblock \href {http://arxiv.org/abs/2006.04982} {\path{arXiv:2006.04982}},
  \href {http://dx.doi.org/10.1140/epjc/s10052-020-8037-0}
  {\path{doi:10.1140/epjc/s10052-020-8037-0}}.

\bibitem{Gao:2021wun}
F.~Gao, J.~Papavassiliou, J.~M. Pawlowski, {Fully coupled functional equations
  for the quark sector of QCD}, Phys. Rev. D 103~(9) (2021) 094013.
\newblock \href {http://arxiv.org/abs/2102.13053} {\path{arXiv:2102.13053}},
  \href {http://dx.doi.org/10.1103/PhysRevD.103.094013}
  {\path{doi:10.1103/PhysRevD.103.094013}}.

\bibitem{El-Bennich:2022obe}
B.~El-Bennich, F.~E. Serna, R.~C. da~Silveira, L.~A.~F. Rangel, A.~Bashir,
  E.~Rojas, {Dressed quark-gluon vertex form factors from gauge symmetry}, Rev.
  Mex. Fis. Suppl. 3~(3) (2022) 0308092.
\newblock \href {http://arxiv.org/abs/2201.04144} {\path{arXiv:2201.04144}},
  \href {http://dx.doi.org/10.31349/SuplRevMexFis.3.0308092}
  {\path{doi:10.31349/SuplRevMexFis.3.0308092}}.

\bibitem{Aguilar:2023mam}
A.~C. Aguilar, M.~.~N. Ferreira, D.~Iba\~nez, J.~Papavassiliou, {Schwinger
  displacement of the quark\textendash{}gluon vertex}, Eur. Phys. J. C 83~(10)
  (2023) 967.
\newblock \href {http://arxiv.org/abs/2308.16297} {\path{arXiv:2308.16297}},
  \href {http://dx.doi.org/10.1140/epjc/s10052-023-12103-8}
  {\path{doi:10.1140/epjc/s10052-023-12103-8}}.

\bibitem{Aguilar:2024ciu}
A.~C. Aguilar, M.~N. Ferreira, B.~M. Oliveira, J.~Papavassiliou, G.~T.
  Linhares, {Infrared properties of the quark-gluon vertex in general
  kinematics}, Eur. Phys. J. C 84~(11) (2024) 1231.
\newblock \href {http://arxiv.org/abs/2408.15370} {\path{arXiv:2408.15370}},
  \href {http://dx.doi.org/10.1140/epjc/s10052-024-13605-9}
  {\path{doi:10.1140/epjc/s10052-024-13605-9}}.

\bibitem{Alkofer:2023lrl}
R.~Alkofer, {Dynamical Chiral Symmetry Breaking in Quantum Chromo Dynamics:
  Delicate and Intricate}, Symmetry 15~(9) (2023) 1787.
\newblock \href {http://arxiv.org/abs/2309.09679} {\path{arXiv:2309.09679}},
  \href {http://dx.doi.org/10.3390/sym15091787}
  {\path{doi:10.3390/sym15091787}}.

\bibitem{Alkofer:2023syz}
R.~Alkofer, F.~J. Llanes-Estrada, A.~Salas-Bernardez, {Spinning pairs:
  Supporting P03 quark-pair creation from Landau-gauge Green\textquoteright{}s
  functions}, Phys. Rev. D 109~(7) (2024) 074015.
\newblock \href {http://arxiv.org/abs/2312.14994} {\path{arXiv:2312.14994}},
  \href {http://dx.doi.org/10.1103/PhysRevD.109.074015}
  {\path{doi:10.1103/PhysRevD.109.074015}}.

\bibitem{Guzman:2025qbq}
V.~M.~B. Guzm{\'a}n, A.~Bashir, {One-loop off-shell quark-gluon vertex in
  arbitrary gauge and dimensions: A streamlined approach through the
  second-order formalism of QCD}, Phys. Rev. D 111~(5) (2025) 056025.
\newblock \href {http://arxiv.org/abs/2503.00350} {\path{arXiv:2503.00350}},
  \href {http://dx.doi.org/10.1103/PhysRevD.111.056025}
  {\path{doi:10.1103/PhysRevD.111.056025}}.

\bibitem{Fu:2025hcm}
W.-j. Fu, C.~Huang, J.~M. Pawlowski, Y.-y. Tan, L.-j. Zhou, {Four-quark
  scatterings in QCD III}, Phys. Rev. D 112~(5) (2025) 054047.
\newblock \href {http://arxiv.org/abs/2502.14388} {\path{arXiv:2502.14388}},
  \href {http://dx.doi.org/10.1103/4sh5-w4yc} {\path{doi:10.1103/4sh5-w4yc}}.

\bibitem{Ferreira:2025wpu}
M.~N. Ferreira, A.~S. Miramontes, J.~M. Morgado, J.~Papavassiliou, J.~M.
  Pawlowski, {Pion physics with dressed quark-gluon vertices}, Eur. Phys. J. C
  86~(4) (2026) 325.
\newblock \href {http://arxiv.org/abs/2512.04853} {\path{arXiv:2512.04853}},
  \href {http://dx.doi.org/10.1140/epjc/s10052-026-15487-5}
  {\path{doi:10.1140/epjc/s10052-026-15487-5}}.

\bibitem{Ferreira:2026gbe}
M.~N. Ferreira, A.~S. Miramontes, J.~M. Morgado, J.~Papavassiliou, {Light
  mesons in the symmetric-vertex approximation,~}\href
  {http://arxiv.org/abs/2604.07221} {\path{arXiv:2604.07221}}.

\bibitem{Alkofer:2003jj}
R.~Alkofer, W.~Detmold, C.~S. Fischer, P.~Maris, {Analytic properties of the
  Landau gauge gluon and quark propagators}, Phys. Rev. D 70 (2004) 014014.
\newblock \href {http://arxiv.org/abs/hep-ph/0309077}
  {\path{arXiv:hep-ph/0309077}}, \href
  {http://dx.doi.org/10.1103/PhysRevD.70.014014}
  {\path{doi:10.1103/PhysRevD.70.014014}}.

\bibitem{Pawlowski:2024kxc}
J.~M. Pawlowski, J.~Wessely, {The causal structure of the quark propagator},
  Eur. Phys. J. C 85~(9) (2025) 970.
\newblock \href {http://arxiv.org/abs/2412.12033} {\path{arXiv:2412.12033}},
  \href {http://dx.doi.org/10.1140/epjc/s10052-025-14683-z}
  {\path{doi:10.1140/epjc/s10052-025-14683-z}}.

\bibitem{MQHgithub}
M.~Q. Huber, {https://github.com/markusqh/YM\_data}.

\bibitem{Huber:2020keu}
M.~Q. Huber, {Correlation functions of Landau gauge Yang-Mills theory}, Phys.
  Rev. D 101 (2020) 114009.
\newblock \href {http://arxiv.org/abs/2003.13703} {\path{arXiv:2003.13703}},
  \href {http://dx.doi.org/10.1103/PhysRevD.101.114009}
  {\path{doi:10.1103/PhysRevD.101.114009}}.

\bibitem{Huber:2018ned}
M.~Q. Huber, {Nonperturbative properties of Yang{\textendash}Mills theories},
  Phys. Rept. 879 (2020) 1--92.
\newblock \href {http://arxiv.org/abs/1808.05227} {\path{arXiv:1808.05227}},
  \href {http://dx.doi.org/10.1016/j.physrep.2020.04.004}
  {\path{doi:10.1016/j.physrep.2020.04.004}}.

\bibitem{Huber:2020ngt}
M.~Q. Huber, C.~S. Fischer, H.~Sanchis-Alepuz, {Spectrum of scalar and
  pseudoscalar glueballs from functional methods}, Eur. Phys. J. C 80~(11)
  (2020) 1077.
\newblock \href {http://arxiv.org/abs/2004.00415} {\path{arXiv:2004.00415}},
  \href {http://dx.doi.org/10.1140/epjc/s10052-020-08649-6}
  {\path{doi:10.1140/epjc/s10052-020-08649-6}}.

\bibitem{Huber:2021yfy}
M.~Q. Huber, C.~S. Fischer, H.~Sanchis-Alepuz, {Higher spin glueballs from
  functional methods}, Eur. Phys. J. C 81~(12) (2021) 1083, [Erratum:
  Eur.Phys.J.C 82, 38 (2022)].
\newblock \href {http://arxiv.org/abs/2110.09180} {\path{arXiv:2110.09180}},
  \href {http://dx.doi.org/10.1140/epjc/s10052-021-09864-5}
  {\path{doi:10.1140/epjc/s10052-021-09864-5}}.

\bibitem{Huber:2025kwy}
M.~Q. Huber, C.~S. Fischer, H.~Sanchis-Alepuz, {Apparent convergence in
  functional glueball calculations}, Eur. Phys. J. C 85~(8) (2025) 859.
\newblock \href {http://arxiv.org/abs/2503.03821} {\path{arXiv:2503.03821}},
  \href {http://dx.doi.org/10.1140/epjc/s10052-025-14590-3}
  {\path{doi:10.1140/epjc/s10052-025-14590-3}}.

\bibitem{Fischer:2008uz}
C.~S. Fischer, A.~Maas, J.~M. Pawlowski, {On the infrared behavior of Landau
  gauge Yang-Mills theory}, Annals Phys. 324 (2009) 2408--2437.
\newblock \href {http://arxiv.org/abs/0810.1987} {\path{arXiv:0810.1987}},
  \href {http://dx.doi.org/10.1016/j.aop.2009.07.009}
  {\path{doi:10.1016/j.aop.2009.07.009}}.

\bibitem{Alkofer:2008jy}
R.~Alkofer, M.~Q. Huber, K.~Schwenzer, {Infrared singularities in Landau gauge
  Yang-Mills theory}, Phys. Rev. D 81 (2010) 105010.
\newblock \href {http://arxiv.org/abs/0801.2762} {\path{arXiv:0801.2762}},
  \href {http://dx.doi.org/10.1103/PhysRevD.81.105010}
  {\path{doi:10.1103/PhysRevD.81.105010}}.

\bibitem{vonSmekal:1997ohs}
L.~von Smekal, R.~Alkofer, A.~Hauck, {The Infrared behavior of gluon and ghost
  propagators in Landau gauge QCD}, Phys. Rev. Lett. 79 (1997) 3591--3594.
\newblock \href {http://arxiv.org/abs/hep-ph/9705242}
  {\path{arXiv:hep-ph/9705242}}, \href
  {http://dx.doi.org/10.1103/PhysRevLett.79.3591}
  {\path{doi:10.1103/PhysRevLett.79.3591}}.

\bibitem{vonSmekal:1997ern}
L.~von Smekal, A.~Hauck, R.~Alkofer, {A Solution to Coupled
  Dyson\textendash{}Schwinger Equations for Gluons and Ghosts in Landau Gauge},
  Annals Phys. 267 (1998) 1--60, [Erratum: Annals Phys. 269, 182 (1998)].
\newblock \href {http://arxiv.org/abs/hep-ph/9707327}
  {\path{arXiv:hep-ph/9707327}}, \href
  {http://dx.doi.org/10.1006/aphy.1998.5806}
  {\path{doi:10.1006/aphy.1998.5806}}.

\bibitem{Zwanziger:2001kw}
D.~Zwanziger, {Nonperturbative Landau gauge and infrared critical exponents in
  QCD}, Phys. Rev. D 65 (2002) 094039.
\newblock \href {http://arxiv.org/abs/hep-th/0109224}
  {\path{arXiv:hep-th/0109224}}, \href
  {http://dx.doi.org/10.1103/PhysRevD.65.094039}
  {\path{doi:10.1103/PhysRevD.65.094039}}.

\bibitem{Lerche:2002ep}
C.~Lerche, L.~von Smekal, {On the infrared exponent for gluon and ghost
  propagation in Landau gauge QCD}, Phys. Rev. D 65 (2002) 125006.
\newblock \href {http://arxiv.org/abs/hep-ph/0202194}
  {\path{arXiv:hep-ph/0202194}}, \href
  {http://dx.doi.org/10.1103/PhysRevD.65.125006}
  {\path{doi:10.1103/PhysRevD.65.125006}}.

\bibitem{Fischer:2002hna}
C.~S. Fischer, R.~Alkofer, {Infrared exponents and running coupling of SU(N)
  Yang-Mills theories}, Phys. Lett. B 536 (2002) 177--184.
\newblock \href {http://arxiv.org/abs/hep-ph/0202202}
  {\path{arXiv:hep-ph/0202202}}, \href
  {http://dx.doi.org/10.1016/S0370-2693(02)01809-9}
  {\path{doi:10.1016/S0370-2693(02)01809-9}}.

\bibitem{Fischer:2005ui}
C.~S. Fischer, B.~Gruter, R.~Alkofer, {Solving coupled Dyson-Schwinger
  equations on a compact manifold}, Annals Phys. 321 (2006) 1918--1938.
\newblock \href {http://arxiv.org/abs/hep-ph/0506053}
  {\path{arXiv:hep-ph/0506053}}, \href
  {http://dx.doi.org/10.1016/j.aop.2005.11.007}
  {\path{doi:10.1016/j.aop.2005.11.007}}.

\bibitem{Huber:2007kc}
M.~Q. Huber, R.~Alkofer, C.~S. Fischer, K.~Schwenzer, {The Infrared behavior of
  Landau gauge Yang-Mills theory in d=2, d=3 and d=4 dimensions}, Phys. Lett. B
  659 (2008) 434--440.
\newblock \href {http://arxiv.org/abs/0705.3809} {\path{arXiv:0705.3809}},
  \href {http://dx.doi.org/10.1016/j.physletb.2007.10.073}
  {\path{doi:10.1016/j.physletb.2007.10.073}}.

\bibitem{Aguilar:2008xm}
A.~C. Aguilar, D.~Binosi, J.~Papavassiliou, {Gluon and ghost propagators in the
  Landau gauge: Deriving lattice results from Schwinger-Dyson equations}, Phys.
  Rev. D 78 (2008) 025010.
\newblock \href {http://arxiv.org/abs/0802.1870} {\path{arXiv:0802.1870}},
  \href {http://dx.doi.org/10.1103/PhysRevD.78.025010}
  {\path{doi:10.1103/PhysRevD.78.025010}}.

\bibitem{Aguilar:2015bud}
A.~C. Aguilar, D.~Binosi, J.~Papavassiliou, {The Gluon Mass Generation
  Mechanism: A Concise Primer}, Front. Phys. (Beijing) 11~(2) (2016) 111203.
\newblock \href {http://arxiv.org/abs/1511.08361} {\path{arXiv:1511.08361}},
  \href {http://dx.doi.org/10.1007/s11467-015-0517-6}
  {\path{doi:10.1007/s11467-015-0517-6}}.

\bibitem{Eichmann:2021zuv}
G.~Eichmann, J.~M. Pawlowski, J.~a.~M. Silva, {Mass generation in Landau-gauge
  Yang-Mills theory}, Phys. Rev. D 104~(11) (2021) 114016.
\newblock \href {http://arxiv.org/abs/2107.05352} {\path{arXiv:2107.05352}},
  \href {http://dx.doi.org/10.1103/PhysRevD.104.114016}
  {\path{doi:10.1103/PhysRevD.104.114016}}.

\bibitem{Maas:2009se}
A.~Maas, {Constructing non-perturbative gauges using correlation functions},
  Phys. Lett. B 689 (2010) 107--111.
\newblock \href {http://arxiv.org/abs/0907.5185} {\path{arXiv:0907.5185}},
  \href {http://dx.doi.org/10.1016/j.physletb.2010.04.052}
  {\path{doi:10.1016/j.physletb.2010.04.052}}.

\bibitem{Blank:2010pa}
M.~Blank, A.~Krassnigg, A.~Maas, {Rho-meson, Bethe-Salpeter equation, and the
  far infrared}, Phys. Rev. D 83 (2011) 034020.
\newblock \href {http://arxiv.org/abs/1007.3901} {\path{arXiv:1007.3901}},
  \href {http://dx.doi.org/10.1103/PhysRevD.83.034020}
  {\path{doi:10.1103/PhysRevD.83.034020}}.

\bibitem{Alkofer:2000wg}
R.~Alkofer, L.~von Smekal, {The Infrared behavior of QCD Green's functions:
  Confinement dynamical symmetry breaking, and hadrons as relativistic bound
  states}, Phys. Rept. 353 (2001) 281.
\newblock \href {http://arxiv.org/abs/hep-ph/0007355}
  {\path{arXiv:hep-ph/0007355}}, \href
  {http://dx.doi.org/10.1016/S0370-1573(01)00010-2}
  {\path{doi:10.1016/S0370-1573(01)00010-2}}.

\bibitem{Fischer:2003rp}
C.~S. Fischer, R.~Alkofer, {Nonperturbative propagators, running coupling and
  dynamical quark mass of Landau gauge QCD}, Phys. Rev. D 67 (2003) 094020.
\newblock \href {http://arxiv.org/abs/hep-ph/0301094}
  {\path{arXiv:hep-ph/0301094}}, \href
  {http://dx.doi.org/10.1103/PhysRevD.67.094020}
  {\path{doi:10.1103/PhysRevD.67.094020}}.

\bibitem{Fischer:2003zc}
C.~S. Fischer, {Nonperturbative propagators, running coupling and dynamical
  mass generation in ghost - anti-ghost symmetric gauges in QCD}, Phd thesis
  (2003).
\newblock \href {http://arxiv.org/abs/hep-ph/0304233}
  {\path{arXiv:hep-ph/0304233}}.

\bibitem{Windisch:2014lce}
A.~Windisch, {Features of strong quark correlations at vanishing and
  non-vanishing density}, Ph.D. thesis, Graz U. (2014).

\bibitem{Hopfer:2014szm}
M.~Hopfer, {Gauge Theories: QCD $\&$ Beyond}: {Aspects of Quark Physics: From
  the Phase Transition to B-Meson Decays}, Ph.D. thesis, Graz U. (2014).

\bibitem{Sanchis-Alepuz:2015tha}
H.~Sanchis-Alepuz, R.~Williams, {Hadronic Observables from Dyson-Schwinger and
  Bethe-Salpeter equations}, J. Phys. Conf. Ser. 631~(1) (2015) 012064.
\newblock \href {http://arxiv.org/abs/1503.05896} {\path{arXiv:1503.05896}},
  \href {http://dx.doi.org/10.1088/1742-6596/631/1/012064}
  {\path{doi:10.1088/1742-6596/631/1/012064}}.

\bibitem{Williams:2015cvx}
R.~Williams, C.~S. Fischer, W.~Heupel, {Light mesons in QCD and unquenching
  effects from the 3PI effective action}, Phys. Rev. D 93~(3) (2016) 034026.
\newblock \href {http://arxiv.org/abs/1512.00455} {\path{arXiv:1512.00455}},
  \href {http://dx.doi.org/10.1103/PhysRevD.93.034026}
  {\path{doi:10.1103/PhysRevD.93.034026}}.

\bibitem{Eichmann:2016yit}
G.~Eichmann, H.~Sanchis-Alepuz, R.~Williams, R.~Alkofer, C.~S. Fischer,
  {Baryons as relativistic three-quark bound states}, Prog. Part. Nucl. Phys.
  91 (2016) 1--100.
\newblock \href {http://arxiv.org/abs/1606.09602} {\path{arXiv:1606.09602}},
  \href {http://dx.doi.org/10.1016/j.ppnp.2016.07.001}
  {\path{doi:10.1016/j.ppnp.2016.07.001}}.

\bibitem{MQHunpublished}
M.~Q. Huber~et al., in preparation (2026).

\bibitem{code_release}
G.~Wieland, R.~Alkofer, {No planar degeneracy for the Landau gauge quark-gluon
  vertex - code release} (2026).
\newblock \href {http://dx.doi.org/10.5281/zenodo.19676273}
  {\path{doi:10.5281/zenodo.19676273}}.

\bibitem{data_release}
G.~Wieland, R.~Alkofer, {No planar degeneracy for the Landau gauge quark-gluon
  vertex - data release} (2026).
\newblock \href {http://dx.doi.org/10.5281/zenodo.19676105}
  {\path{doi:10.5281/zenodo.19676105}}.

\bibitem{Pagels:1979hd}
H.~Pagels, S.~Stokar, {The Pion Decay Constant, Electromagnetic Form-Factor and
  Quark Electromagnetic Selfenergy in QCD}, Phys. Rev. D 20 (1979) 2947.
\newblock \href {http://dx.doi.org/10.1103/PhysRevD.20.2947}
  {\path{doi:10.1103/PhysRevD.20.2947}}.

\bibitem{Oliveira:2025boh}
O.~Oliveira, T.~Frederico, W.~de~Paula, {On the momentum space structure of the
  quark propagator}, Eur. Phys. J. C 85~(3) (2025) 280.
\newblock \href {http://arxiv.org/abs/2502.18335} {\path{arXiv:2502.18335}},
  \href {http://dx.doi.org/10.1140/epjc/s10052-025-13984-7}
  {\path{doi:10.1140/epjc/s10052-025-13984-7}}.

\bibitem{Alkofer:2026vux}
R.~Alkofer, C.~S. Fischer, F.~Zierler, {Chiral symmetry restoration effects
  onto the meson spectrum from a Dyson-Schwinger/Bethe-Salpeter approach,
  }\href {http://arxiv.org/abs/2602.17456} {\path{arXiv:2602.17456}}.

\bibitem{Tripolt:2018xeo}
R.-A. Tripolt, P.~Gubler, M.~Ulybyshev, L.~Von~Smekal, {Numerical analytic
  continuation of Euclidean data}, Comput. Phys. Commun. 237 (2019) 129--142.
\newblock \href {http://arxiv.org/abs/1801.10348} {\path{arXiv:1801.10348}},
  \href {http://dx.doi.org/10.1016/j.cpc.2018.11.012}
  {\path{doi:10.1016/j.cpc.2018.11.012}}.

\bibitem{Huber:2023uzd}
M.~Q. Huber, W.~J. Kern, R.~Alkofer, {How to Determine the Branch Points of
  Correlation Functions in Euclidean Space II: Three-Point Functions}, Symmetry
  15~(2) (2023) 414.
\newblock \href {http://arxiv.org/abs/2302.01350} {\path{arXiv:2302.01350}},
  \href {http://dx.doi.org/10.3390/sym15020414}
  {\path{doi:10.3390/sym15020414}}.

\bibitem{Hopfer:2014zna}
M.~Hopfer, C.~S. Fischer, R.~Alkofer, {Running coupling in the conformal window
  of large-Nf QCD}, JHEP 11 (2014) 035.
\newblock \href {http://arxiv.org/abs/1405.7031} {\path{arXiv:1405.7031}},
  \href {http://dx.doi.org/10.1007/JHEP11(2014)035}
  {\path{doi:10.1007/JHEP11(2014)035}}.

\bibitem{Zierler:2023qvz}
F.~Zierler, R.~Alkofer, {Dependence of the Landau gauge ghost-gluon-vertex on
  the number of flavors}, Phys. Rev. D 109~(7) (2024) 074024.
\newblock \href {http://arxiv.org/abs/2312.06463} {\path{arXiv:2312.06463}},
  \href {http://dx.doi.org/10.1103/PhysRevD.109.074024}
  {\path{doi:10.1103/PhysRevD.109.074024}}.

\bibitem{Eichmann:2018ytt}
G.~Eichmann, G.~Ramalho, {Nucleon resonances in Compton scattering}, Phys. Rev.
  D 98~(9) (2018) 093007.
\newblock \href {http://arxiv.org/abs/1806.04579} {\path{arXiv:1806.04579}},
  \href {http://dx.doi.org/10.1103/PhysRevD.98.093007}
  {\path{doi:10.1103/PhysRevD.98.093007}}.

\bibitem{Kizilersu:1995iz}
A.~Kizilersu, M.~Reenders, M.~R. Pennington, {One loop QED vertex in any
  covariant gauge: Its complete analytic form}, Phys. Rev. D 52 (1995)
  1242--1259.
\newblock \href {http://arxiv.org/abs/hep-ph/9503238}
  {\path{arXiv:hep-ph/9503238}}, \href
  {http://dx.doi.org/10.1103/PhysRevD.52.1242}
  {\path{doi:10.1103/PhysRevD.52.1242}}.

\end{thebibliography}

\end{document}